\documentclass[a4paper,11pt]{article}
\pdfoutput=1
\PassOptionsToPackage{hyphens}{url}
\usepackage{jheppub}
\usepackage[T1]{fontenc}
\usepackage[ddmmyy,24hr]{datetime}
\usepackage{bigdelim}
\usepackage{booktabs}
\usepackage{dcolumn}
\usepackage{multirow}
\usepackage{subfigure}
\usepackage{cancel}
\usepackage{stackrel}
\usepackage{paralist}
\usepackage{xspace}
\newcommand{\nua}[1]{\ensuremath{\rlap{\kern-2.5pt\ensuremath{\overset{\scriptscriptstyle(-)}{\phantom{\nu}}}}{\ensuremath{{\nu}_{#1}}}}\xspace}

\title{\boldmath Updated Global 3+1 Analysis of Short-BaseLine Neutrino Oscillations}

\author[a]{S. Gariazzo,}
\author[b]{C. Giunti,}
\author[c,d]{M. Laveder,}
\author[e]{and Y.F. Li}

\affiliation[a]{Instituto de F\'isica Corpuscular (CSIC-Universitat de Val\`encia), Paterna (Valencia), Spain}
\affiliation[b]{INFN, Sezione di Torino, Via P. Giuria 1, I--10125 Torino, Italy}
\affiliation[c]{Dipartimento di Fisica e Astronomia ``G. Galilei'', Universit\`a di Padova, Italy}
\affiliation[d]{INFN, Sezione di Padova, Via F. Marzolo 8, I--35131 Padova, Italy}
\affiliation[e]{Institute of High Energy Physics, Chinese Academy of Sciences, Beijing 100049, China}

\emailAdd{gariazzo@ific.uv.es}
\emailAdd{giunti@to.infn.it}
\emailAdd{laveder@pd.infn.it}
\emailAdd{liyufeng@ihep.ac.cn}

\abstract{
We present the results of an updated fit of short-baseline neutrino oscillation data
in the framework of 3+1 active-sterile neutrino mixing.
We first consider
$\nu_{e}$ and $\bar\nu_{e}$ disappearance
in the light of the Gallium and reactor anomalies.
We discuss the implications of the recent
measurement of the reactor $\bar\nu_{e}$ spectrum in the NEOS experiment,
which shifts the allowed regions of the parameter space towards smaller values of
$|U_{e4}|^2$.
The $\beta$-decay constraints of the
Mainz
and
Troitsk
experiments allow us to limit the oscillation length
between about
$2$ cm
and
$7$ m
at $3\sigma$
for neutrinos with an energy of 1 MeV.
The corresponding oscillations can be discovered in a model-independent way
in ongoing reactor and source experiments
by measuring $\nu_{e}$ and $\bar\nu_{e}$ disappearance
as a function of distance.
We then consider the global fit of the data on short-baseline
$\nua{\mu}\to\nua{e}$ transitions
in the light of the LSND anomaly,
taking into account the constraints from
$\nua{e}$ and $\nua{\mu}$ disappearance experiments,
including the recent data of the MINOS and IceCube experiments.
The combination of
the NEOS constraints on $|U_{e4}|^2$
and
the MINOS and IceCube constraints on $|U_{\mu4}|^2$
lead to an unacceptable appearance-disappearance tension
which becomes tolerable only in a pragmatic fit
which neglects the MiniBooNE low-energy anomaly.
The minimization of the global $\chi^2$ in the
space of the four mixing parameters
$\Delta{m}^2_{41}$,
$|U_{e4}|^2$,
$|U_{\mu4}|^2$, and
$|U_{\tau4}|^2$
leads to three allowed regions
with narrow
$\Delta{m}^{2}_{41}$ widths at
$ \Delta{m}^{2}_{41}
\approx
1.7
\, (\text{best-fit})
,\,
1.3
\, (\text{at}\,2\sigma)
,\,
2.4
\, (\text{at}\,3\sigma)
\, \text{eV}^2 $.
The effective amplitude of short-baseline
$\nua{\mu}\to\nua{e}$ oscillations is limited by
$ 0.00048
\lesssim
\sin^22\vartheta_{e\mu}
\lesssim
0.0020 $
at $3\sigma$.
The restrictions of the allowed regions of the mixing parameters
with respect to our previous global fits
are mainly due to the NEOS constraints.
We present a comparison of the allowed regions of the mixing parameters
with the sensitivities of ongoing experiments,
which show that it is likely that
these experiments will determine in a definitive way if
the reactor, Gallium and LSND anomalies are due to active-sterile neutrino oscillations or not.
}

\begin{document}

\maketitle

\section{Introduction}
\label{sec:introduction}

Neutrino physics is a powerful probe of new physics beyond the Standard Model.
The
LSND \cite{Athanassopoulos:1995iw,Aguilar:2001ty},
Gallium \cite{Abdurashitov:2005tb,Laveder:2007zz,Giunti:2006bj,Giunti:2010zu,Giunti:2012tn}
and
reactor \cite{Mention:2011rk}
anomalies
are intriguing indications in favor of the existence of light sterile neutrinos
connected with low-energy new physics.
In order to assess the viability of the light sterile neutrino hypothesis,
it is necessary to perform a global fit of neutrino oscillation data
which takes into account not only the
LSND, Gallium and reactor anomalies,
but also the data of many other experiments
which constrain active-sterile neutrino mixing
(see the reviews in Refs.~\cite{Bilenky:1998dt,GonzalezGarcia:2007ib,Conrad:2012qt,Gariazzo:2015rra}).

In this paper we consider 3+1 active-sterile neutrino mixing,
in which the three standard active neutrinos
$\nu_{e}$,
$\nu_{\mu}$,
$\nu_{\tau}$
are mainly composed of three sub-eV massive neutrinos
$\nu_{1}$,
$\nu_{2}$,
$\nu_{3}$
and there is a sterile neutrino
$\nu_{s}$
which is mainly composed of a fourth massive neutrino
$\nu_{4}$
at the eV scale.
This is the only allowed four-neutrino mixing scheme after the demise of the 2+2 scheme
\cite{Maltoni:2004ei}
and the fact that the 1+3 scheme with three massive neutrinos at the eV scale
is strongly disfavored by cosmological measurements
\cite{Ade:2015xua}
and by the experimental bounds on
neutrinoless double-$\beta$ decay
if massive neutrinos are Majorana particles
(see Refs.~\cite{Bilenky:2014uka,DellOro:2016tmg}).
We do not consider neutrino mixing schemes with more than one sterile neutrino,
which are not necessary to explain the current data
(see the discussions in Refs.~\cite{Gariazzo:2015rra,Giunti:2015mwa}).

In the framework of 3+1 active-sterile mixing,
short-baseline (SBL) experiments are sensitive only to the oscillations
generated by the squared-mass difference
$\Delta m^2_{41} \simeq \Delta m^2_{42} \simeq \Delta m^2_{43} \gtrsim 1 \, \text{eV}^2$,
with
$\Delta m^2_{jk} \equiv m_{j}^2 - m_{k}^2$,
that is much larger than the
the solar squared-mass difference
$\Delta m^2_{\text{SOL}} = \Delta m^2_{21} \approx 7.4 \times 10^{-5} \, \text{eV}^2$
and the atmospheric squared-mass difference
$\Delta m^2_{\text{ATM}} = |\Delta m^2_{31}| \simeq |\Delta m^2_{32}| \approx 2.5 \times 10^{-3} \, \text{eV}^2$,
which generate the observed solar, atmospheric and long-baseline neutrino oscillations
explained by the standard three-neutrino mixing
(see Refs.~\cite{Capozzi:2016rtj,Esteban:2016qun}).
The 3+1 active-sterile mixing scheme is a perturbation of
the standard three-neutrino mixing
in which the $3\times3$ unitary mixing matrix $U$ is extended to a $4\times4$ unitary mixing matrix
with
$|U_{e4}|^2 , \, |U_{\mu4}|^2 , \, |U_{\tau4}|^2 \ll 1 $.
The effective oscillation probabilities of the flavor neutrinos in short-baseline experiments
are given by~\cite{Bilenky:1996rw}
\begin{equation}
P_{\alpha\beta}^{(\text{SBL})}
\simeq
\left|
\delta_{\alpha\beta}
-
\sin^2 2\vartheta_{\alpha\beta}
\sin^{2}\!\left( \frac{\Delta{m}^2_{41}L}{4E} \right)
\right|,
\label{prb3p1}
\end{equation}
where $\alpha,\beta =e,\mu,\tau,s$,
$L$ is the source-detector distance and $E$ is the neutrino energy.
The short-baseline oscillation amplitudes depend only on the absolute values of the
elements in the fourth column of the mixing matrix:
\begin{equation}
\sin^2 2\vartheta_{\alpha\beta}
=
4
|U_{\alpha 4}|^2
\left| \delta_{\alpha\beta} -  |U_{\beta 4}|^2 \right|
.
\label{amp3p1}
\end{equation}
Hence, the transition probabilities of neutrinos and antineutrinos are equal
and it is not possible to measure in short-baseline experiments
any CP-violating effect generated by the complex phases
in the mixing matrix\footnote{%
CP violating effects due to active-sterile neutrino mixing
can, however, be observed in
long-baseline
\cite{deGouvea:2014aoa,Klop:2014ima,Berryman:2015nua,Gandhi:2015xza,Palazzo:2015gja,Agarwalla:2016mrc,Agarwalla:2016xxa,Choubey:2016fpi,Agarwalla:2016xlg,Capozzi:2016vac}
and
solar
\cite{Long:2013hwa}
neutrino experiments.
}.

In this paper we update the analysis of short-baseline neutrino oscillation data
\cite{Giunti:2012tn,Giunti:2012bc,Giunti:2013aea,Gariazzo:2015rra}
revising the analysis of the rates measured in reactor neutrino experiments
according to Ref.~\cite{Giunti:2016elf}
and
taking into account the recent measurements of the
MINOS \cite{MINOS:2016viw},
IceCube \cite{TheIceCube:2016oqi}, and
NEOS \cite{Ko:2016owz}
experiments.
The MINOS and IceCube constraints on
$\nu_{\mu}$ and $\bar\nu_{\mu}$ disappearance
are expected \cite{Giunti:2016oan}
to disfavor
the low-$\Delta{m}^{2}_{41}$--high-$\sin^{2}2\vartheta_{\mu\mu}$
and
the low-$\Delta{m}^{2}_{41}$--high-$\sin^{2}2\vartheta_{e\mu}$
parts of the allowed region that we found in our previous analyses
\cite{Giunti:2012tn,Giunti:2012bc,Giunti:2013aea,Gariazzo:2015rra},
as was found in the 3+1 global fit presented in Ref.~\cite{Collin:2016aqd},
which updated Ref.~\cite{Collin:2016rao}
with the addition of the IceCube data.
The NEOS \cite{Ko:2016owz} collaboration
measured the spectrum of reactor $\bar\nu_{e}$'s
at a distance of 24 m and normalized their data to the
Daya Bay spectrum
\cite{An:2016srz}
measured at the large distance of about 550 m,
where short-baseline oscillations are averaged out.
Analyzing this normalized spectrum with short-baseline oscillations they found the best fit at
$\Delta{m}^{2}_{41} = 1.73 \, \text{eV}^2$
and
$\sin^22\vartheta_{ee} = 0.05$,
with a $\chi^2$ which is lower by 6.5 with respect to the
standard case of three-neutrino mixing without short-baseline oscillations.
This is a $2.1\sigma$ indication in favor of
short-baseline oscillations
and it is intriguing to note that the
best-fit value of $\Delta{m}^{2}_{41}$ is close to the best-fit value
found in our previous global analysis of short-baseline data
\cite{Gariazzo:2015rra},
$\Delta{m}^{2}_{41} = 1.6 \, \text{eV}^2$,
albeit with a larger
$\sin^22\vartheta_{ee} = 0.11$.
However, as one can see from Table~5 of Ref.~\cite{Gariazzo:2015rra},
the lower bound of the $3\sigma$ allowed range of $\sin^22\vartheta_{ee}$
was 0.046,
which is below the NEOS best-fit value.
Hence, the NEOS data are not incompatible with the global indications
of short-baseline oscillations and
we expect that their inclusion in the fit
will lead to a shift of the allowed region towards smaller values
of $\sin^22\vartheta_{ee}$ and, consequently, of $|U_{e4}|^2$.

It is well known
\cite{Okada:1996kw,Bilenky:1996rw,Kopp:2011qd,Giunti:2011gz,Giunti:2011hn,Giunti:2011cp,Conrad:2012qt,Archidiacono:2012ri,Archidiacono:2013xxa,Kopp:2013vaa,Giunti:2013aea,Gariazzo:2015rra}
that the global fits of short-baseline data
are affected by the so-called
``appearance-disappearance'' tension
which is present
\cite{Giunti:2015mwa}
for any number $N_{s}$ of sterile neutrinos
in 3+$N_{s}$ mixing schemes which are perturbations
of the standard three-neutrino mixing
required for the explanation of
the observation of solar, atmospheric and long-baseline neutrino oscillations
(see Refs.~\cite{Capozzi:2016rtj,Esteban:2016qun}).
We expect that the inclusion in the global fit of the
recent measurements of the
MINOS \cite{MINOS:2016viw},
IceCube \cite{TheIceCube:2016oqi}, and
NEOS \cite{Ko:2016owz}
experiment will increase somewhat the appearance-disappearance tension.
In Ref.~\cite{Giunti:2013aea}
we proposed a ``pragmatic approach''
in which the appearance-disappearance tension is alleviated
by excluding from the global fit
the low-energy bins of the MiniBooNE experiment
\cite{AguilarArevalo:2008rc,Aguilar-Arevalo:2013pmq}
which have an anomalous excess of $\nu_{e}$-like events
that is under investigation in the
MicroBooNE
experiment at Fermilab
\cite{Gollapinni:2015lca}.
In this paper we will discuss
the effect of MINOS, IceCube and NEOS data on
the appearance-disappearance tension
and how much it is alleviated in the
pragmatic approach.

The plan of the paper is as follows.
In section~\ref{sec:nuedis}
we consider the experimental data on
short-baseline $\nu_{e}$ and $\bar\nu_{e}$ disappearance,
motivated by the Gallium and reactor anomalies.
In section~\ref{sec:fits}
we consider the global fit of appearance and disappearance data,
which is motivated by the addition of the LSND anomaly to the Gallium and reactor anomalies.
We draw our conclusions
in section~\ref{sec:conclusions}.

\section{$\boldsymbol{\nu_{e}}$ and $\boldsymbol{\bar\nu_{e}}$ disappearance}
\label{sec:nuedis}

In this section we consider only the experimental data
on short-baseline $\nu_{e}$ and $\bar\nu_{e}$ disappearance,
which include
the Gallium neutrino anomaly
\cite{Abdurashitov:2005tb,Laveder:2007zz,Giunti:2006bj,Giunti:2010zu,Giunti:2012tn}
and
the reactor antineutrino anomaly~\cite{Mention:2011rk}.
First, we discuss in subsection~\ref{sub:rates}
our evaluation of the reactor antineutrino anomaly
by considering the updated results of the
reactor $\bar\nu_{e}$ rates measured in several
reactor neutrino experiments.
In subsection~\ref{sub:spectra}
we add the constraints of the spectra
measured in the old Bugey-3 experiment \cite{Declais:1995su}
and in the recent NEOS experiment \cite{Ko:2016owz}.
Finally,
in subsection~\ref{sub:nuedis}
we present our results
for the combined fit of reactor and Gallium data
and
for the global fit of all the
$\nu_{e}$ and $\bar\nu_{e}$ disappearance
data.

\subsection{Reactor rates}
\label{sub:rates}

The reactor neutrino experiments
which measured the absolute antineutrino flux
that are considered in our analysis\footnote{%
We do not consider the still preliminary data of the
Neutrino-4 experiment~\cite{Serebrov:2017nxa}.
}
are listed in Table~\ref{tab:rat}.
For each experiment labeled with the index $a$,
we listed the corresponding four fission fractions $f^{a}_{k}$,
the ratio of measured and predicted rates $R_{a}^{\text{exp}}$,
the corresponding relative experimental uncertainty
$\sigma_{a}^{\text{exp}}$,
the relative uncertainty
$\sigma_{a}^{\text{cor}}$
which is correlated in each group of experiments
indicated by the braces,
and
the relative theoretical uncertainty
$\sigma_{a}^{\text{the}}$
which is correlated among all the experiments.
The ratios
$R_{a}^{\text{exp}}$
and the uncertainties
$\sigma_{a}^{\text{exp}}$
and
$\sigma_{a}^{\text{cor}}$
are the same as those in Ref.~\cite{Giunti:2016elf}.
In the following we repeat for convenience\footnote{%
We also correct, in Table~\ref{tab:rat},
the misprints of the Rovno88 correlations in Tab.~2 of Ref.~\cite{Giunti:2016elf}.
}
their derivation and
we explain the derivation of the relative theoretical uncertainty
$\sigma_{a}^{\text{the}}$.

The ratios of measured and predicted rates of the short-baseline experiments
Bugey-4 \cite{Declais:1994ma},
Rovno91 \cite{Kuvshinnikov:1990ry},
Bugey-3 \cite{Declais:1995su},
Gosgen \cite{Zacek:1986cu},
ILL \cite{Kwon:1981ua,Hoummada:1995zz},
Krasnoyarsk87 \cite{Vidyakin:1987ue},
Krasnoyarsk94 \cite{Vidyakin:1990iz,Vidyakin:1994ut},
Rovno88 \cite{Afonin:1988gx}, and
SRP \cite{Greenwood:1996pb}
have been calculated by the Saclay group in Ref.~\cite{Mention:2011rk}.
The calculation of the
${}^{235}\text{U}$,
${}^{239}\text{Pu}$, and
${}^{241}\text{Pu}$
antineutrino fluxes was subsequently improved by P. Huber in \cite{Huber:2011wv}.
We took into account this correction
with the following rescaling of the Saclay ratios\footnote{%
The missing index $a=18$ corresponds to the
Krasnoyarsk99-34
experiment discussed below.
}:
\begin{equation}
R_{a}^{\text{exp}}
=
R_{a,\text{S}}^{\text{exp}}
\dfrac{\sum_{k} f^{a}_{k} \sigma_{f,k}^{\text{S}}}{\sum_{k} f^{a}_{k} \sigma_{f,k}^{\text{SH}}}
\quad
(a=1,\ldots,17,19,20)
,
\label{rescaling}
\end{equation}
where
$\sigma_{f,k}^{\text{S}}$
and
$\sigma_{f,k}^{\text{SH}}$
are, respectively,
the
Saclay \cite{Mention:2011rk}
and
Saclay+Huber \cite{Huber:2011wv}
cross sections per fission
given in Tab.~\ref{tab:csf}.
The index $k = 235, 238, 239, 241$
indicates the four fissionable isotopes
$^{235}\text{U}$,
$^{238}\text{U}$,
$^{239}\text{Pu}$, and
$^{241}\text{Pu}$
which constitute the reactor fuel.

We considered the Krasnoyarsk99-34 experiment \cite{Kozlov:1999ct}
that was not considered in Refs.~\cite{Mention:2011rk,Zhang:2013ela},
by rescaling the value of the corresponding experimental cross section per fission
in comparison with the Krasnoyarsk94-57 result.
For the long-baseline experiments
Chooz \cite{Apollonio:2002gd}
and
Palo Verde \cite{Boehm:2001ik},
we applied the rescaling in Eq.~(\ref{rescaling})
with the ratios $R_{a,\text{S}}^{\text{exp}}$
given in Ref.~\cite{Zhang:2013ela},
divided by the corresponding
survival probability $P_{\text{sur}}$
caused by $\vartheta_{13}$.
For
Nucifer
\cite{Boireau:2015dda},
Daya Bay
\cite{An:2016srz},
RENO
\cite{RENO-AAP2016},
and
Double Chooz\footnote{Double Chooz Collaboration, Private Communication.}
we use the ratios provided by the respective experimental collaborations.

The experimental uncertainties and their correlations listed in Table~\ref{tab:rat}
have been obtained from the corresponding experimental papers.
In particular:
\begin{itemize}

\item
The Bugey-4 and Rovno91 experiments
have a correlated 1.4\% uncertainty,
because they used the same detector
\cite{Declais:1994ma}.

\item
The Rovno88 experiments
have a correlated 2.2\% reactor-related uncertainty
\cite{Afonin:1988gx}.
In addition,
each of the two groups
of integral (Rovno88-1I and Rovno88-2I)
and spectral (Rovno88-1S, Rovno88-2S, and Rovno88-3S)
measurements
have a correlated 3.1\% detector-related uncertainty
\cite{Afonin:1988gx}.

\item
The Bugey-3 experiments
have a correlated 4.0\% uncertainty
obtained from Tab.~9 of \cite{Declais:1994ma}.

\item
The Gosgen and ILL experiments
have a correlated 3.8\% uncertainty,
because they used the same detector
\cite{Zacek:1986cu}.
In addition, the Gosgen experiments
have a correlated 2.0\% reactor-related uncertainty
\cite{Zacek:1986cu}.

\item
The 1987 Krasnoyarsk87-33 and Krasnoyarsk87-92 experiments
have a correlated 4.1\% uncertainty, because they used the same detector
at 32.8 and 92.3 m from two reactors \cite{Vidyakin:1987ue}.
The Krasnoyarsk94-57 experiment was performed in 1990-94 with a different detector at
57.0 and 57.6 m from the same two reactors
\cite{Vidyakin:1990iz}.
The Krasnoyarsk99-34 experiment was performed in 1997-99 with a new integral-type detector
at 34 m from the same reactor of the Krasnoyarsk87-33 experiment
\cite{Kozlov:1999cs}.
There may be reactor-related uncertainties correlated among the four
Krasnoyarsk experiments,
but,
taking into account the time separations and the absence of any information,
we conservatively neglected them.

\item
Following Ref.~\cite{Zhang:2013ela},
we considered the two SRP measurements
as uncorrelated,
because the two measurements would be incompatible
with the correlated uncertainty estimated in Ref.~\cite{Greenwood:1996pb}.

\end{itemize}

For each experiment labeled with the index $a$,
the relative theoretical uncertainty
$\sigma_{a}^{\text{the}}$ in Table~\ref{tab:rat}
is given by
\begin{equation}
\sigma_{a}^{\text{the}}
=
\dfrac
{\sqrt{\sum_{k,j} f^{a}_{k} \rho^{\text{SH}}_{kj} f^{a}_{j} }}
{\sum_{k} f^{a}_{k} \sigma_{f,k}^{\text{SH}}}
,
\label{reltheunc}
\end{equation}
where $\rho^{\text{SH}}$
is the covariance matrix of the
cross sections per fission of the four fissionable isotopes
given in Tab.~\ref{tab:cov}.
In this covariance matrix,
$\sigma_{f,238}^{\text{SH}}$
is uncorrelated from the other
cross sections per fission
and the corresponding uncertainty
is that given in Ref.~\cite{Mention:2011rk}
(we neglected the correlation due to the cross section uncertainty,
which is of the order of $0.1\%$).
The other three cross sections per fission
have been calculated using the Huber \cite{Huber:2011wv}
antineutrino fluxes which have been obtained
by inverting the spectra of the electrons
emitted by the $\beta$ decays of the products of the thermal fission of
$^{235}\text{U}$,
$^{239}\text{Pu}$, and
$^{241}\text{Pu}$
which have been measured at ILL in the 80's
\cite{Schreckenbach:1985ep,Hahn:1989zr,Haag:2014kia}.
As explained in Ref.~\cite{Huber:2011wv},
the values of the three antineutrino fluxes
are correlated.
We calculated the uncertainties and correlations of
$\sigma_{f,235}^{\text{SH}}$,
$\sigma_{f,239}^{\text{SH}}$, and
$\sigma_{f,241}^{\text{SH}}$
using the information given in Ref.~\cite{Huber:2011wv}.
The square roots of the diagonal elements of the covariance matrix
$\rho^{\text{SH}}$
in Tab.~\ref{tab:cov} give the uncertainties of the cross sections per fission
reported in Tab.~\ref{tab:csf}.
One can see that the uncertainties of
$\sigma_{f,235}^{\text{SH}}$,
$\sigma_{f,239}^{\text{SH}}$, and
$\sigma_{f,241}^{\text{SH}}$
are slightly larger than those calculated
by Saclay group in Ref.~\cite{Mention:2011rk}.

Let us note that after the discovery of the unexpected ``5 MeV bump''
in the spectrum of the
RENO \cite{RENO:2015ksa},
Double Chooz \cite{Abe:2014bwa},
and
Daya Bay \cite{An:2015nua}
experiments
it is believed
\cite{Huber:2016fkt,Hayes:2016qnu}
that the theoretical uncertainties of the
reactor antineutrino fluxes
may be larger than those calculated in Refs.~\cite{Mueller:2011nm,Huber:2011wv}.
However,
since there is no well-motivated quantitative estimation of how much the
theoretical uncertainties should be increased,
we are compelled to use the uncertainties calculated in Refs.~\cite{Mueller:2011nm,Huber:2011wv}.

\begin{figure}[!t]
\centering
\includegraphics*[width=\linewidth]{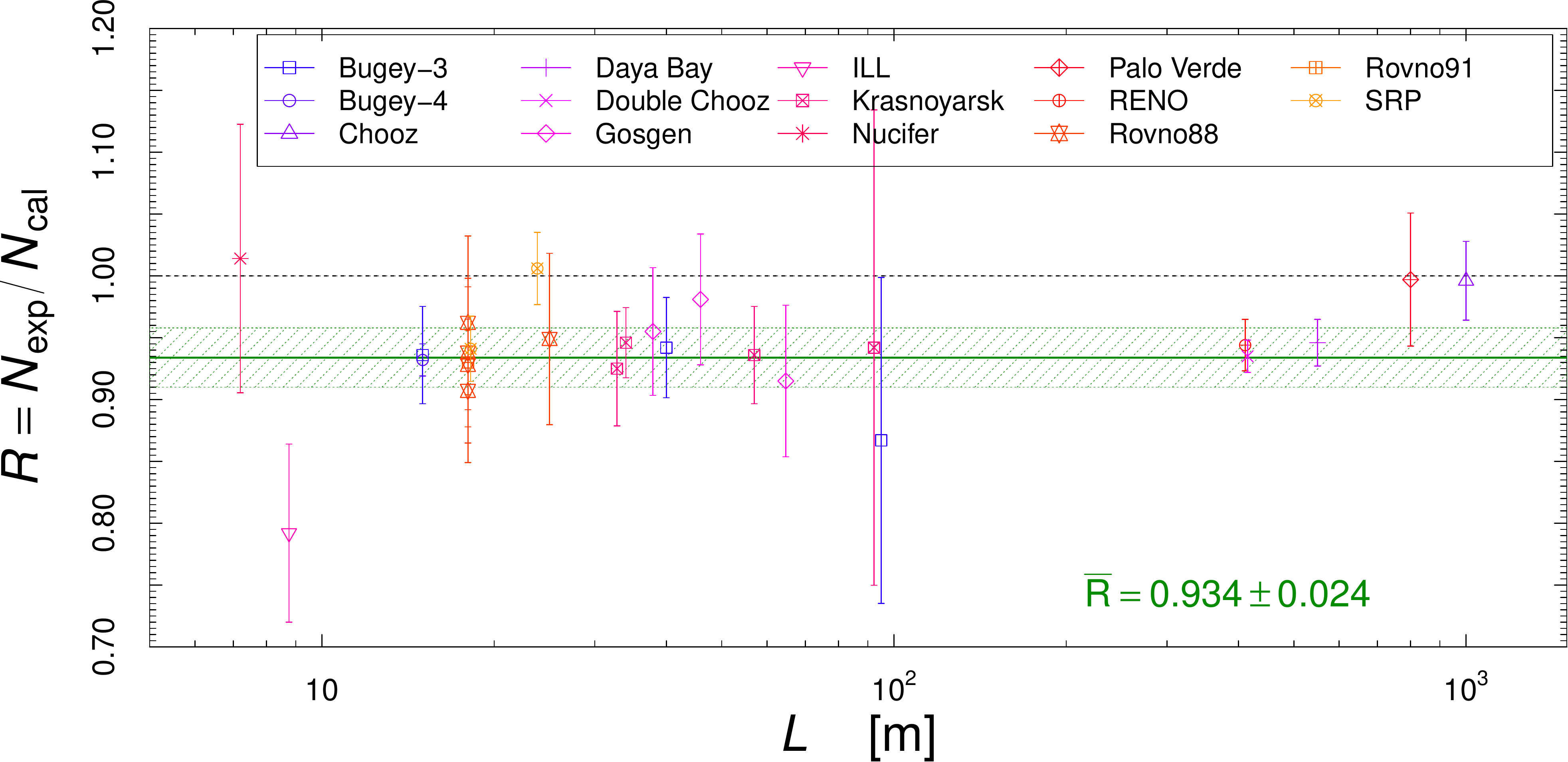}
\caption{ \label{fig:rea-avg}
Ratios $R$ of the reactor experiments considered in our analysis
as functions of the reactor-detector distance $L$.
The horizontal band shows the average ratio $\overline{R}$ and its uncertainty.
The error bars show the experimental uncertainties.
}
\end{figure}

Figure~\ref{fig:rea-avg}
shows the experimental ratios as functions of the reactor-detector distance $L$.
The horizontal band shows the average ratio $\overline{R}$ and its uncertainty,
\begin{equation}
\overline{R}
=
0.934 \pm 0.024
,
\label{rea-avg}
\end{equation}
which has been obtained by summing in quadrature the experimental and theoretical
uncertainties.
Hence, the reactor antineutrino anomaly is at the level of about
$2.8\sigma$.

The slight difference of the value of $\overline{R}$ in Eq.~(\ref{rea-avg})
with respect to our previous estimates in Refs.~\cite{Giunti:2012tn,Gariazzo:2015rra}
is due to the following three changes in our analysis:

\begin{enumerate}

\item
The revaluation \cite{Giunti:2016elf} of the experimental ratios
$R_{a}^{\text{exp}}$
listed in Table~\ref{tab:rat}.

\item
The new treatment of the theoretical uncertainties $\sigma_{a}^{\text{the}}$ according to
Eq.~(\ref{reltheunc})
instead of considering an unrealistic common 2.0\% \cite{Mention:2011rk}.

\item
The new data of the
Nucifer, Daya Bay, RENO and Double Chooz experiments
and the consideration for the first time of the
Krasnoyarsk99-34 experiment.

\end{enumerate}

\begin{figure}[!t]
\centering
\setlength{\tabcolsep}{0pt}
\begin{tabular}{cc}
\subfigure[]{\label{fig:rea-rat}
\includegraphics*[width=0.49\linewidth]{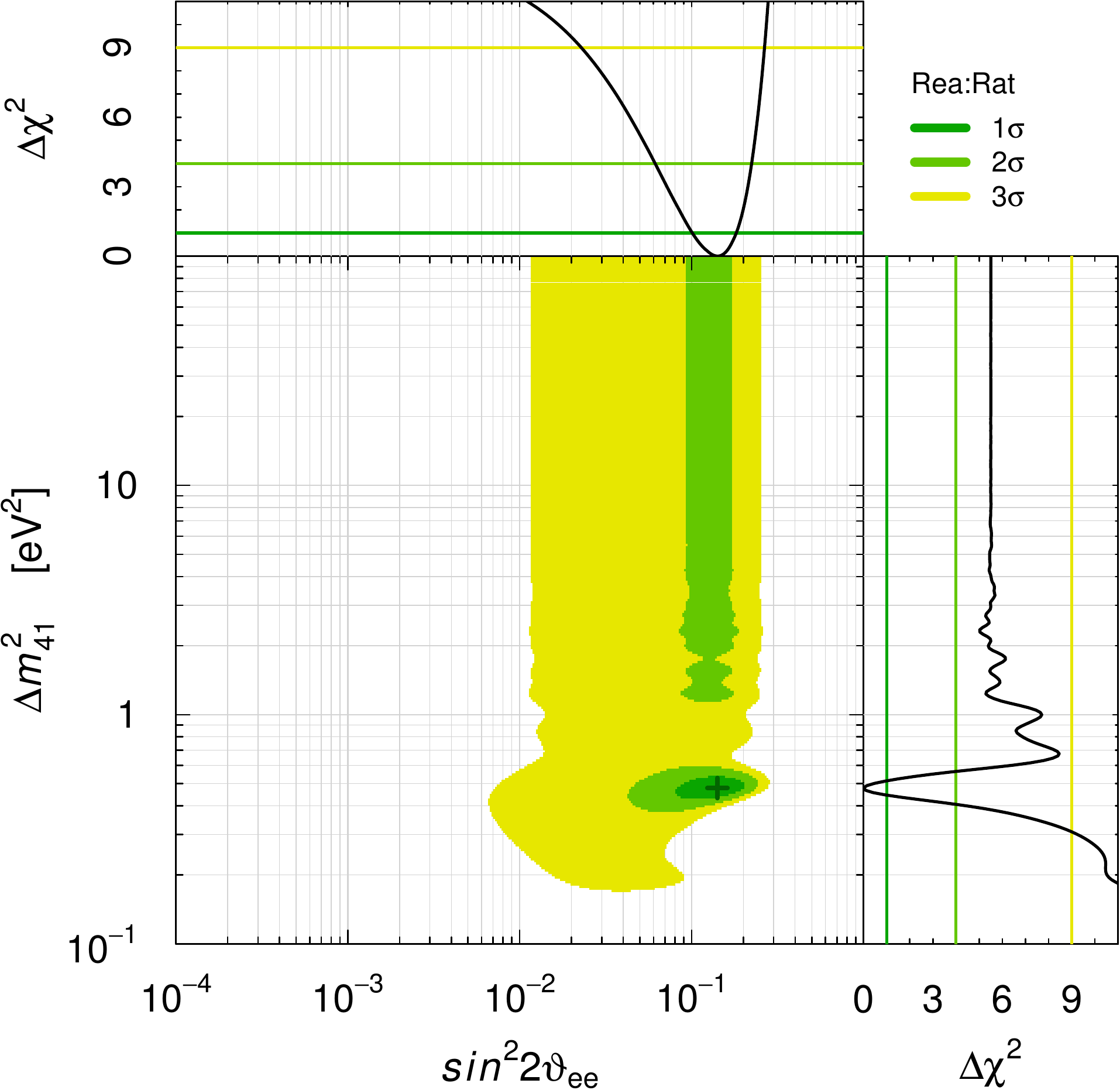}
}
&
\subfigure[]{\label{fig:rea-spe}
\includegraphics*[width=0.49\linewidth]{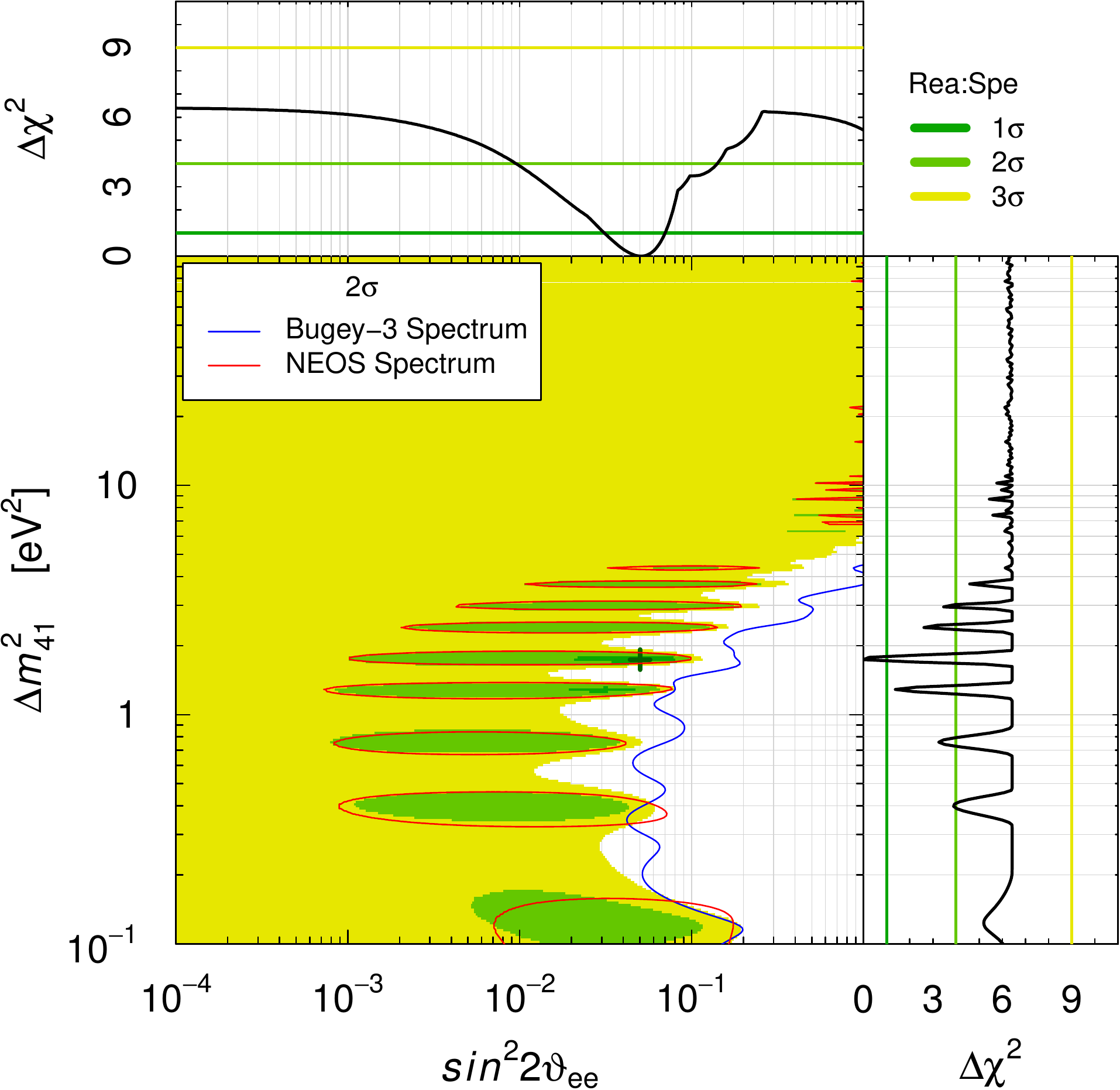}
}
\end{tabular}
\caption{ \label{fig:nuedis1}
Allowed regions in the
$\sin^{2}2\vartheta_{ee}$--$\Delta{m}^{2}_{41}$ plane
and
marginal $\Delta\chi^{2}$'s
for
$\sin^{2}2\vartheta_{ee}$ and $\Delta{m}^{2}_{41}$
obtained from:
\subref{fig:rea-rat}
the combined fit of the rates of the reactor neutrino experiments in Tab.~\ref{tab:rat};
\subref{fig:rea-spe}
the combined fit of the spectra of Bugey-3 \cite{Declais:1995su}
and
NEOS \cite{Ko:2016owz}
reactor antineutrino experiments;
The best-fit points corresponding to $\chi^2_{\text{min}}$
in Table~\ref{tab:nuedis} are indicated by crosses.
}
\end{figure}

The reactor antineutrino anomaly can be explained in the framework of 3+1 neutrino mixing
through neutrino oscillations generated by
the effective mixing angle
$
\sin^2 2\vartheta_{ee}
=
4
|U_{e4}|^2
\left( 1 -  |U_{e4}|^2 \right)
$,
which determines the survival probability of
$\nu_{e}$'s and $\bar\nu_{e}$'s
according to Eq.~(\ref{prb3p1}).
The result of the fit of the reactor rates are given in the first column of Table~\ref{tab:nuedis}
and in Fig.~\ref{fig:rea-rat},
where we have drawn the allowed regions
in the $\sin^{2}2\vartheta_{ee}$--$\Delta{m}^{2}_{41}$ plane.

From Fig.~\ref{fig:rea-rat}
one can see that the allowed $1\sigma$ region\footnote{%
In all the paper we consider allowed regions at
$1\sigma$,
$2\sigma$, and
$3\sigma$,
which correspond, respectively,
to
68.27\%,
95.45\%, and
99.73\%
confidence level.
The allowed regions in two-parameter planes are drawn considering two degrees of freedom,
which correspond, respectively, to
$\Delta\chi^2 = 2.30$, $6.18$, and $11.83$
from the minimum $\chi^2_{\text{min}}$.
}
in the $\sin^{2}2\vartheta_{ee}$--$\Delta{m}^{2}_{41}$ plane
is at a rather low value of
$\Delta{m}^{2}_{41}$,
but the allowed regions at $2\sigma$ and $3\sigma$ extend to higher values of $\Delta{m}^{2}_{41}$, without an upper bound.
The favorite values of the amplitude
$\sin^{2}2\vartheta_{ee}$
of $\nu_{e}$-disappearance oscillations
are around 0.1,
but the allowed $3\sigma$ region
in the $\sin^{2}2\vartheta_{ee}$--$\Delta{m}^{2}_{41}$ plane
covers the range
$
0.0066
\lesssim
\sin^{2}2\vartheta_{ee}
\lesssim
0.28
$,
which corresponds to
$
0.0017
\lesssim
|U_{e4}|^2
\lesssim
0.076
$.

Table~\ref{tab:nuedis}
gives the $\chi^{2}$ difference $\Delta\chi^{2}_{\text{NO}}$ between the $\chi^{2}$ of no oscillations and $\chi^{2}_{\text{min}}$, and
the corresponding number of $\sigma$'s for the two degrees of freedom
corresponding to the two fitted parameters
$\Delta{m}^2_{41}$
and
$\sin^22\vartheta_{ee}$.
The case of no oscillations turns out to be disfavored at the level of
$3.2\sigma$.

\subsection{Reactor spectra}
\label{sub:spectra}

In our previous analyses
\cite{Giunti:2012tn,Giunti:2012bc,Giunti:2013aea,Gariazzo:2015rra}
we considered the ratio of the spectra measured at 40 m and 15 m from the source
in the
Bugey-3 experiment \cite{Declais:1995su}.
These data provide robust information on short-baseline
$\bar\nu_{e}$ disappearance,
which is independent of any theoretical calculation of the spectrum
and of the solution of the
``5 MeV bump'' problem mentioned in subsection~\ref{sub:rates}.

In this paper we add the constraints obtained
in the recent NEOS experiment
by taking into account the $\chi^2$
corresponding to Fig.~4 of Ref.~\cite{Ko:2016owz},
which has been kindly provided to us by
the NEOS Collaboration\footnote{NEOS Collaboration, Private Communication.}.
The NEOS constraints are mostly independent of theoretical calculations of the spectrum
and of the solution of the
``5 MeV bump'' problem,
because the NEOS $\chi^2$ has been obtained by fitting the NEOS spectrum
normalized to the Daya Bay spectrum
\cite{An:2016srz}
measured at the large distance of about 550 m,
where the short-baseline oscillations due to $\Delta{m}^{2}_{41}$ are averaged out.
A small dependence on the theoretical calculation of the spectrum \cite{Mueller:2011nm,Huber:2011wv}
comes from the corrections due to the differences of the fission fractions of the NEOS and Daya Bay reactors \cite{An:2016srz,Ko:2016owz}
and a small dependence on the ``5 MeV bump'' problem
comes from a possible dependence of the
``5 MeV bump''
on the different fission fractions of NEOS and Daya Bay
\cite{Huber:2016xis}.
We neglect these possible small effects.

The results of the fit of the Bugey-3 and NEOS spectra are given in the second column of Table~\ref{tab:nuedis}
and in Fig.~\ref{fig:rea-spe},
where one can see that the NEOS constraints are dominating.
There are closed allowed islands at $2\sigma$
which are determined mainly by the NEOS data
and
the best-fit values of the oscillation parameters in Table~\ref{tab:nuedis}
correspond to the best fit reported in Ref.~\cite{Ko:2016owz}.
Hence,
the NEOS constraints can be interpreted as a weak indication in favor
of short-baseline oscillations
which may be compatible with the reactor antineutrino anomaly
based on the reactor rates discussed in subsection~\ref{sub:rates}.
This is confirmed by the disfavoring of the case of no oscillations
at the level of
$2.1\sigma$,
as shown in Table~\ref{tab:nuedis}.

The third column of Table~\ref{tab:nuedis}
and Fig.~\ref{fig:rea-all}
show the results of the combined fit of the rate and spectral data of
reactor antineutrino experiments.
As reported in Table~\ref{tab:nuedis},
the combined fit disfavors
the case of no oscillations at the level of
$2.9\sigma$,
which is about the same level
obtained from the analysis of the reactor rates alone.
Hence,
the NEOS constraints do not exclude the reactor antineutrino anomaly.
However,
in spite of the weak NEOS indication in favor
of short-baseline oscillations discussed above,
the statistical significance of the anomaly does not increase
by including the NEOS data because there is a mild
tension with the reactor rates
which is illustrated by the $2\sigma$ contours
in Fig.~\ref{fig:rea-all}.
Indeed, the rates-spectra parameter goodness-of-fit is only
$ 2\%$
($
\Delta\chi^2
/
\text{NDF}
=
8.3
/
2
$).

\subsection{Global $\boldsymbol{\nu_{e}}$ and $\boldsymbol{\bar\nu_{e}}$ disappearance}
\label{sub:nuedis}

In this subsection we discuss the combination of the reactor data with the
data of the Gallium neutrino anomaly,
other
$\nu_{e}$ and $\bar\nu_{e}$ disappearance data
and
the $\beta$-decay constraints of the
Mainz
\cite{Kraus:2012he}
and
Troitsk
\cite{Belesev:2012hx,Belesev:2013cba}
experiments.

\begin{figure}[!t]
\centering
\setlength{\tabcolsep}{0pt}
\begin{tabular}{cc}
\subfigure[]{\label{fig:rea-all}
\includegraphics*[width=0.49\linewidth]{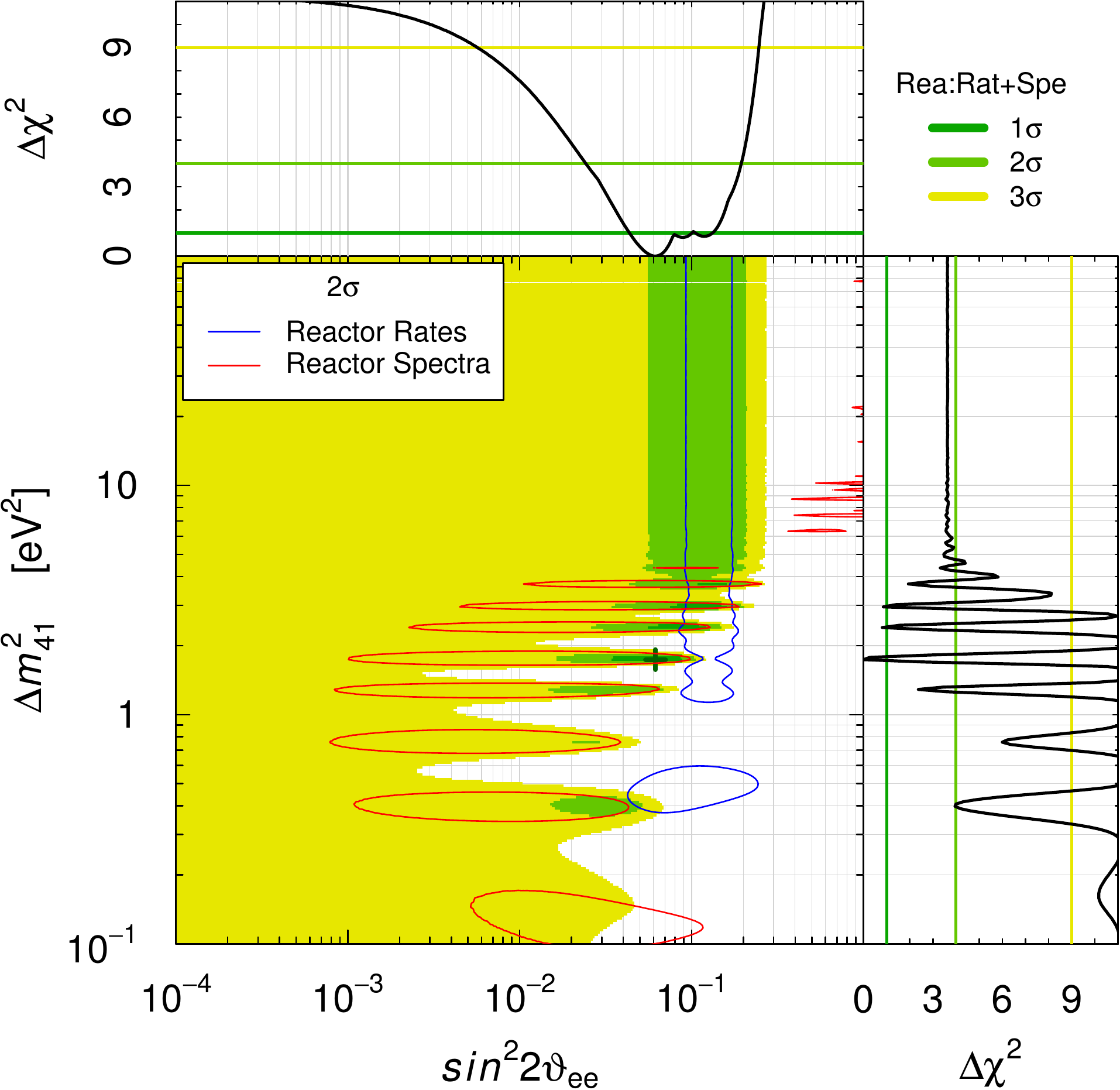}
}
&
\subfigure[]{\label{fig:rea-gal}
\includegraphics*[width=0.49\linewidth]{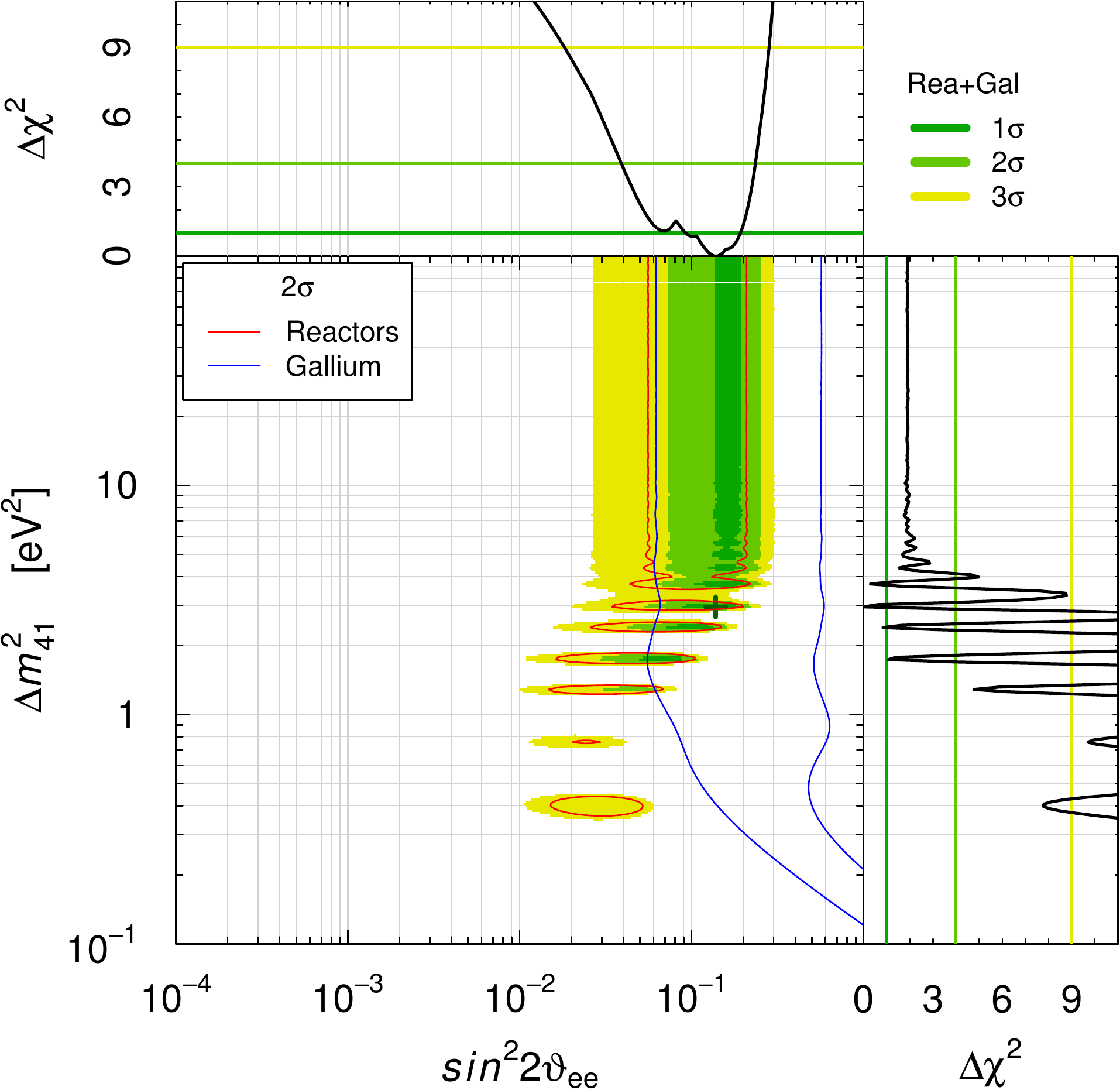}
}
\end{tabular}
\caption{ \label{fig:nuedis2}
Allowed regions in the
$\sin^{2}2\vartheta_{ee}$--$\Delta{m}^{2}_{41}$ plane
and
marginal $\Delta\chi^{2}$'s
for
$\sin^{2}2\vartheta_{ee}$ and $\Delta{m}^{2}_{41}$
obtained from:
\subref{fig:rea-all}
the combined fit of the rate and spectral data of
reactor antineutrino experiments;
\subref{fig:rea-gal}
the combined fit of the reactor and Gallium data.
The best-fit points corresponding to $\chi^2_{\text{min}}$
in Table~\ref{tab:nuedis} are indicated by crosses.
}
\end{figure}

The fourth column of Table~\ref{tab:nuedis}
and Fig.~\ref{fig:rea-gal}
show the results of the combined fit of reactor and Gallium data.
Since both sets of data indicate short-baseline
$\nu_{e}$ and $\bar\nu_{e}$ disappearance,
the statistical significance of active-sterile neutrino oscillations increases to
$3.6\sigma$
and the $3\sigma$ allowed regions
in the $\sin^{2}2\vartheta_{ee}$--$\Delta{m}^{2}_{41}$ plane
are confined to
$
0.010
\lesssim
\sin^{2}2\vartheta_{ee}
\lesssim
0.30
$
and
$
\Delta{m}^2_{41}
\gtrsim
0.35 \, \text{eV}^2
$.

Besides the reactor and Gallium data,
short-baseline
$\nu_{e}$ disappearance\footnote{%
We work in the framework of a local quantum field theory
in which the CPT symmetry implies that
the survival probabilities of neutrinos and antineutrinos
are equal
(see Ref.~\cite{Giunti:2007ry}).
}
is constrained by solar and KamLAND
neutrino data \cite{Giunti:2009xz,Palazzo:2011rj,Palazzo:2012yf,Giunti:2012tn,Palazzo:2013me},
by
the
KARMEN~\cite{Armbruster:1998uk}
and
LSND~\cite{Auerbach:2001hz}
$\nu_{e} + {}^{12}\text{C} \to {}^{12}\text{N}_{\text{g.s.}} + e^{-}$
scattering data
\cite{Conrad:2011ce,Giunti:2011cp}
and
by the T2K near detector constraints
\cite{Abe:2014nuo}.

\begin{figure}[!t]
\centering
\includegraphics*[width=0.49\linewidth]{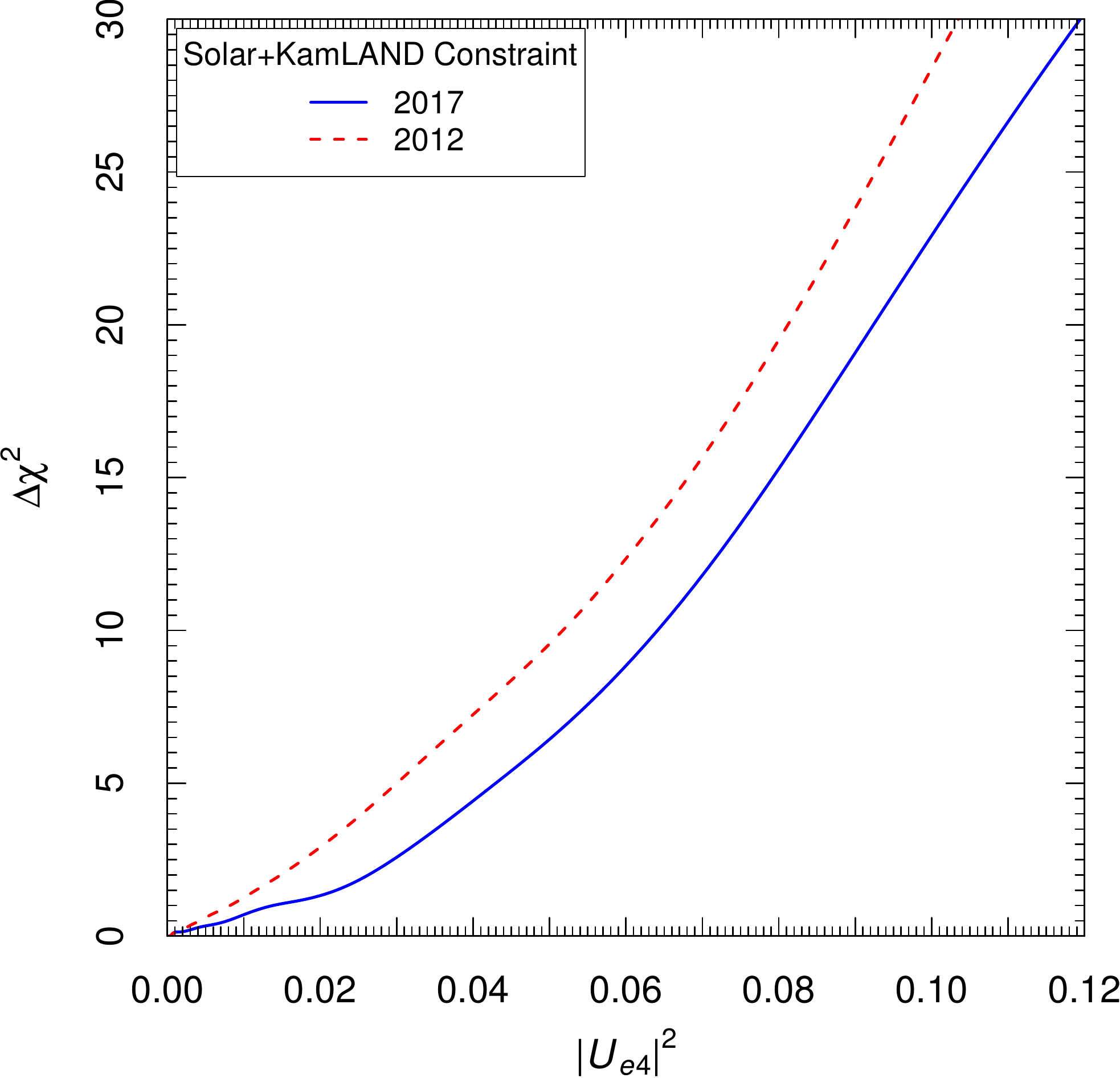}
\caption{ \label{fig:see-sun}
Marginal $\Delta\chi^2 = \chi^2 - \chi^2_{\text{min}}$
as a function of
$\sin^{2}2\vartheta_{ee}$
obtained from the fit of current solar+KamLAND neutrino data (2017)
compared with the one obtained in 2012 in Ref.~\cite{Giunti:2012tn}.
}
\end{figure}

We updated our 2012 solar+KamLAND constraint~\cite{Giunti:2012tn} by
including the latest solar data:
the new results from the fourth phase of the Super-Kamiokande experiment~\cite{Abe:2016nxk}
and the final results of Borexino phase-I~\cite{Bellini:2013lnn}.
We also updated the KamLAND data analysis by using the Saclay+Huber cross sections per fission~\cite{Giunti:2016elf}.
Finally, we used the updated value of $\vartheta_{13}$
in the 2016 Review of Particle Physics~\cite{Olive:2016xmw}.
We obtained the new marginal $\Delta\chi^2$ shown in Fig.~\ref{fig:see-sun},
where it is confronted with the old one obtained in Ref.~\cite{Giunti:2012tn}.
The new solar+KamLAND constraint is weaker than the 2012 one because of
the larger Saclay+Huber reactor rate prediction used in the analysis of KamLAND data
and because the new value of $\vartheta_{13}$ is smaller than that in 2012.

\begin{figure}[!t]
\centering
\setlength{\tabcolsep}{0pt}
\begin{tabular}{cc}
\subfigure[]{\label{fig:nue-dis}
\includegraphics*[width=0.49\linewidth]{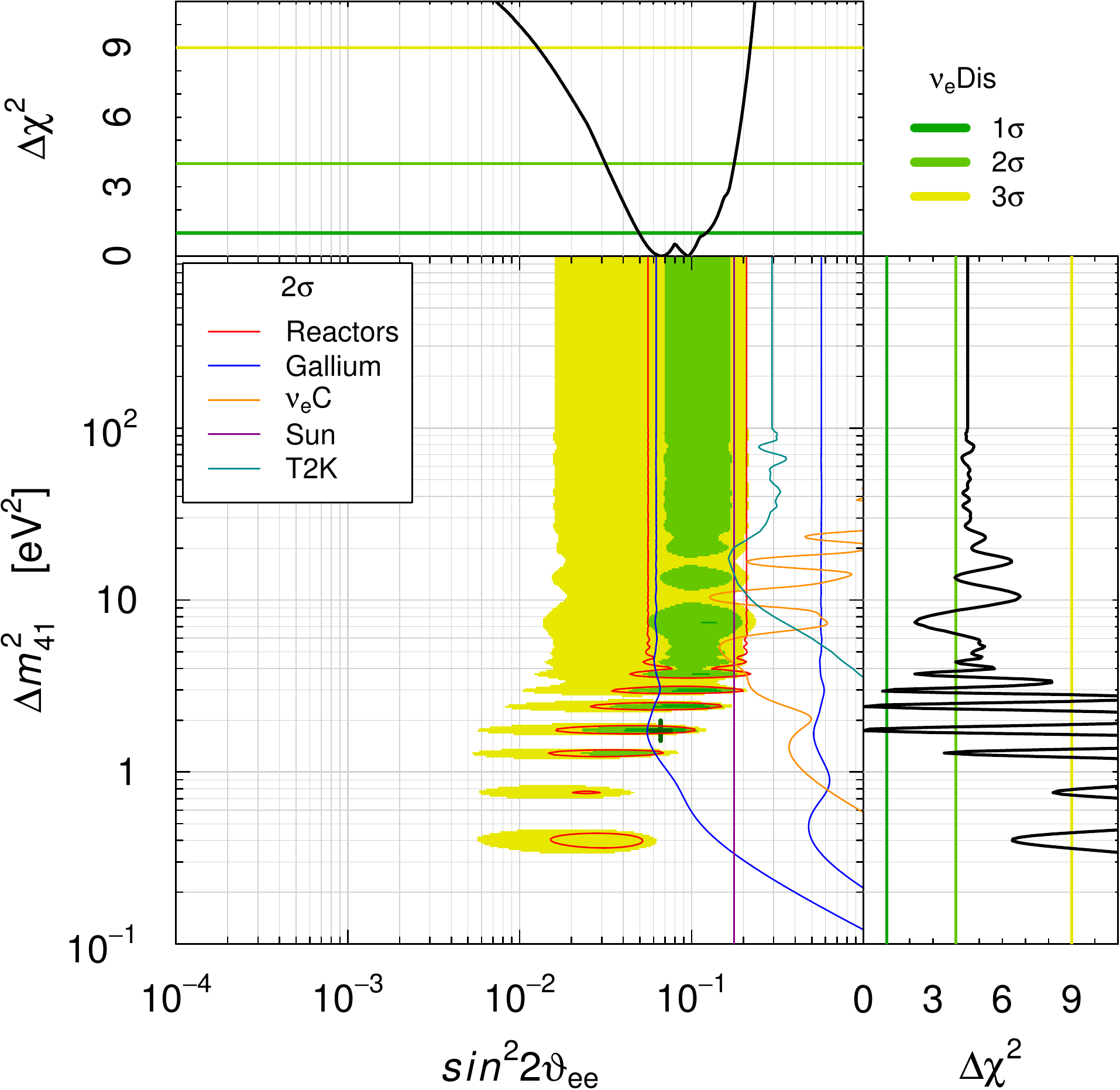}
}
&
\subfigure[]{\label{fig:nue-mbt}
\includegraphics*[width=0.49\linewidth]{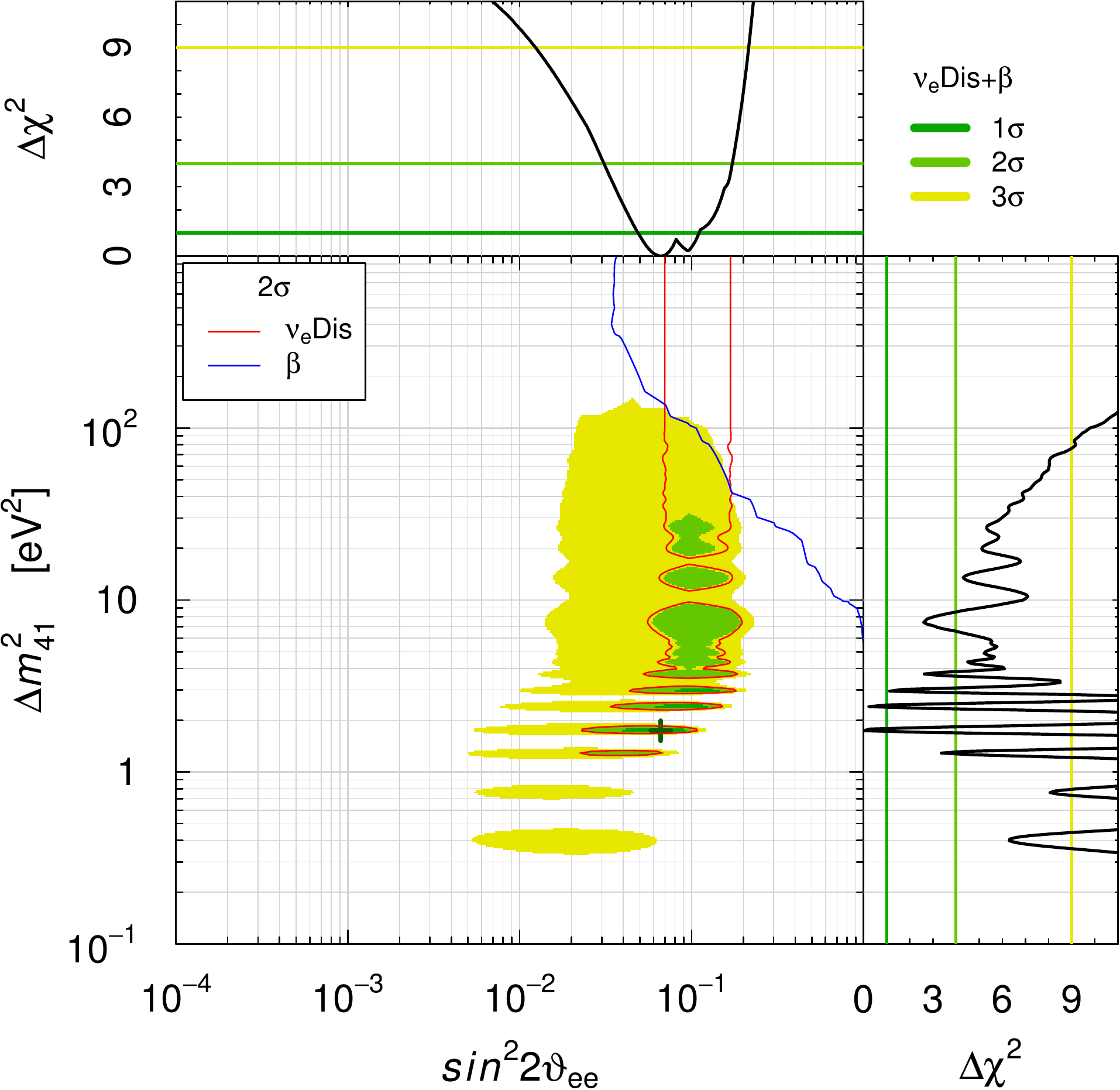}
}
\end{tabular}
\caption{ \label{fig:nuedis3}
Allowed regions in the
$\sin^{2}2\vartheta_{ee}$--$\Delta{m}^{2}_{41}$ plane
and
marginal $\Delta\chi^{2}$'s
for
$\sin^{2}2\vartheta_{ee}$ and $\Delta{m}^{2}_{41}$
obtained from:
\subref{fig:nue-dis}
the
combined fit of
$\nu_{e}$ and $\bar\nu_{e}$ disappearance data;
\subref{fig:nue-mbt}
the
combined fit of
$\nu_{e}$ and $\bar\nu_{e}$ disappearance data and
the $\beta$-decay constraints of the
Mainz
\cite{Kraus:2012he}
and
Troitsk
\cite{Belesev:2012hx,Belesev:2013cba}
experiments.
The best-fit points corresponding to $\chi^2_{\text{min}}$
in Table~\ref{tab:nuedis} are indicated by crosses.
}
\end{figure}

The results of the combined analysis of all
$\nu_{e}$ and $\bar\nu_{e}$ disappearance data
are shown in the fifth column of Table~\ref{tab:nuedis}
and Fig.~\ref{fig:nue-dis}.
Since the analysis of
solar+KamLAND, $\nu_{e}$-${}^{12}\text{C}$, and T2K data
do not show any indication of
short-baseline
$\nu_{e}$ disappearance,
the combination with the reactor and Gallium data
shifts the allowed regions in the
$\sin^{2}2\vartheta_{ee}$--$\Delta{m}^{2}_{41}$ plane
in Fig.~\ref{fig:nue-dis} to smaller values of
$\sin^{2}2\vartheta_{ee}$
with respect to Fig.~\ref{fig:rea-gal}:
$
0.0054
\lesssim
\sin^{2}2\vartheta_{ee}
\lesssim
0.23
$.
On the other hand,
the $3\sigma$ range of $\Delta{m}^2_{41}$ in Figs.~\ref{fig:rea-gal} and \ref{fig:nue-dis} is similar,
with the lower bound
$
\Delta{m}^2_{41}
\gtrsim
0.35 \, \text{eV}^2
$
and no upper bound.

Large values of $\Delta{m}^2_{41}$ can be constrained
with the data of $\beta$-decay experiments
(see Ref.~\cite{Gariazzo:2015rra}).
As in Ref.~\cite{Giunti:2012bc},
we use
the $\beta$-decay constraints of the
Mainz
\cite{Kraus:2012he}
and
Troitsk
\cite{Belesev:2012hx,Belesev:2013cba}
experiments,
which give the allowed regions
in the
$\sin^{2}2\vartheta_{ee}$--$\Delta{m}^{2}_{41}$ plane
shown in Fig.~\ref{fig:nue-mbt}.
One can see
that the allowed regions are confined to the range
\begin{equation}
1.3
\,
(0.33)
\, \text{eV}^2
\lesssim
\Delta{m}^{2}_{41}
\lesssim
32
\,
(148)
\, \text{eV}^2
\quad
\text{at}
\quad
2\sigma
\,
(3\sigma)
.
\label{nue-mbt-dm2}
\end{equation}
For the oscillation length
$
L^{\text{osc}}_{41}
=
4 \pi E / \Delta{m}^{2}_{41}
$
we have
\begin{equation}
8
\,
(2)
\, \text{cm}
\lesssim
\frac{L^{\text{osc}}_{41}}{E\,[\text{MeV}]}
\lesssim
2
\,
(7)
\, \text{m}
\quad
\text{at}
\quad
2\sigma
\,
(3\sigma)
.
\label{nue-mbt-Losc}
\end{equation}
This is a range of oscillation lengths which can be explored
in a model independent way
in the new short-baseline reactor neutrino experiments
(DANSS \cite{Alekseev:2016llm},
Neutrino-4 \cite{Serebrov:2017nxa},
PROSPECT \cite{Ashenfelter:2015uxt},
SoLid \cite{Michiels:2016qui},
STEREO \cite{Manzanillas:2017rta})
and in the SOX \cite{Borexino:2013xxa}
and BEST \cite{Barinov:2016znv}
radioactive source experiments
by measuring the reactor antineutrino rate as a function of distance.
However, they will need to be sensitive to small oscillations with the amplitude
\begin{equation}
0.022
\,
(0.0050)
\lesssim
\sin^{2}2\vartheta_{ee}
\lesssim
0.19
\,
(0.23)
\quad
\text{at}
\quad
2\sigma
\,
(3\sigma)
.
\label{nue-mbt-st2}
\end{equation}

\begin{figure}[!t]
\centering
\setlength{\tabcolsep}{0pt}
\begin{tabular}{cc}
\subfigure[]{\label{fig:fut-rea}
\includegraphics*[width=0.49\linewidth]{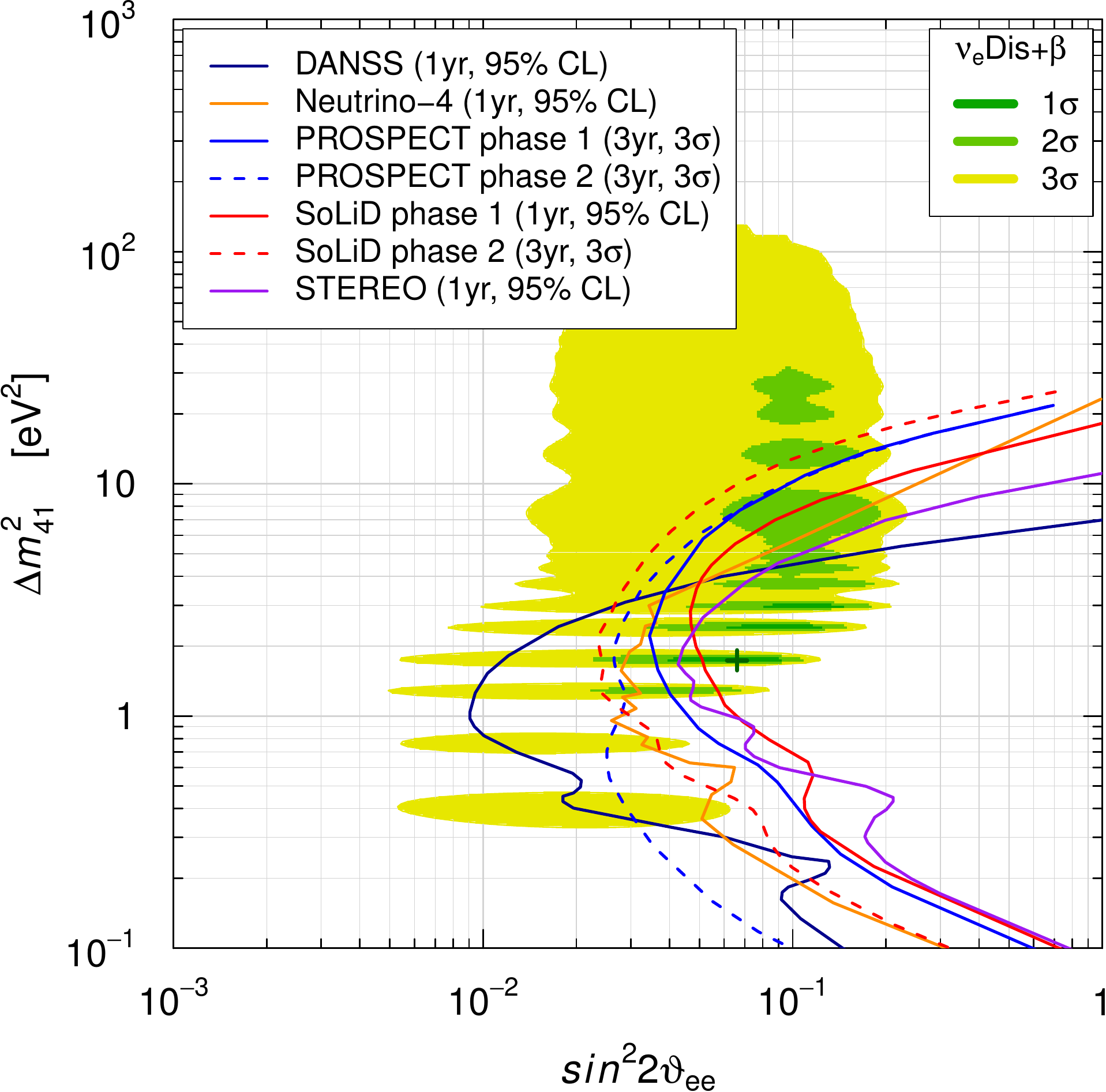}
}
&
\subfigure[]{\label{fig:fut-rad}
\includegraphics*[width=0.49\linewidth]{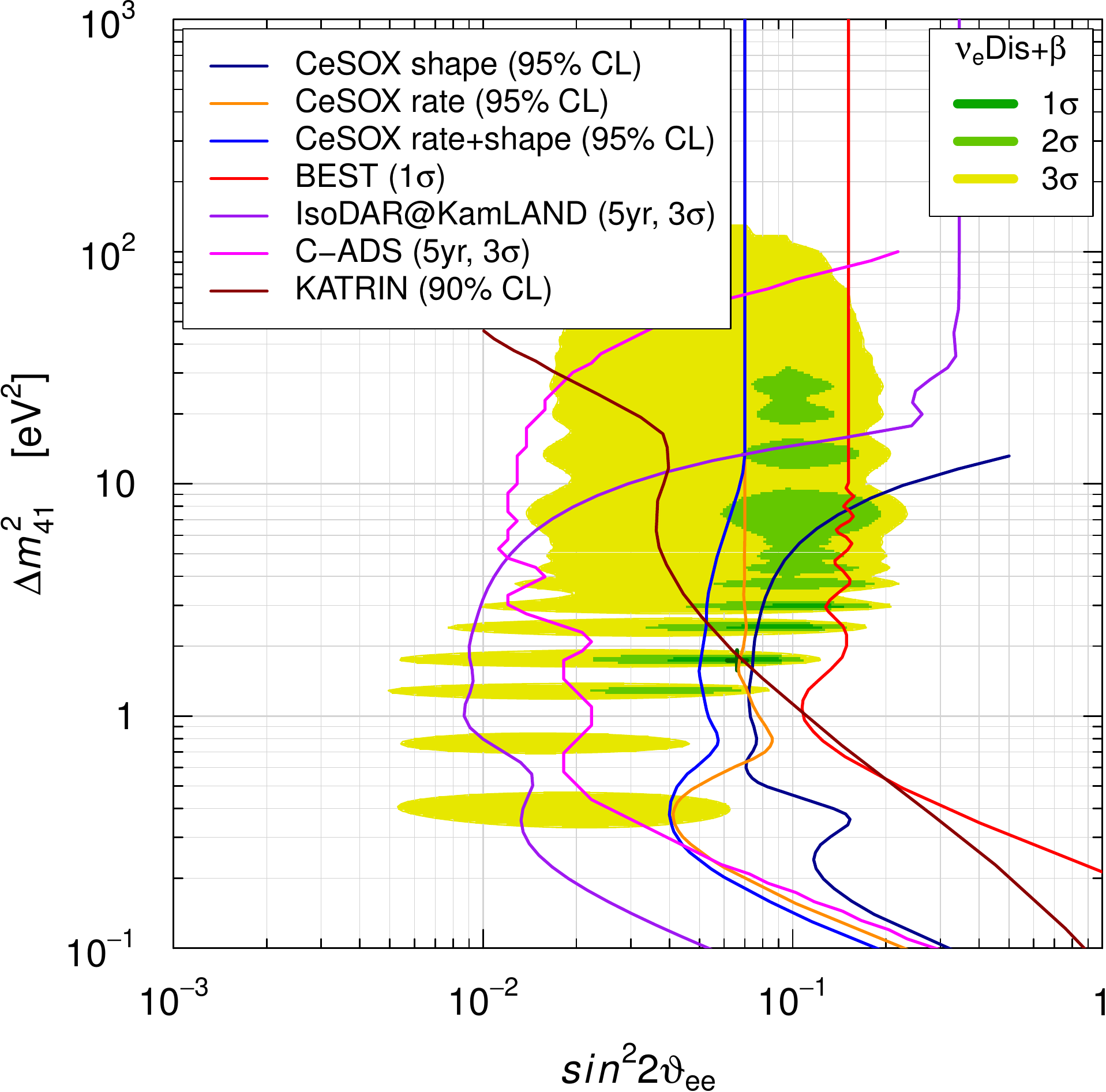}
}
\end{tabular}
\caption{ \label{fig:nuedisfut}
Sensitivities of future
reactor \subref{fig:fut-rea}
and
source \subref{fig:fut-rad}
experiments
compared with the allowed regions in the
$\sin^{2}2\vartheta_{ee}$--$\Delta{m}^{2}_{41}$ plane
in Fig.~\ref{fig:nue-mbt}.
}
\end{figure}

Figure~\ref{fig:fut-rea}
shows the sensitivities of the short-baseline reactor antineutrino experiments
DANSS \cite{Alekseev:2016llm},
Neutrino-4 \cite{Serebrov:2012sq},
PROSPECT \cite{Ashenfelter:2015uxt},
SoLid \cite{Ryder:2015sma}, and
STEREO \cite{Helaine:2016bmc}
in comparison with the allowed regions in the
$\sin^{2}2\vartheta_{ee}$--$\Delta{m}^{2}_{41}$ plane
in Fig.~\ref{fig:nue-mbt}.
One can see that they will cover most of the allowed regions for
$ \Delta{m}^{2}_{41} \lesssim 10 \, \text{eV}^2 $
and not too small
$\sin^{2}2\vartheta_{ee}$.
Figure~\ref{fig:fut-rad}
shows the sensitivities of the
CeSOX \cite{Borexino:2013xxa} and BEST \cite{Barinov:2016znv} source experiments,
of IsoDAR@KamLAND \cite{Abs:2015tbh} and C-ADS \cite{Ciuffoli:2015uta},
and of the
KATRIN \cite{Drexlin-NOW2016}) electron neutrino mass experiment\footnote{%
See also the studies in Refs.~\cite{Riis:2010zm,Formaggio:2011jg,SejersenRiis:2011sj,Esmaili:2012vg}.
There are also promising possibilities to
observe the effects of eV-scale neutrinos in
Holmium electron-capture experiments
\cite{Gastaldo:2016kak}.
}.
The source experiments will cover the large-$\sin^{2}2\vartheta_{ee}$ parts
of the allowed regions,
the IsoDAR@KamLAND and C-ADS experiments can cover almost all the allowed regions,
except the large-$\Delta{m}^{2}_{41}$ part
and the
small-$\sin^{2}2\vartheta_{ee}$--small-$\Delta{m}^{2}_{41}$ parts,
and KATRIN will cover
the large-$\Delta{m}^{2}_{41}$ part.
Hence,
there are favorable perspectives for a definitive solution of the
short-baseline $\nua{e}$ disappearance problem
in the near future.

\section{Fits of appearance and disappearance data}
\label{sec:fits}

In this section we present the results of 3+1 fits of short-baseline
neutrino oscillation data which
include
$\nu_{\mu}\to\nu_{e}$ and $\bar\nu_{\mu}\to\bar\nu_{e}$
appearance data
and
$\nu_{\mu}$ and $\bar\nu_{\mu}$ disappearance data,
in addition to the
$\nu_{e}$ and $\bar\nu_{e}$ disappearance data
considered in section~\ref{sec:nuedis}.
Our fits are based on a $\chi^2$ analysis in the four-dimensional space of the mixing parameters
$\Delta{m}^2_{41}$,
$|U_{e4}|^2$,
$|U_{\mu4}|^2$, and
$|U_{\tau4}|^2$.

We consider the following short-baseline
$\nu_{\mu}\to\nu_{e}$ and $\bar\nu_{\mu}\to\bar\nu_{e}$
appearance data:
the LSND signal in favor of
$\bar\nu_{\mu}\to\bar\nu_{e}$
transitions
\cite{Athanassopoulos:1995iw,Aguilar:2001ty},
the data of the
MiniBooNE \cite{AguilarArevalo:2008rc,Aguilar-Arevalo:2013pmq}
experiment,
and the constraints of the
BNL-E776 \cite{Borodovsky:1992pn},
KARMEN \cite{Armbruster:2002mp},
NOMAD \cite{Astier:2003gs},
ICARUS \cite{Antonello:2013gut}
and
OPERA \cite{Agafonova:2013xsk}
experiments.

There is no indication in favor of short-baseline
$\nu_{\mu}$ and $\bar\nu_{\mu}$ disappearance
from any experiment.
Therefore,
the current $\nu_{\mu}$ and $\bar\nu_{\mu}$ disappearance data
lead to constraints on $|U_{\mu4}|^2$.
We consider the constraints obtained from
the CDHSW experiment \cite{Dydak:1983zq},
from the analysis \cite{Maltoni:2007zf} of
the data of
atmospheric neutrino oscillation experiments,
from the analysis of the
SciBooNE-MiniBooNE
neutrino \cite{Mahn:2011ea} and antineutrino \cite{Cheng:2012yy} data,
which were included in our previous fits
\cite{Giunti:2013aea,Gariazzo:2015rra,Giunti:2016oan}.
In addition,
we take into account the recent data of the
MINOS \cite{MINOS:2016viw}
and
IceCube \cite{TheIceCube:2016oqi}
experiments.
The MINOS constraint was easily included in our numerical computation
by using the ROOT program in the MINOS data release\footnote{%
\url{http://www-numi.fnal.gov/PublicInfo/forscientists.html}
},
which computes the $\chi^2$ for input values of the 3+1 mixing parameters.
On the other hand,
we had to calculate the IceCube $\chi^2$,
as described in the following subsection~\ref{sub:icecube}.

\subsection{Analysis of IceCube data}
\label{sub:icecube}

The IceCube detector measures the incoming (anti)muons
generated by the interaction of atmospheric muon (anti)neutrinos
with the surrounding earth and ice,
as a function of the neutrino energy and of the zenith angle.
For high-energy, up-going atmospheric neutrinos that reach the detector
after having crossed the Earth,
the ratio $L/E$ is of the same order of that in SBL experiments.
In this case, the oscillations arising from the
usual atmospheric and solar squared mass differences
have a very long wavelength and can be neglected,
but the SBL squared mass difference
$\Delta{m}^2_{41}$
plays an active role.
The sterile neutrino influence on the observed flux is given by the 
matter effects that modify the oscillation patterns inside the Earth.
This happens because the matter potential is different
for the different active neutrino flavors,
for which the charged and neutral current interactions are not the same
\cite{GonzalezGarcia:2005xw},
and there is no potential for the sterile neutrinos.

We use the 20,145 released IceCube events
in the approximate energy range between
320~GeV and 20~TeV,
detected over 343.7 days
in the 86-strings configuration~\cite{Aartsen:2015rwa}
for constraining the active-sterile mixing parameters.
The 99.9\% of the IceCube events is expected to come from
neutrino-induced muon events,
where the neutrinos originate from the decays
of atmospheric pions and kaons.
The contribution from charmed meson decays
is negligible~\cite{Aartsen:2014muf,Aartsen:2015rwa}.

The calculation of the $\chi^2$ contribution from IceCube
is divided into three parts:
the calculation of the theoretical flux for each set of mixing parameters,
for which one needs to propagate the atmospheric neutrinos through the Earth,
the estimate of the expected number of events in the detector,
for which we use the IceCube Monte Carlo data,
and finally the computation of the $\chi^2$,
obtained comparing theoretical and observed events.
For all these parts we use the data%
\footnote{\url{http://icecube.wisc.edu/science/data/IC86-sterile-neutrino/}}
and we follow the prescriptions presented in
Ref.~\cite{TheIceCube:2016oqi}.


To obtain the predicted neutrino flux at the detector, we use
the \texttt{$\nu$-SQuIDS} code%
\footnote{\url{http://github.com/Arguelles/nuSQuIDS}},
a C++ package based on
the Simple Quantum Integro-Differential Solver (\texttt{SQuIDS})%
\footnote{\url{http://github.com/jsalvado/SQuIDS}}
\cite{Delgado:2014kpa},
that contains all the necessary tools to numerically solve
the master equation that rules the neutrino evolution in the Earth%
~\cite{GonzalezGarcia:2005xw}.

The initial flux we consider is the unoscillated HKKM flux
\cite{Honda:1995hz,Honda:2004yz,Sanuki:2006yd,Honda:2006qj}
with the H3a knee correction \cite{Gaisser:2013bla},
that we use for obtaining the initial spectrum
of neutrinos from pion and kaon decays.
This model is usually referred to as the ``Honda-Gaisser'' model.
We do not employ here the other six atmospheric
flux variants that are considered in Ref.~\cite{TheIceCube:2016oqi},
but we tested that our results do not change significantly
if another model is used instead of the Honda-Gaisser one.
Since our analysis is based not only on the IceCube data,
our final result would be almost unaffected.

The unoscillated flux is propagated inside the Earth
with the \texttt{$\nu$-SQuIDS} code, which uses the
Preliminary Reference Earth Model \cite{Dziewonski:1981xy}
for the inner density profile of the Earth.
For the neutrino-matter interactions, the charged current cross section
is dominated by deep inelastic scattering,
which involves neutrino-nucleon scattering.
The main uncertainty in this case is
in the parton distribution functions.
In the \texttt{$\nu$-SQuIDS} code,
the perturbative QCD calculation
in Refs.~\cite{Arguelles:2015wba}
are used for the neutrino-nucleon cross-section calculation.
We do not treat the uncertainties on the Earth density profile and
on the deep inelastic scattering cross section.

The full expression for the (anti)neutrino flux at the detector
is given by \cite{TheIceCube:2016oqi}
(see also Refs.~\cite{Arguelles:2015a,Jones:2015bya})
\begin{equation}
 \label{eq:atm_flux}
 \phi^{\nu(\bar\nu)}_{\text{atm}}(E_{\nu(\bar\nu)},\cos\theta)
 =
 N_0^{\nu(\bar\nu)}
  \mathcal{F}(\delta)
  \left(
  \phi_K^{\nu(\bar\nu)}
  +
  R_{\pi/K}^{}
  \phi_\pi^{\nu(\bar\nu)}
 \right)
 \left(
  \frac{E_{\nu(\bar\nu)}}{E_m}
 \right)^{-\Delta\gamma}\,.
\end{equation}
Here,
$\theta$ is the zenith angle and $E_{\nu(\bar\nu)}$ the energy
of the incoming (anti)neutrino, while
$\phi_{\pi(K)}^{\nu(\bar\nu)}$ is the oscillated
(anti)neutrino flux from pion (kaon) decays.
The free parameters in the above equation are: 
the neutrino and antineutrino flux normalizations,
$N_0^{\nu}$ and $N_0^{\bar\nu}$;
the pion-to-kaon ratio, $R_{\pi/K}^{}$;
the spectral index correction, $\Delta\gamma$.
These are treated as continuous nuisance parameters in our analysis,
as explained in Ref.~\cite{TheIceCube:2016oqi}.
The pivot energy $E_m$ is fixed to be approximately near the median
of the energy distribution of the measured events,
being $E_m=2$~TeV.

The function $\mathcal{F}(\delta)$ parameterizes
the atmospheric density uncertainties.
This function is assumed to be linear and it is obtained by 
imposing the AIRS constraints on the atmospheric temperature\footnote{\url{https://climatedataguide.ucar.edu/climate-data/airs-and-amsu-tropospheric-air-temperature-and-specific-humidity}}.
The expression reads as \cite{Arguelles:2015a}:
\begin{equation}
 \label{eq:f_delta}
 \mathcal{F}(\delta)
 =
 1+(\cos\theta+\cos\theta_0)
 \,\delta
 \left[
  1+\frac{E_{\nu(\bar\nu)}-E_0}{E_1}\cdot
  \frac{1}{1+\exp
    \left(
      -\kappa(\cos\theta+\cos\theta_0)
    \right)}
 \right]\,,
\end{equation}
where $E_0 = 360$~GeV,
$E_1 = 11.279$~TeV,
$\kappa=200$ and
$\cos\theta_0=0.4$.
The parameter $\delta$ represents the last one of our nuisance parameters.

The theoretical flux is converted into a number of expected events
using the Monte Carlo (MC) data released by the IceCube collaboration%
~\cite{TheIceCube:2016oqi}.
The MC data are needed to model the detector capabilities to measure
the incoming events as a function of the real energy and zenith angle
of the muon (anti)neutrino, and
of the corresponding quantities for the reconstructed (anti)muon event.
For each combination of mixing and nuisance parameters,
we use the MC data to convert the obtained theoretical flux
into the expected number of events that we compare with the data
as explained below.
Since IceCube cannot distinguish a muon from an antimuon,
neutrinos and antineutrinos events are summed up together.
It is however important to treat properly both the components,
since the matter oscillation patterns for neutrinos
and antineutrinos are different,
with the consequence that the disappearance of
neutrinos and antineutrinos is not the same.


We build our $\chi^2$ using a binned Poissonian likelihood,
written as
\begin{equation}
 \label{eq:llh}
 \chi^2=-2\ln\mathcal{L}(\boldsymbol\theta)
 =
 2\sum_{i=1}
 \left[
  \mu_i(\boldsymbol\theta)-n_i
  +n_i\ln\frac{n_i}{\mu_i(\boldsymbol\theta)}
 \right]\,,
\end{equation}
where
$n_i$ represents the number of observed events in the bin $i$ and
$\mu_i(\boldsymbol\theta)$ the corresponding
number of expected events as a function
of the model parameters $\boldsymbol\theta$,
that includes both mixing and nuisance parameters.
Following Ref.~\cite{TheIceCube:2016oqi},
we consider a grid with 10 logarithmic bins in the reconstructed energy,
with $400\text{ GeV}\le E^{\text{reco}}_{\mu(\bar\mu)}\le20\text{ TeV}$,
and 21 linear bins for the cosine of the reconstructed zenith angle,
in the range $-1\le\cos\theta^{\text{reco}}_{\mu(\bar\mu)}\le0.2$.

For each combination of the mixing parameters,
we minimize the $\chi^2$ over the five nuisance parameters
described above
($N_0^{\nu}$, $N_0^{\bar\nu}$, $R_{\pi/K}^{}$, $\Delta\gamma$, $\delta$).
We adopt a standard Nelder-Mead algorithm for the minimization
\cite{Nelder:1965a}.
It is important to note that for each point in the mixing parameter space
we needed to minimize independently over the nuisance parameters.
We checked that the preferred values of the nuisance parameters
do not vary significantly outside the adopted
Gaussian priors~\cite{TheIceCube:2016oqi}.

\begin{figure}[!t]
\centering
\includegraphics*[width=0.49\linewidth]{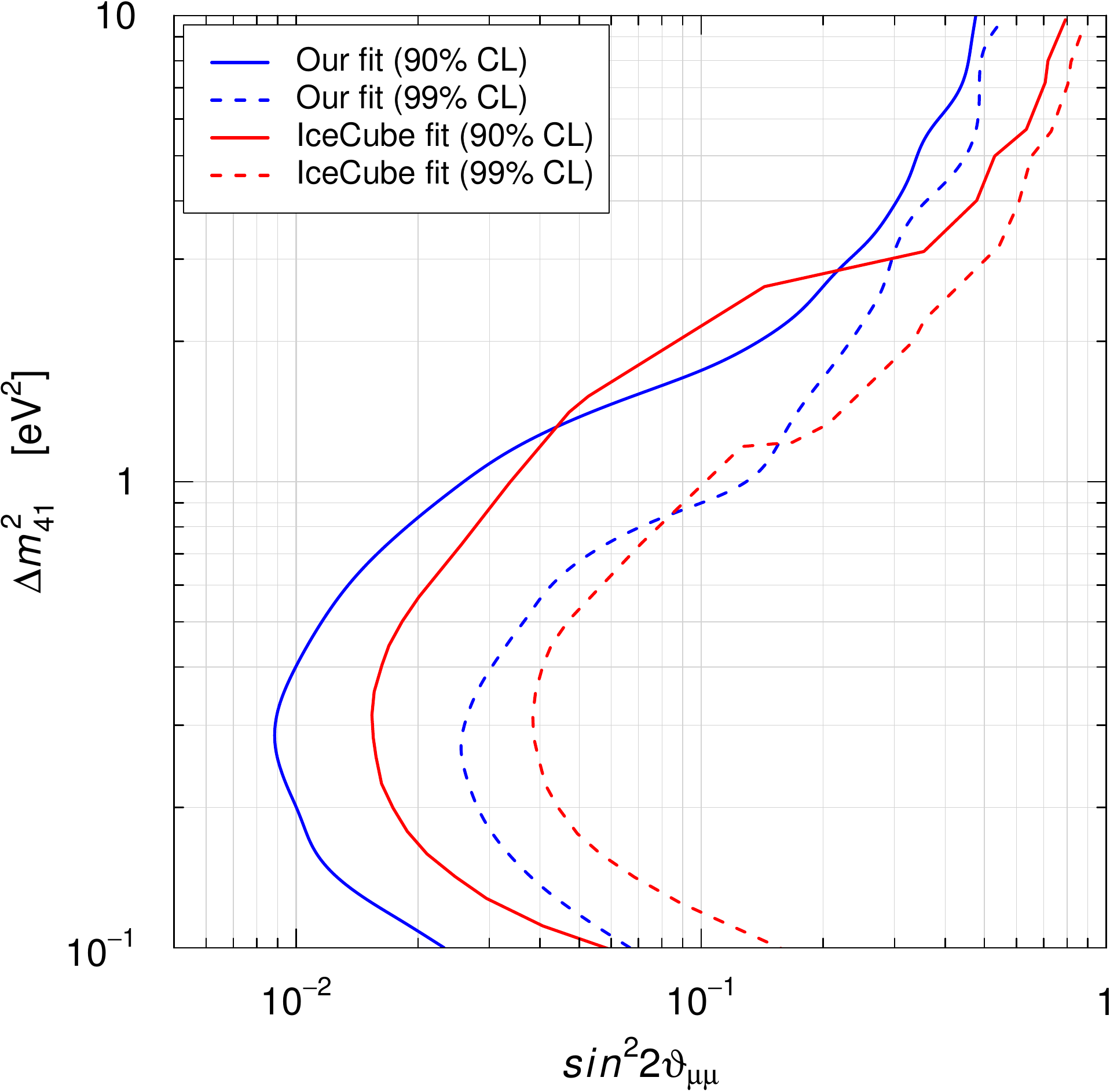}
\caption{ \label{fig:icecube}
Comparison of the official IceCube
90\% and 99\% CL exclusion curves in the
$\sin^{2}2\vartheta_{\mu\mu}$--$\Delta{m}^{2}_{41}$ plane
\cite{TheIceCube:2016oqi}
with our results.
All curves have been obtained assuming $|U_{e4}|^2=|U_{\tau4}|^2=0$.
}
\end{figure}

We show in Fig.~\ref{fig:icecube} the comparison of the official IceCube
90\% and 99\% CL exclusion curves in the
$\sin^{2}2\vartheta_{\mu\mu}$--$\Delta{m}^{2}_{41}$ plane
for $|U_{e4}|^2=|U_{\tau4}|^2=0$
\cite{TheIceCube:2016oqi}
with our results.
In our analysis
of IceCube data we do not vary the efficiency of the
Digital Optical Modules
because we do not have sufficient information.
Despite this fact,
one can see from Fig.~\ref{fig:icecube}
that the results of our analysis are in good agreement with those of the IceCube collaboration.
Moreover,
we emphasize that the IceCube data are just one of the datasets in our global analyses,
and small differences in the IceCube analysis do not play a significant role
when computing the global fit.

Since the calculation of the $\chi^2$ given a set of mixing parameters
is a highly demanding computational task,
it is impossible to directly include the code that calculates the $\chi^2$ of the IceCube data
in our complete fitting routine
without slowing it down in an unacceptable way.
Therefore,
we adopted the following
method.
Since we are more interested in scanning the region near
the expected 3+1 mixing best-fit,
we employed the results of the 3+1 fit of SBL data without IceCube
in order to generate with a Markov Chain Monte Carlo (MCMC) 3,000 random points whose distribution
covers an area of the parameter space around the expected best-fit region.
In order to cover the rest of the full four-dimensional parameter space
($\Delta m_{41}^2$;
$|U_{e4}|^2$,
$|U_{\mu4}|^2$,
$|U_{\tau4}|^2$),
we generated uniformly 21,000 more random points.
We end up with a set P of 24,000 points
for which we computed the IceCube $\chi^2$
in an affordable time.
In the complete fitting routine,
we computed
the IceCube contribution to the $\chi^2$ in each point
in the full four-dimensional parameter space
with a linear interpolation
of the $\chi^2$'s of the nearest points
in the set P obtained with a Delaunay triangulation.

\subsection{Fit of the 2016 data set without MINOS and IceCube}
\label{sub:Glo16A}

In this subsection we present the results of the 3+1 global fit ``Glo16A''
of the appearance and disappearance SBL data available in 2016\footnote{%
We consider all the $\nu_{e}$ and $\bar\nu_{e}$ disappearance data
discussed in section~\ref{sec:nuedis},
with the exceptions of the T2K near detector constraints
\cite{Abe:2014nuo}
on $\sin^{2}2\vartheta_{ee}$,
which unfortunately cannot be included in the global fit
because they have been obtained under the assumption
$|U_{\mu4}|^2=0$,
and of the Mainz
\cite{Kraus:2012he}
and
Troitsk
\cite{Belesev:2012hx,Belesev:2013cba}
$\beta$-decay constraints,
which are not needed because the value of $\Delta{m}^{2}_{41}$
is constrained within a few $\text{eV}^2$
by the combination of appearance and disappearance data.
},
except
MINOS \cite{MINOS:2016viw}
and
IceCube \cite{TheIceCube:2016oqi},
which will be added in subsection~\ref{sub:Glo16B}
in order to clarify their effects on the results of the analysis.
In the Glo16A fit we also do not take into account
the NEOS \cite{Ko:2016owz} data
which have been available to us in the beginning of 2017
and will be considered in subsection~\ref{sub:Glo17}.

\begin{figure}[!t]
\centering
\setlength{\tabcolsep}{0pt}
\begin{tabular}{ccc}
\subfigure[]{\label{fig:glo16a-sem}
\includegraphics*[width=0.3\linewidth]{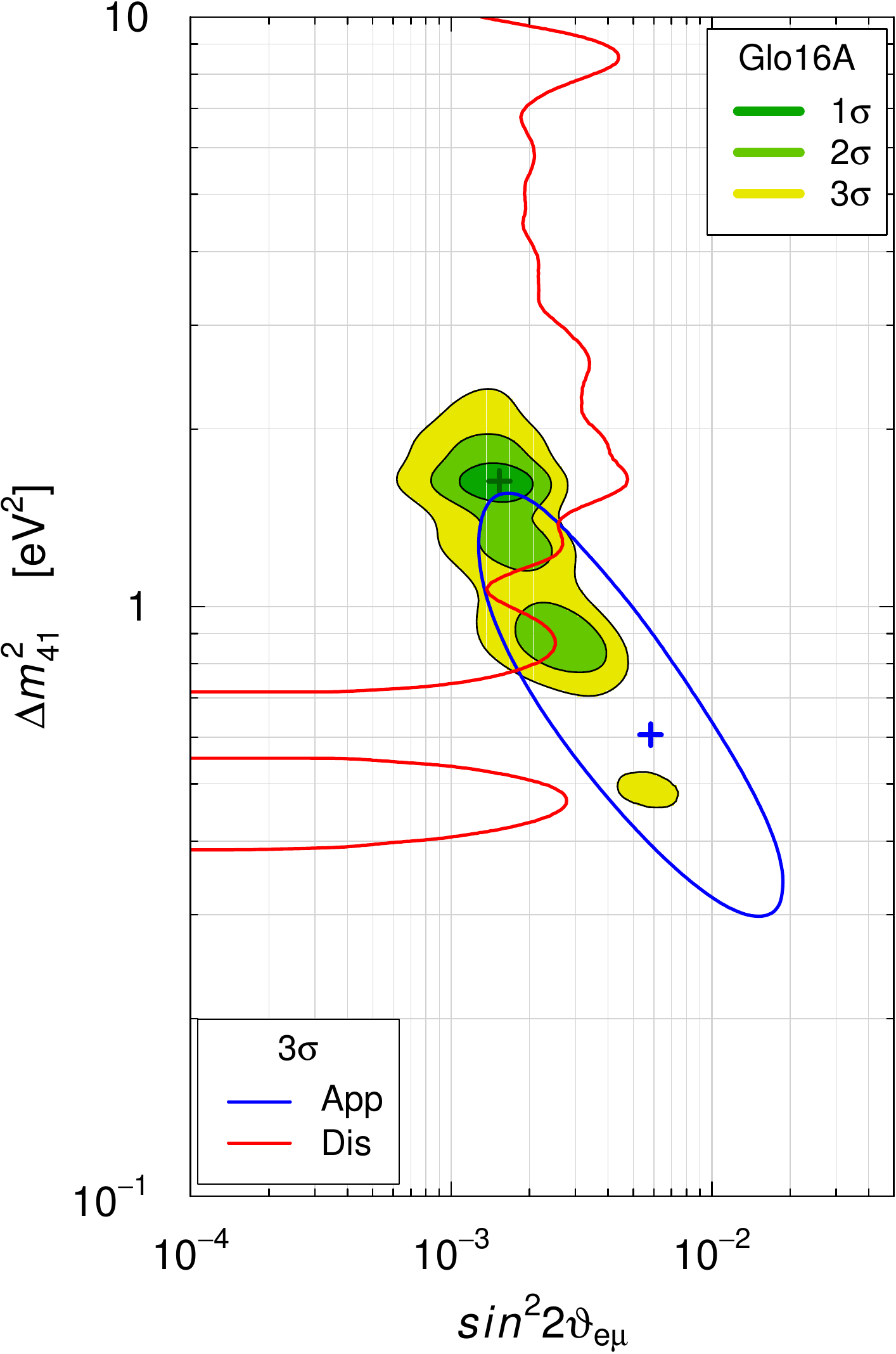}
}
&
\subfigure[]{\label{fig:glo16a-see}
\includegraphics*[width=0.3\linewidth]{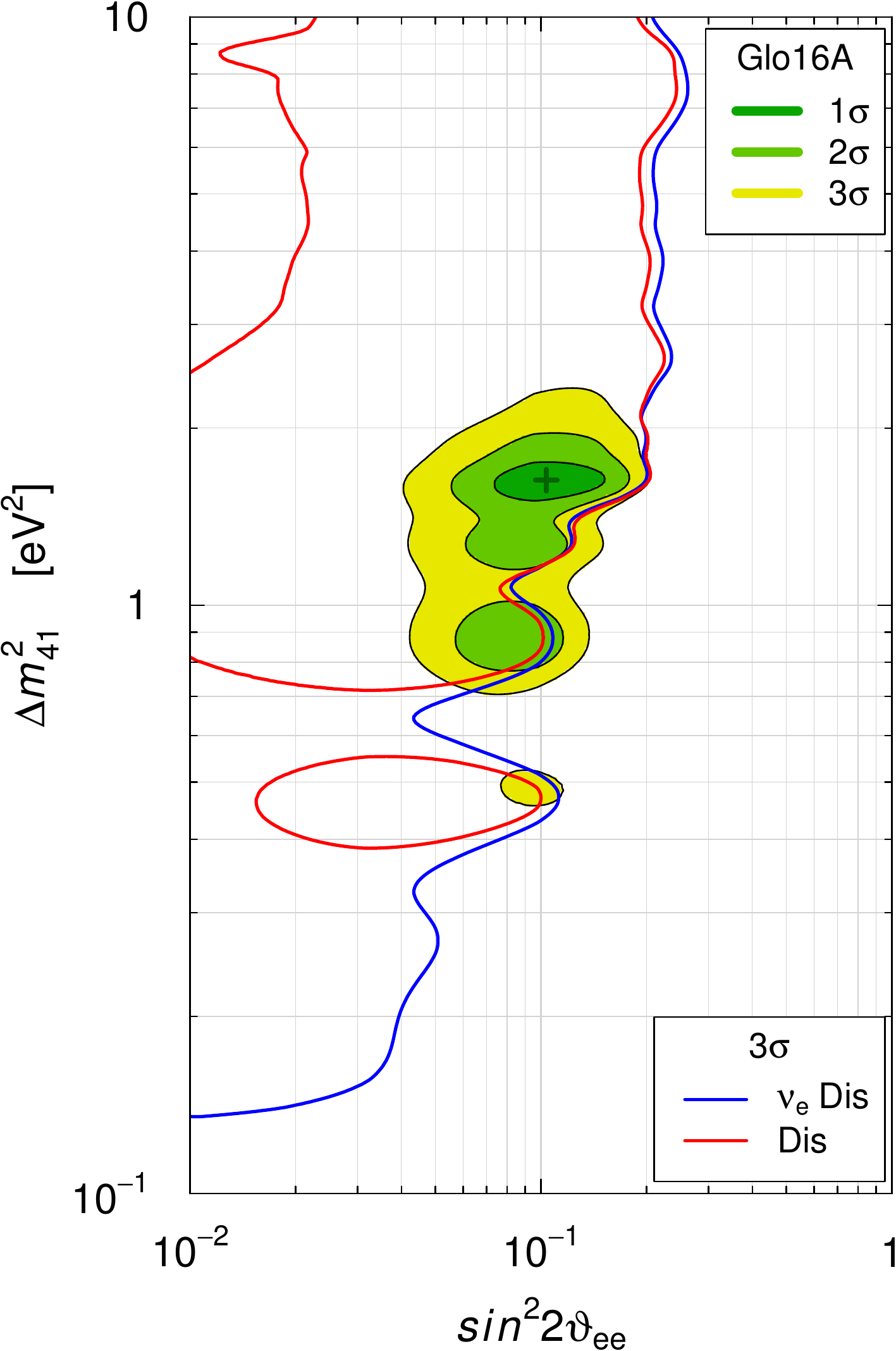}
}
&
\subfigure[]{\label{fig:glo16a-smm}
\includegraphics*[width=0.3\linewidth]{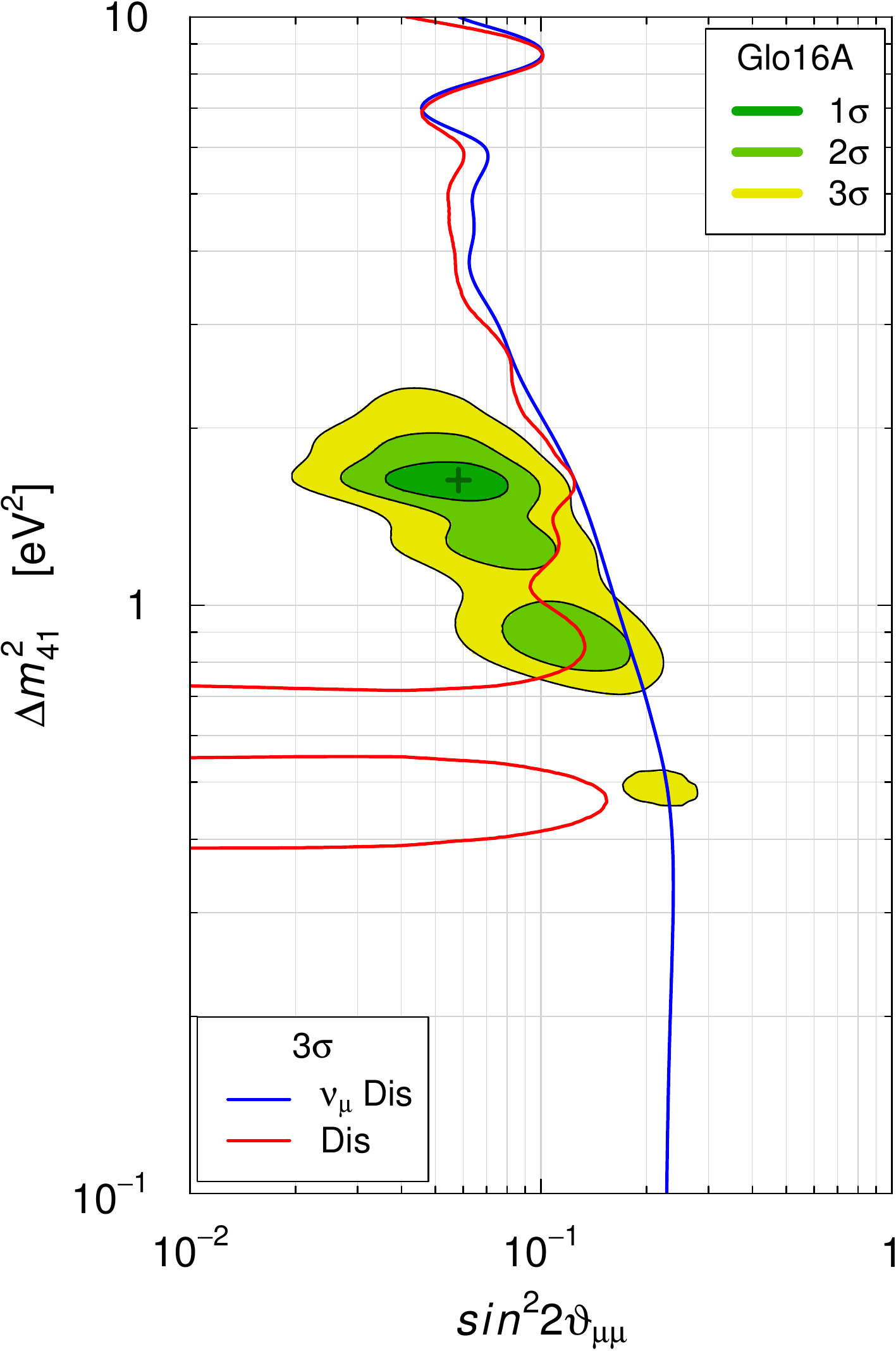}
}
\end{tabular}
\caption{ \label{fig:glo16a}
Allowed regions in the
$\sin^{2}2\vartheta_{e\mu}$--$\Delta{m}^{2}_{41}$ \subref{fig:glo16a-sem},
$\sin^{2}2\vartheta_{ee}$--$\Delta{m}^{2}_{41}$ \subref{fig:glo16a-see},
and
$\sin^{2}2\vartheta_{\mu\mu}$--$\Delta{m}^{2}_{41}$ \subref{fig:glo16a-smm}
planes
obtained in the 3+1 global fit
``Glo16A''
of the 2016 SBL data without the
MINOS \cite{MINOS:2016viw} and
IceCube \cite{TheIceCube:2016oqi}
data.
There is a comparison with the $3\sigma$ allowed regions
obtained from
$\protect\nua{\mu}\to\protect\nua{e}$
SBL appearance data (App)
and the $3\sigma$ constraints obtained from
$\protect\nua{e}$
SBL disappearance data ($\nu_{e}$ Dis),
$\protect\nua{\mu}$
SBL disappearance data ($\nu_{\mu}$ Dis)
and the
combined $\protect\nua{e}$ and $\protect\nua{\mu}$ SBL disappearance data (Dis).
The best-fit points of the Glo16A and App fits are indicated by crosses.
}
\end{figure}

The results of the Glo16A fit are
shown by the first column of Tab.~\ref{tab:all},
by Fig.~\ref{fig:glo16a}, and
by the solid purple curves in Fig.~\ref{fig:mar},
which gives the marginal $\Delta\chi^{2}$
as a function of the mixing parameters
$\Delta{m}^{2}_{41}$,
$|U_{e4}|^2$, and
$|U_{\mu4}|^2$,
from which one can
obtain the corresponding marginal allowed intervals at different confidence levels.

The global goodness of fit of
$4.8\%$
is acceptable,
but there is a relevant
appearance-disappearance tension quantified by a parameter goodness of fit of
$0.13\%$.
If one is willing to accept such appearance-disappearance tension,
one can adopt the allowed regions of the oscillation parameters
shown in Fig.~\ref{fig:glo16a}.

The Glo16A fit is an update of the GLO fit presented in Ref.~\cite{Gariazzo:2015rra},
with a similar set of data.
It can also be compared with the global fit in
Ref.~\cite{Collin:2016rao},
where a similar set of data was considered.
With respect to Ref.~\cite{Collin:2016rao},
we find larger allowed regions for $\Delta{m}^2 \lesssim 2 \, \text{eV}^2$
and we do not have the allowed region at $\Delta{m}^2 \approx 6 \, \text{eV}^2$
found in Ref.~\cite{Collin:2016rao} at 99\% CL.
However, there is an approximate agreement of our results with those of Ref.~\cite{Collin:2016rao},
with a remarkable closeness of the best-fit point
in the mixing parameter space.

\subsection{Effects of MINOS and IceCube}
\label{sub:Glo16B}

In this subsection
we present the 3+1 global fit ``Glo16B'' with the addition of the 2016 data of the
MINOS \cite{MINOS:2016viw}
and
IceCube \cite{TheIceCube:2016oqi}
experiments.
The results are shown by the second column of Tab.~\ref{tab:all},
by Fig.~\ref{fig:glo16b},
and
by the solid blue curves in Fig.~\ref{fig:mar}.

Comparing Fig.~\ref{fig:glo16b-smm} with Fig.~\ref{fig:glo16a-smm},
one can see that the addition of the MINOS and IceCube data
leads to the exclusion of the
low-$\Delta{m}^{2}_{41}$--high-$\sin^{2}2\vartheta_{\mu\mu}$
part of the allowed region,
as expected
(see the discussion in section~\ref{sec:introduction}).
On the other hand,
the high-$\Delta{m}^{2}_{41}$--low-$\sin^{2}2\vartheta_{\mu\mu}$
part of the allowed region is practically unaffected
by the MINOS and IceCube constraints.
As a consequence,
also the low-$\Delta{m}^{2}_{41}$--high-$\sin^{2}2\vartheta_{e\mu}$
part of the allowed region in Fig.~\ref{fig:glo16a-sem}
is excluded in Fig.~\ref{fig:glo16b-sem},
whereas
the high-$\Delta{m}^{2}_{41}$--low-$\sin^{2}2\vartheta_{e\mu}$
part of the allowed region is practically unaffected.

From Tab.~\ref{tab:all} one can see that
including the MINOS and IceCube data increases the appearance-disappearance tension by lowering the
parameter goodness of fit from
$0.13\%$
to
$0.075\%$.
This is a consequence of the reduction of the upper limit of the allowed range of
$|U_{\mu4}|^2$
in the Glo16B fit
with respect to the Glo16A fit
shown in Fig.~\ref{fig:mar-um4}.

Figure~\ref{fig:cmp-sem}
shows the effect of adding to the data set of the Glo16A fit
the MINOS and IceCube data
separately and together.
One can see that the IceCube data are slightly more effective than the
MINOS data in reducing the
low-$\Delta{m}^{2}_{41}$--high-$\sin^{2}2\vartheta_{e\mu}$
part of the allowed region.

The MINOS and IceCube data give information not only on
the 3+1 mixing parameters
$\Delta{m}^{2}_{41}$,
$|U_{e4}|^2$, and
$|U_{\mu4}|^2$ that we have considered so far,
but also on $|U_{\tau4}|^2$.
The sensitivity to $|U_{\tau4}|^2$ is due
in MINOS to the neutral-current event sample
\cite{MINOS:2016viw}
and in IceCube
to the matter effects for high-energy neutrinos propagating in the Earth,
which depend on all the elements of the mixing matrix
\cite{Nunokawa:2003ep,Choubey:2007ji,Razzaque:2011ab,Razzaque:2012tp,Esmaili:2012nz,Esmaili:2013vza,Esmaili:2013cja,Lindner:2015iaa}.
Limits on the value of $|U_{\tau4}|^2$
have been obtained in the analyses of the atmospheric neutrino data of the
Super-Kamiokande
\cite{Abe:2014gda}
and
IceCube DeepCore
\cite{Aartsen:2017bap}
experiments,
in the analysis of the data of the MINOS experiment \cite{Adamson:2010wi,Adamson:2011ku,MINOS:2016viw},
and in the phenomenological fits in Refs.~\cite{Kopp:2013vaa,Collin:2016aqd}.
There is also a bound on
$\sin^22\vartheta_{\mu\tau} = 4 |U_{\mu4}|^2 |U_{\tau4}|^2$
given by the absence of a 3+1 excess of
$\nu_{\mu}\to\nu_{\tau}$
oscillations in the OPERA experiment
\cite{Agafonova:2015neo}.

Figure~\ref{fig:mar-ut4}
shows the marginal $\Delta\chi^{2}$
as a function of $|U_{\tau4}|^2$
in our Glo16B fit,
from which one can see that we obtain the stringent upper bound
\begin{equation}
|U_{\tau4}|^2
\lesssim
0.022
\,
(0.071)
\quad
\text{at}
\quad
2\sigma
\,
(3\sigma)
.
\label{ut4glo16b}
\end{equation}
At 90\% CL we have
$
|U_{\tau4}|^2
\lesssim
0.014
$
and
$
\vartheta_{34} \lesssim 7.4^{\circ}
$
in the common parameterization of the $4\times4$ unitary mixing matrix
used in Ref.~\cite{Collin:2016aqd}.
This bound on $\vartheta_{34}$ is about the same as that reported in
Ref.~\cite{Collin:2016aqd}
for $\Delta{m}^{2}_{41} \approx 6 \, \text{eV}^2$.
However,
we do not find a 90\% CL allowed region of the mixing parameters
at $\Delta{m}^{2}_{41} \approx 6 \, \text{eV}^2$
and our bound on $\vartheta_{34}$ applies for any value of
$\Delta{m}^{2}_{41}$.

Figure~\ref{fig:cmp-ut4}
shows the correlated bounds on
$|U_{\tau4}|^2$
and
$\Delta{m}^{2}_{41}$
that we obtain considering
the MINOS and IceCube data
separately and together.
One can see that the IceCube data give more stringent constraints on
$|U_{\tau4}|^2$
than the MINOS data for
$\Delta{m}^{2}_{41} \lesssim 1.5 \, \text{eV}^2$.

\subsection{Effects of NEOS}
\label{sub:Glo17}

We finally consider also the NEOS \cite{Ko:2016owz} data
and obtain the 3+1 global fit ``Glo17'' which includes all data available so far in 2017.
The results are shown by
the third column of Tab.~\ref{tab:all},
by Fig.~\ref{fig:all},
and by the solid orange curves in Fig.~\ref{fig:mar}.

Comparing Fig.~\ref{fig:all} with Fig.~\ref{fig:glo16b},
it is evident that the inclusion of the NEOS constraints
has a dramatic effect on the allowed regions,
leading to the fragmentation of the allowed region
in three islands with narrow
$\Delta{m}^{2}_{41}$ widths.
The best-fit island is at
$\Delta{m}^{2}_{41} \approx 1.7 \, \text{eV}^2$.
There is an island allowed at $2\sigma$ at
$\Delta{m}^{2}_{41} \approx 1.3 \, \text{eV}^2$,
and
an island allowed at $3\sigma$ at
$\Delta{m}^{2}_{41} \approx 2.4 \, \text{eV}^2$.
Moreover,
the NEOS constraints shifts
the $3\sigma$ allowed range of $|U_{e4}|^2$
from
$ 0.014 - 0.051 $
in the Glo16B fit
to
$ 0.011 - 0.032 $
in the Glo17 fit,
as shown in Fig.~\ref{fig:mar}.
Therefore,
the appearance-disappearance tension is increased,
as shown by the
$0.019\%$
parameter goodness of fit in Tab.~\ref{tab:all}.
Since this low value of the appearance-disappearance
parameter goodness of fit is hardly acceptable,
we are led to consider, in the next subsection,
the ``pragmatic approach''
proposed in Ref.~\cite{Giunti:2013aea}.

\subsection{Pragmatic fit}
\label{sub:PrGlo17}

In this section we consider the ``pragmatic approach'' \cite{Giunti:2013aea}
in which
the low-energy bins of the MiniBooNE experiment
\cite{AguilarArevalo:2008rc,Aguilar-Arevalo:2013pmq}
which have an anomalous excess of $\nua{e}$-like events
are omitted from the global fit.
As shown in Fig.~1b of Ref.~\cite{Giunti:2016oan},
the region allowed by the appearance data
shifts towards larger values of $\Delta{m}^2_{41}$
and smaller values of $\sin^22\vartheta_{e\mu}$
when the MiniBooNE low-energy bins are omitted from the fit.
As a result, the overlap of the appearance and disappearance allowed regions
increases,
relieving the appearance-disappearance tension.

One can question the scientific correctness of the data selection
in the pragmatic approach,
but we note that the MiniBooNE low-energy excess
is widely considered to be suspicious\footnote{%
Part of the MiniBooNE low-energy anomaly
may be explained by taking into account nuclear effects in the energy reconstruction
\cite{Martini:2012fa,Martini:2012uc},
but this effect is not sufficient to solve the problem
\cite{Ericson:2016yjn}.
}
because of the large background.
Some of this background can be due to photon events which
are indistinguishable from $\nua{e}$ events in the
MiniBooNE liquid scintillator detector.
These photons can be generated by the decays of $\pi^{0}$'s
produced by the neutral-current interactions of the $\nua{\mu}$ beam.
When only one of the two photons emitted in the $\pi^{0}$ decay
is visible,
its signal cannot be distinguished from a $\nua{e}$ event in a liquid-scintillator detector.
The suspicion that this photon background may be responsible for the
MiniBooNE low-energy excess motivated the realization of the
MicroBooNE
experiment at Fermilab
\cite{Gollapinni:2015lca},
which is able to distinguish between photon and $\nua{e}$
events by using a Liquid Argon Time Projection Chamber (LArTPC).
Waiting for the results of this experiment,
we think that it is reasonable to
adopt the pragmatic approach
of omitting from the global fit
the MiniBooNE low-energy data.

\begin{figure}[!t]
\centering
\setlength{\tabcolsep}{0pt}
\begin{tabular}{cc}
\subfigure[]{\label{fig:mar-d41}
\includegraphics*[width=0.49\linewidth]{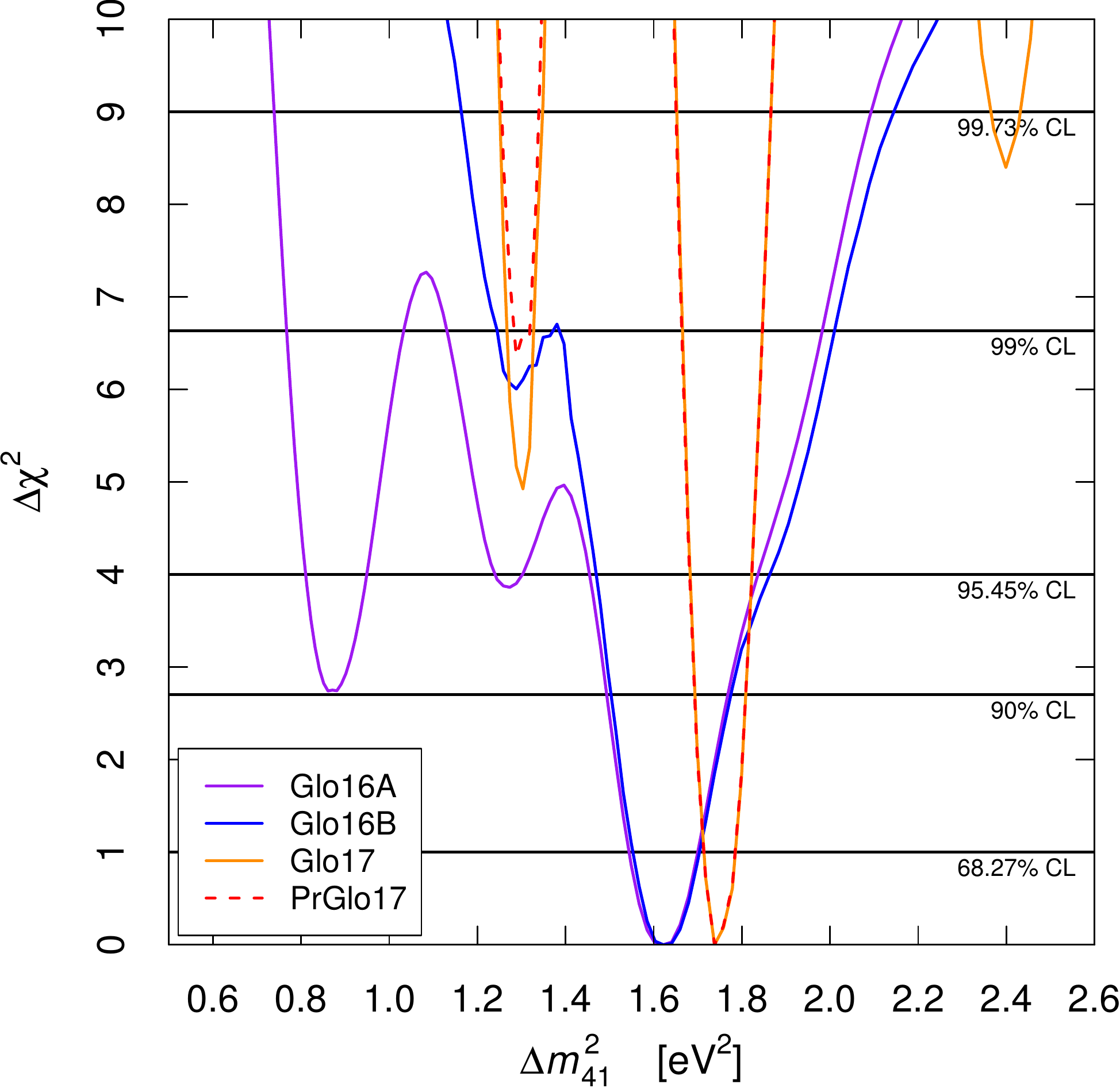}
}
&
\subfigure[]{\label{fig:mar-ue4}
\includegraphics*[width=0.49\linewidth]{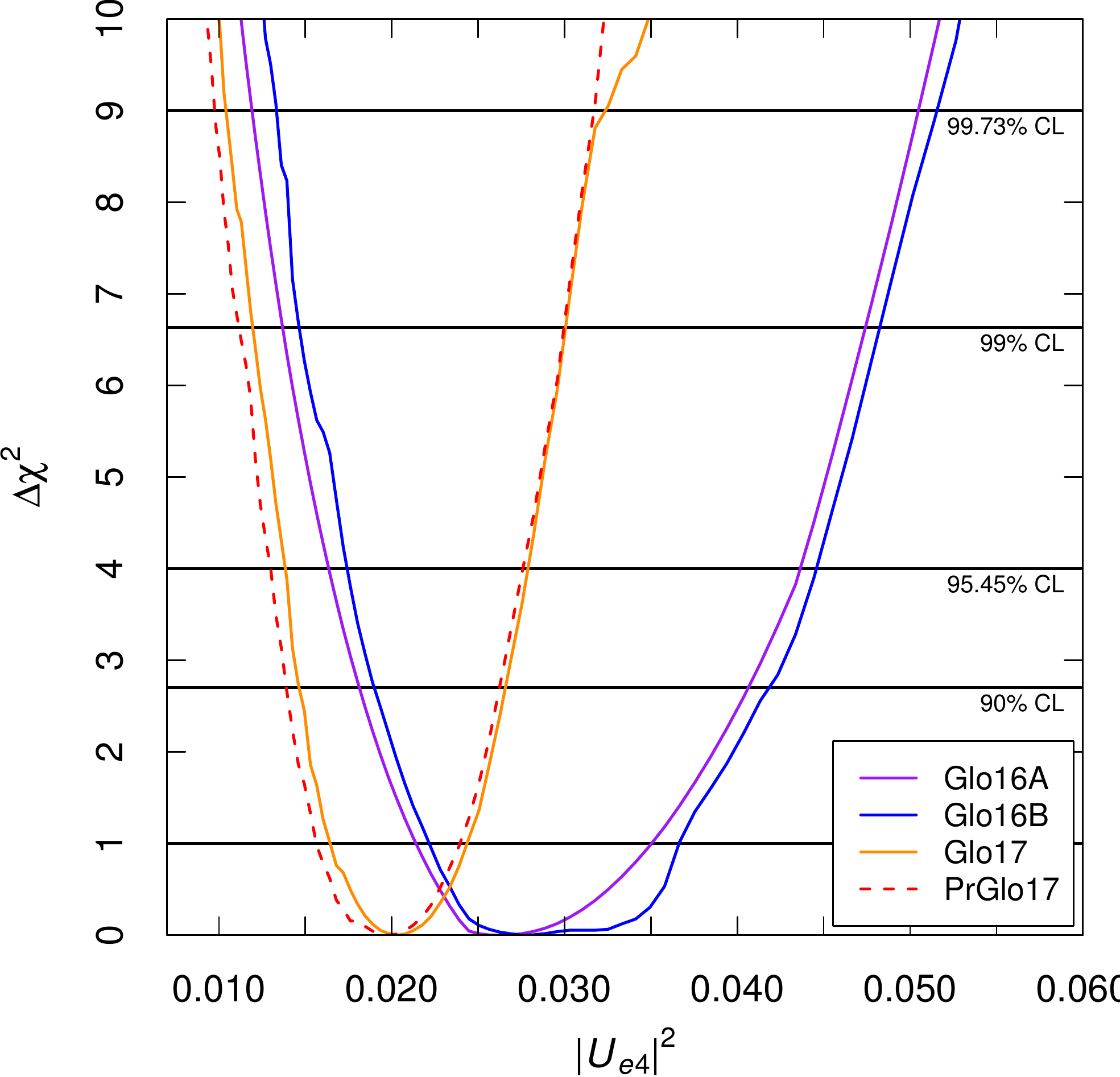}
}
\\
\subfigure[]{\label{fig:mar-um4}
\includegraphics*[width=0.49\linewidth]{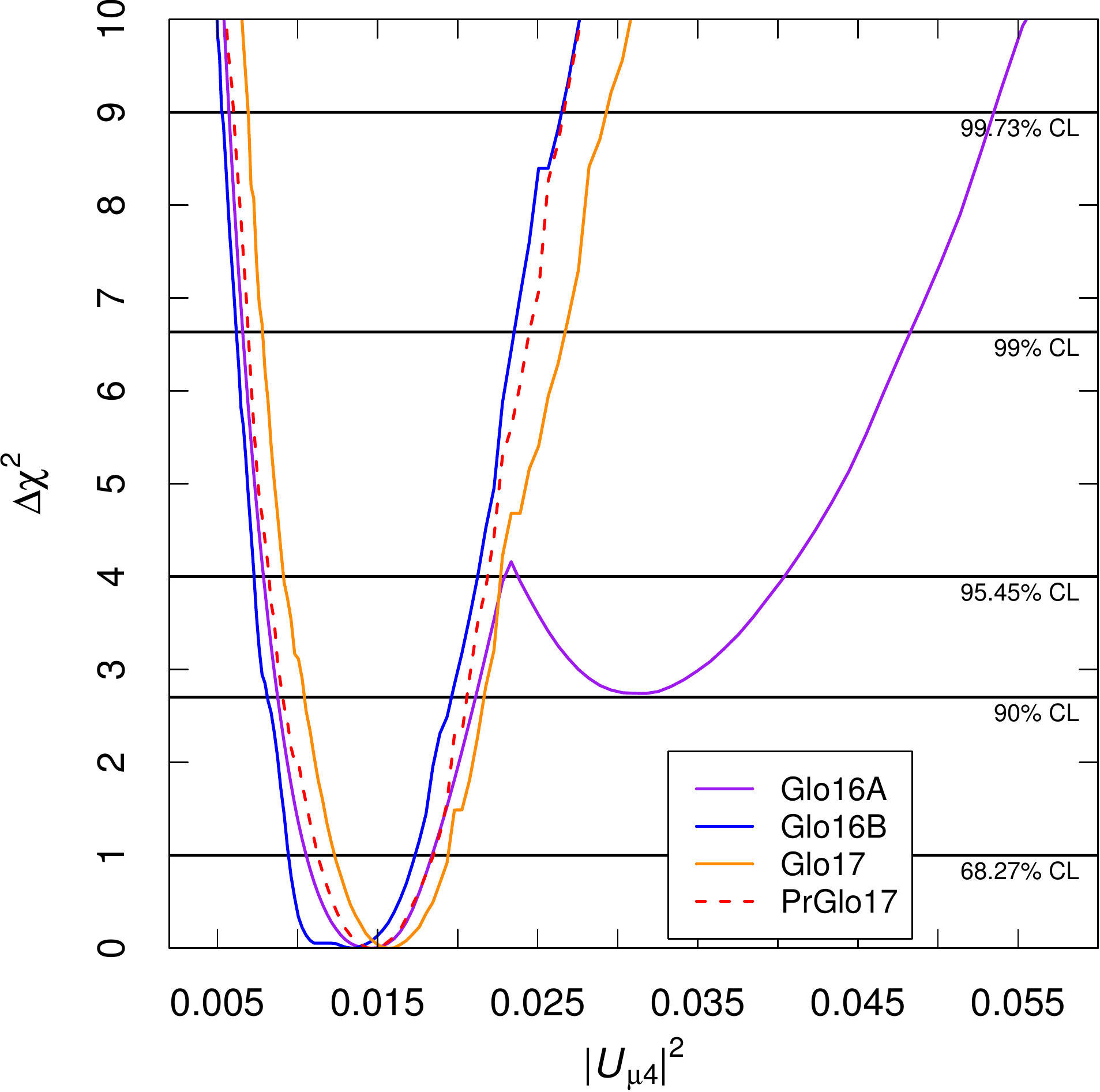}
}
&
\subfigure[]{\label{fig:mar-ut4}
\includegraphics*[width=0.49\linewidth]{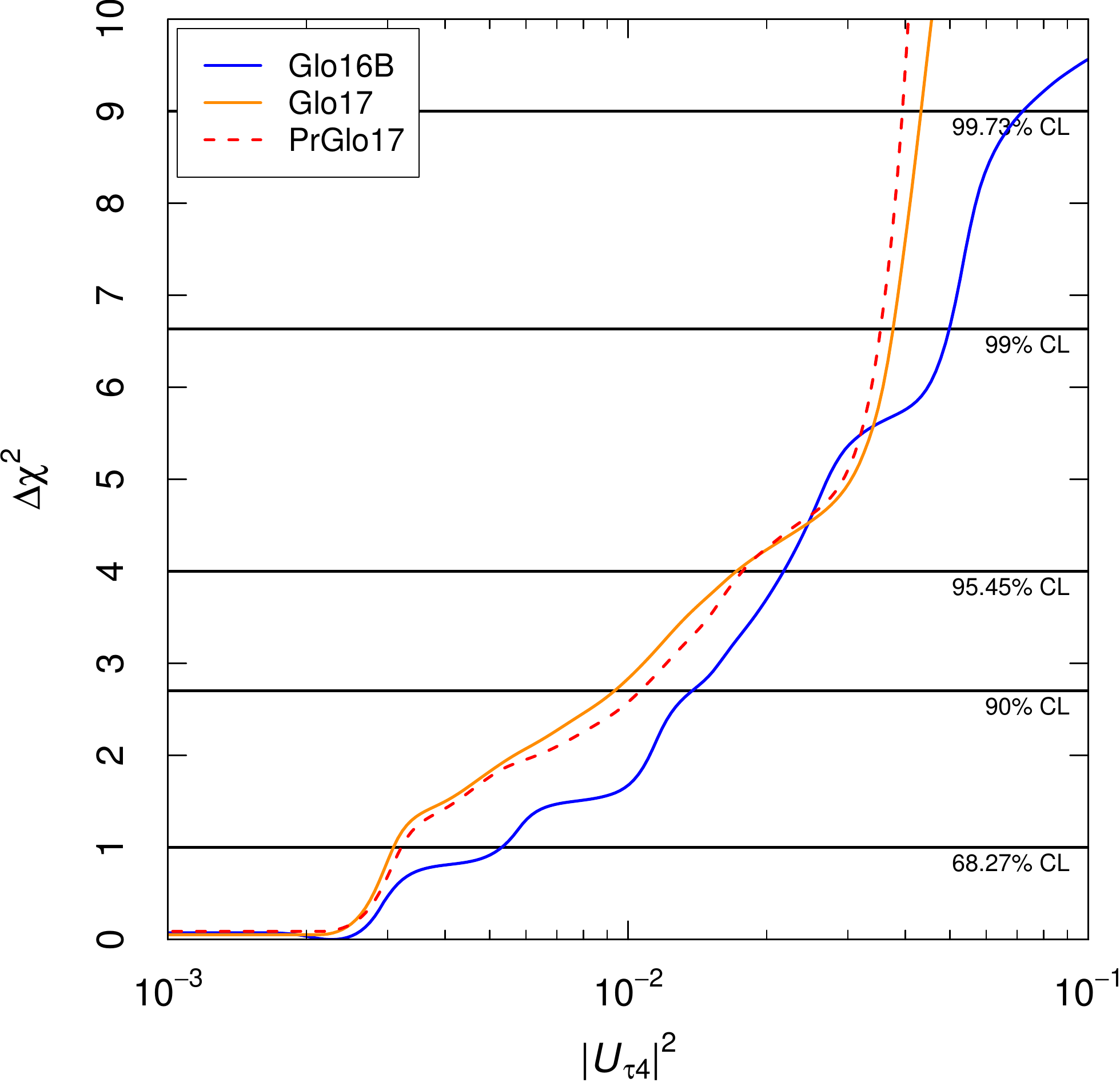}
}
\end{tabular}
\caption{ \label{fig:mar}
Marginal $\Delta\chi^{2} = \chi^2 - \chi^2_{\text{min}}$
as a function of the mixing parameters
$\Delta{m}^{2}_{41}$ \subref{fig:mar-d41},
$|U_{e4}|^2$ \subref{fig:mar-ue4},
$|U_{\mu4}|^2$ \subref{fig:mar-um4}, and
$|U_{\tau4}|^2$ \subref{fig:mar-ut4}.
The black horizontal lines show the
$\Delta\chi^{2}$
for one degree of freedom corresponding to the indicated confidence level (CL).
}
\end{figure}

\begin{figure}[!t]
\centering
\setlength{\tabcolsep}{0pt}
\begin{tabular}{ccc}
\subfigure[]{\label{fig:glo16b-sem}
\includegraphics*[width=0.3\linewidth]{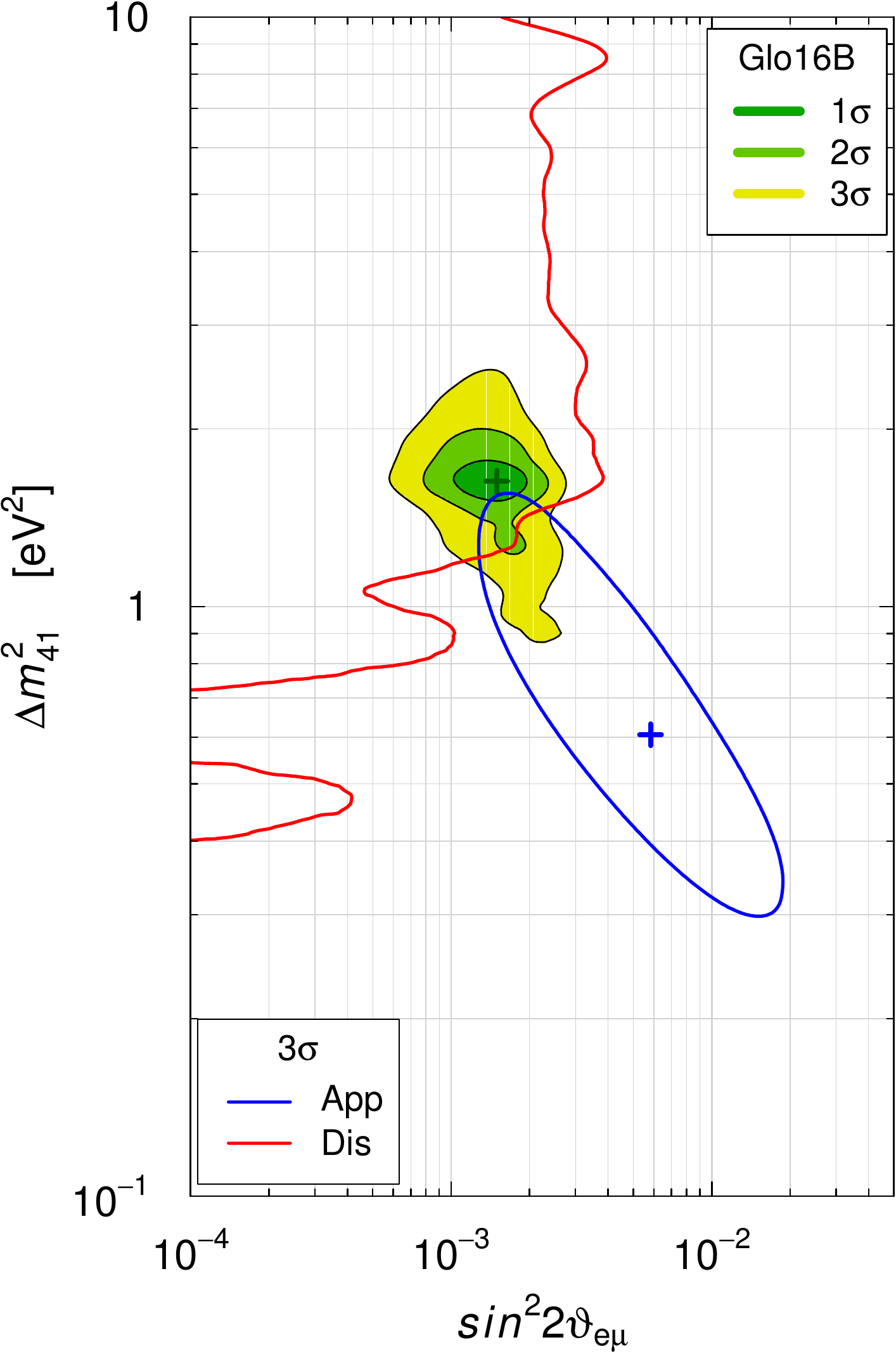}
}
&
\subfigure[]{\label{fig:glo16b-see}
\includegraphics*[width=0.3\linewidth]{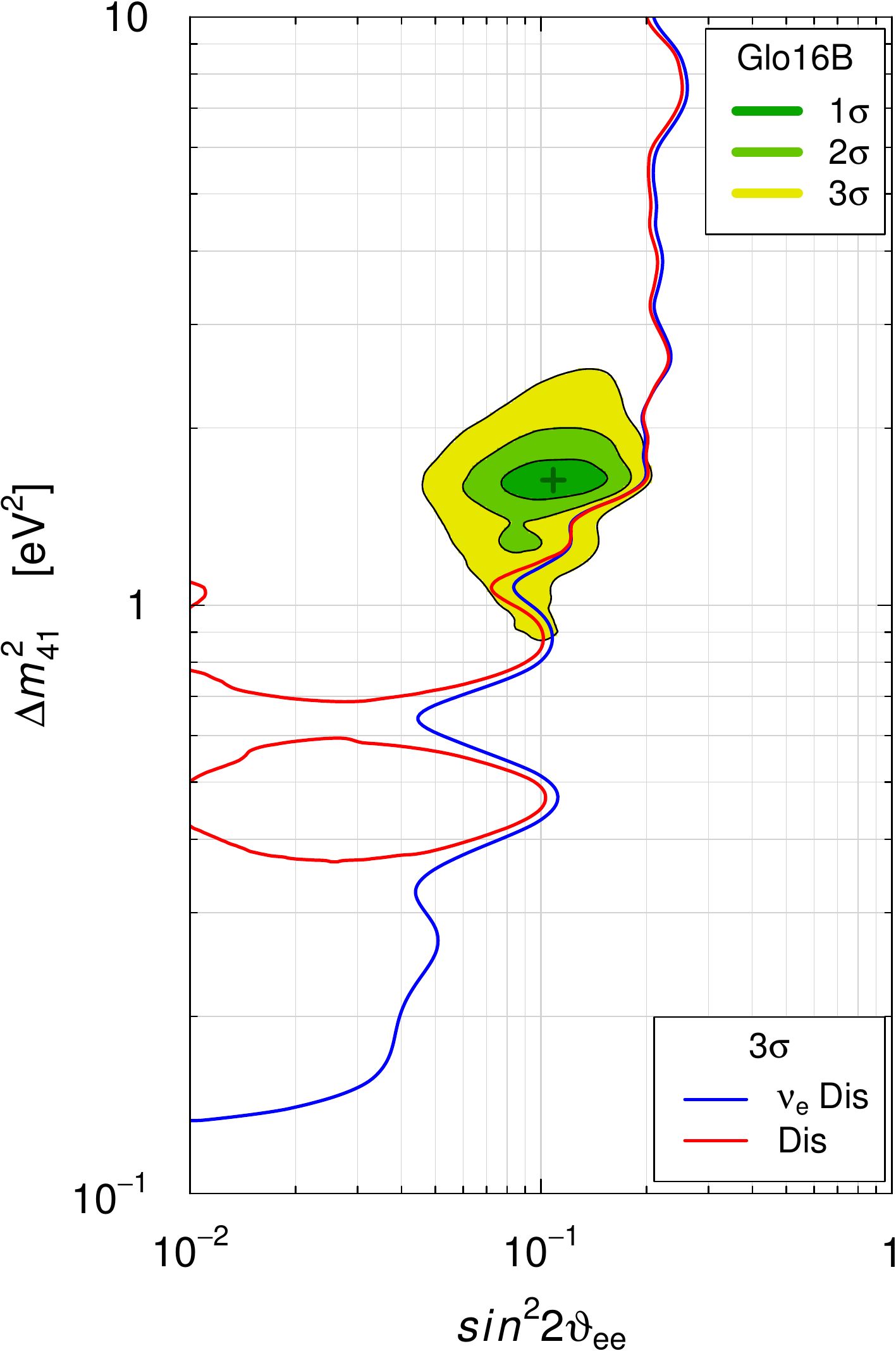}
}
&
\subfigure[]{\label{fig:glo16b-smm}
\includegraphics*[width=0.3\linewidth]{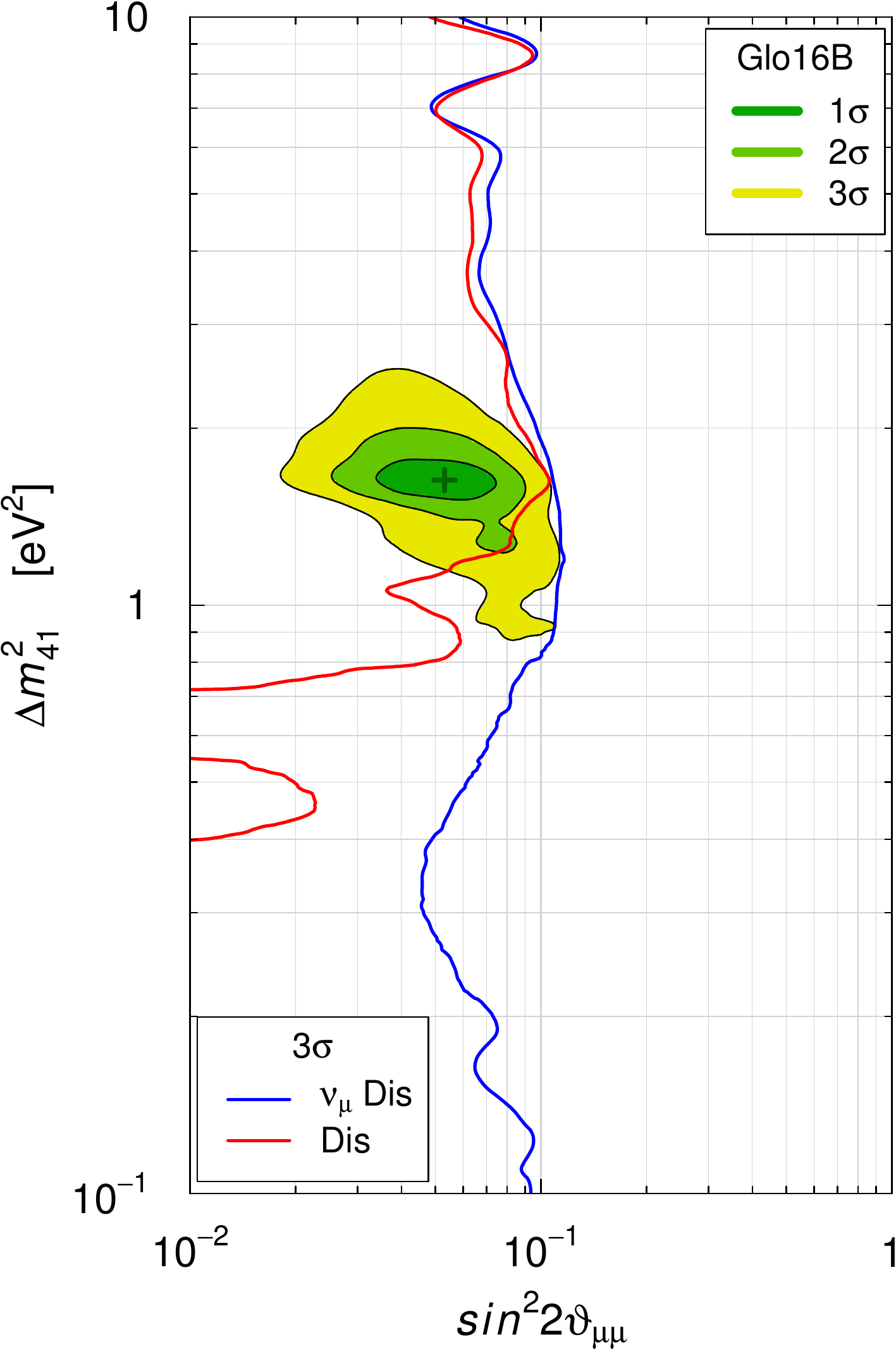}
}
\end{tabular}
\caption{ \label{fig:glo16b}
Allowed regions in the
$\sin^{2}2\vartheta_{e\mu}$--$\Delta{m}^{2}_{41}$ \subref{fig:glo16b-sem},
$\sin^{2}2\vartheta_{ee}$--$\Delta{m}^{2}_{41}$ \subref{fig:glo16b-see}, and
$\sin^{2}2\vartheta_{\mu\mu}$--$\Delta{m}^{2}_{41}$ \subref{fig:glo16b-smm},
planes
obtained in the 3+1 global fit
``Glo16B''
of all 2016 SBL data.
There is a comparison with the $3\sigma$ allowed regions
obtained from
$\protect\nua{\mu}\to\protect\nua{e}$
SBL appearance data (App)
and the $3\sigma$ constraints obtained from
$\protect\nua{e}$
SBL disappearance data ($\nu_{e}$ Dis),
$\protect\nua{\mu}$
SBL disappearance data ($\nu_{\mu}$ Dis)
and the
combined $\protect\nua{e}$ and $\protect\nua{\mu}$ SBL disappearance data (Dis).
The best-fit points of the Glo16B and App fits are indicated by crosses.
}
\end{figure}

\begin{figure}[!t]
\centering
\setlength{\tabcolsep}{0pt}
\begin{tabular}{cc}
\subfigure[]{\label{fig:cmp-sem}
\includegraphics*[width=0.49\linewidth]{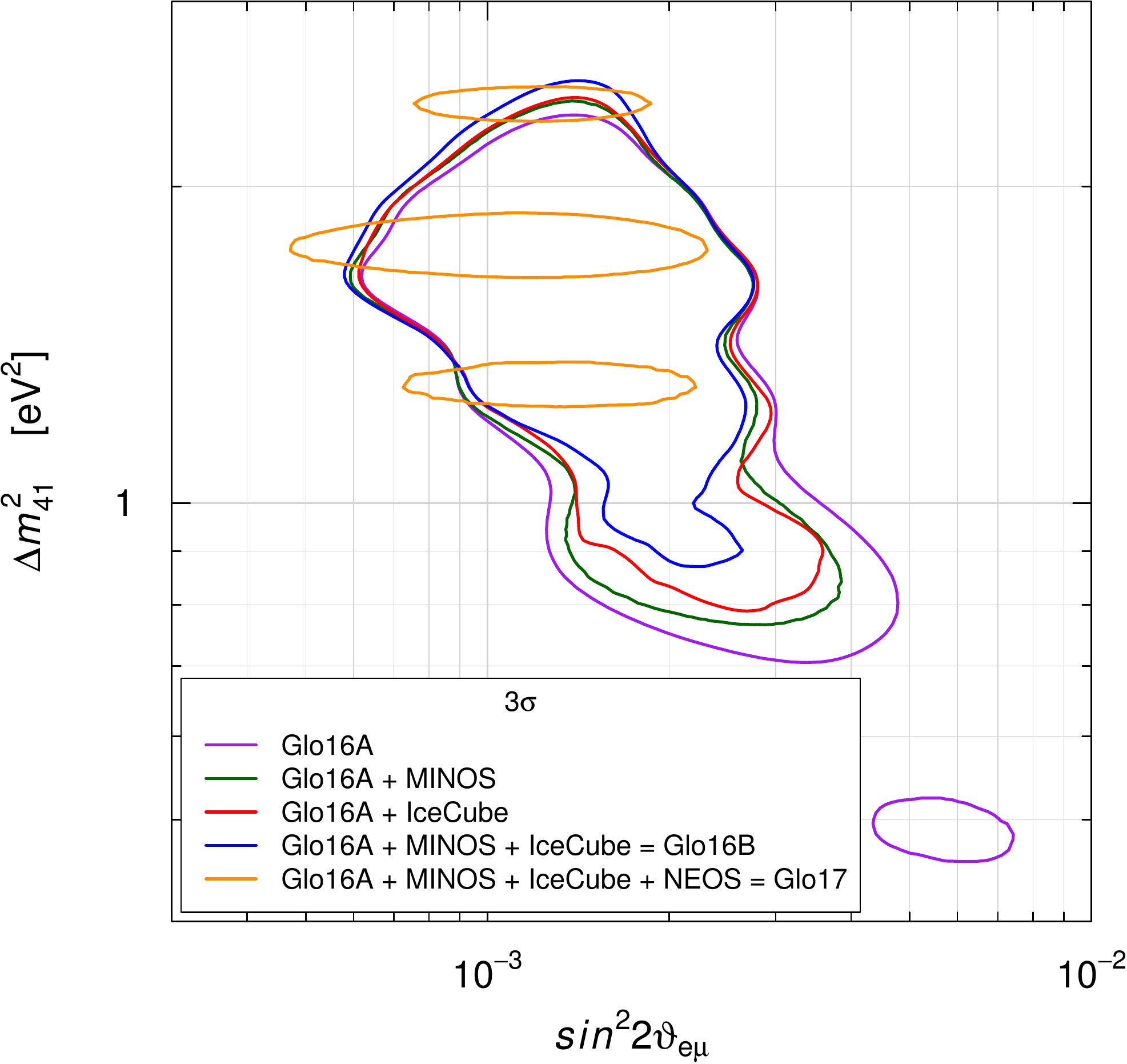}
}
&
\subfigure[]{\label{fig:cmp-ut4}
\includegraphics*[viewport=0 0 568 536, width=0.49\linewidth]{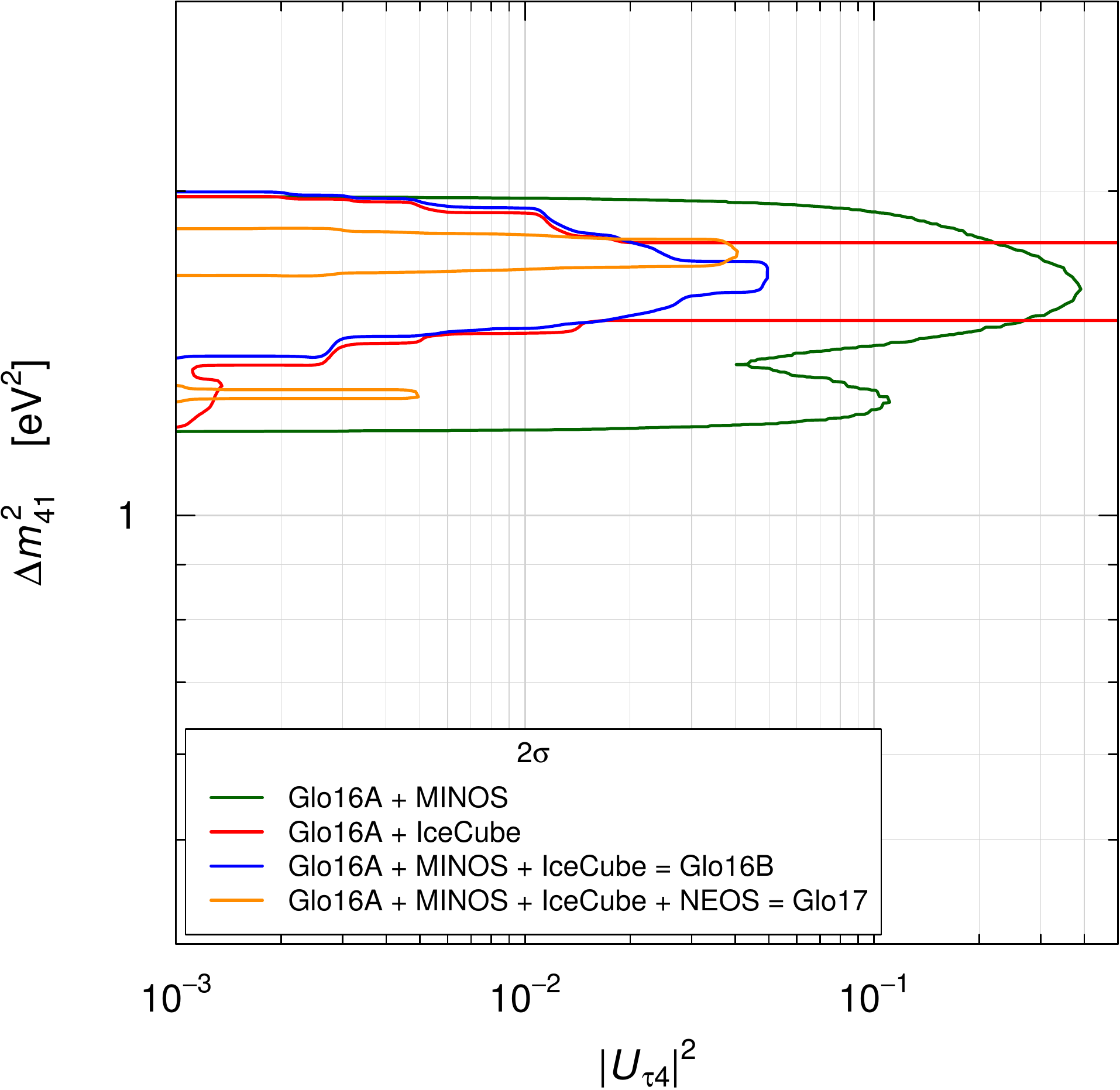}
}
\end{tabular}
\caption{ \label{fig:cmp}
Comparison of
\subref{fig:cmp-sem}
the $3\sigma$ allowed regions in the
$\sin^{2}2\vartheta_{e\mu}$--$\Delta{m}^{2}_{41}$
plane
and
\subref{fig:cmp-ut4}
the $2\sigma$ allowed regions in the
$|U_{\tau4}|^2$--$\Delta{m}^{2}_{41}$ 
plane
obtained by adding to the data set
of the Glo16A fit
the MINOS and IceCube data separately and together,
and by adding also the NEOS data.
}
\end{figure}

\begin{figure}[!t]
\centering
\setlength{\tabcolsep}{0pt}
\begin{tabular}{ccc}
\subfigure[]{\label{fig:all-sem}
\includegraphics*[width=0.3\linewidth]{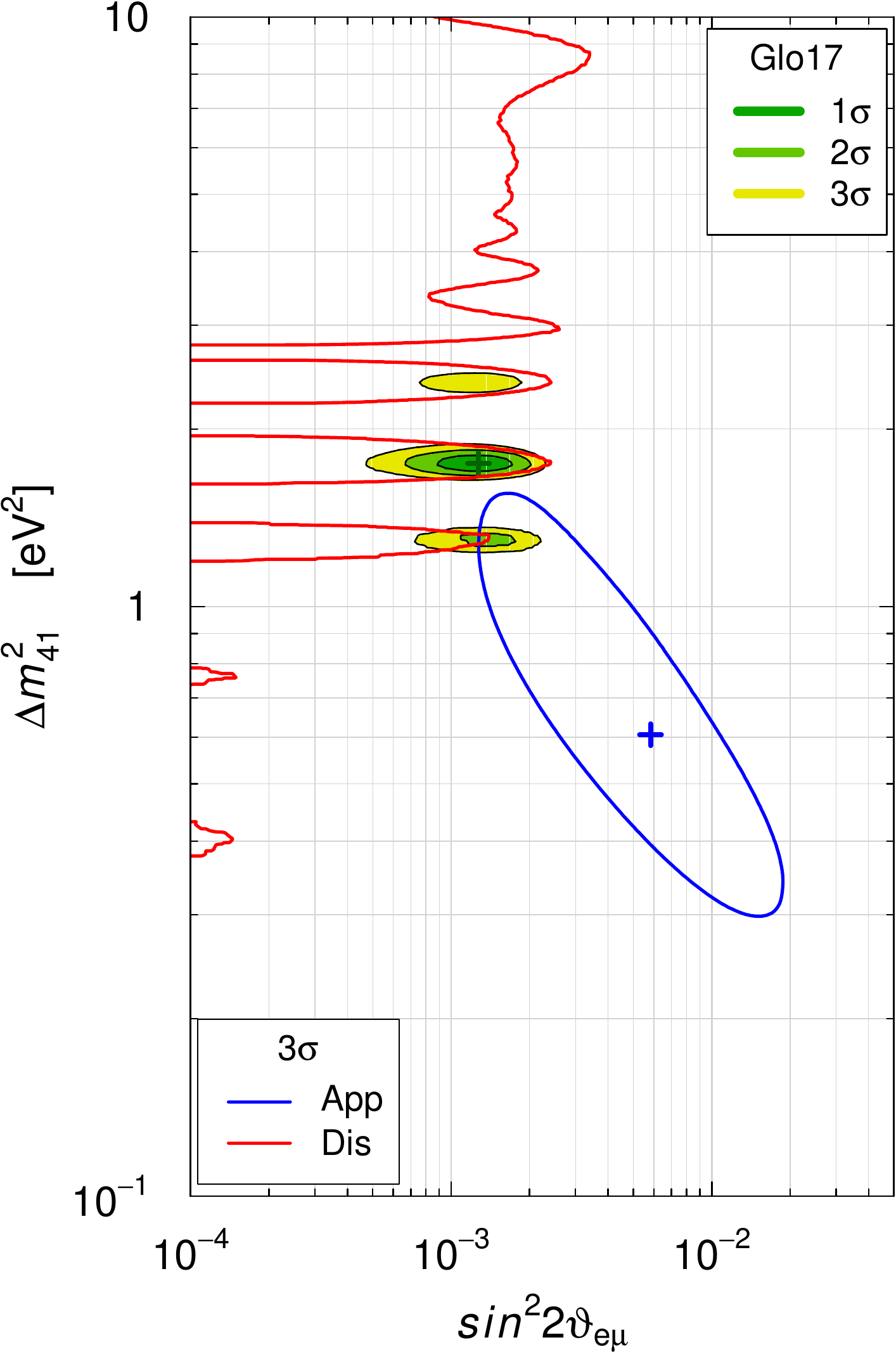}
}
&
\subfigure[]{\label{fig:all-see}
\includegraphics*[width=0.3\linewidth]{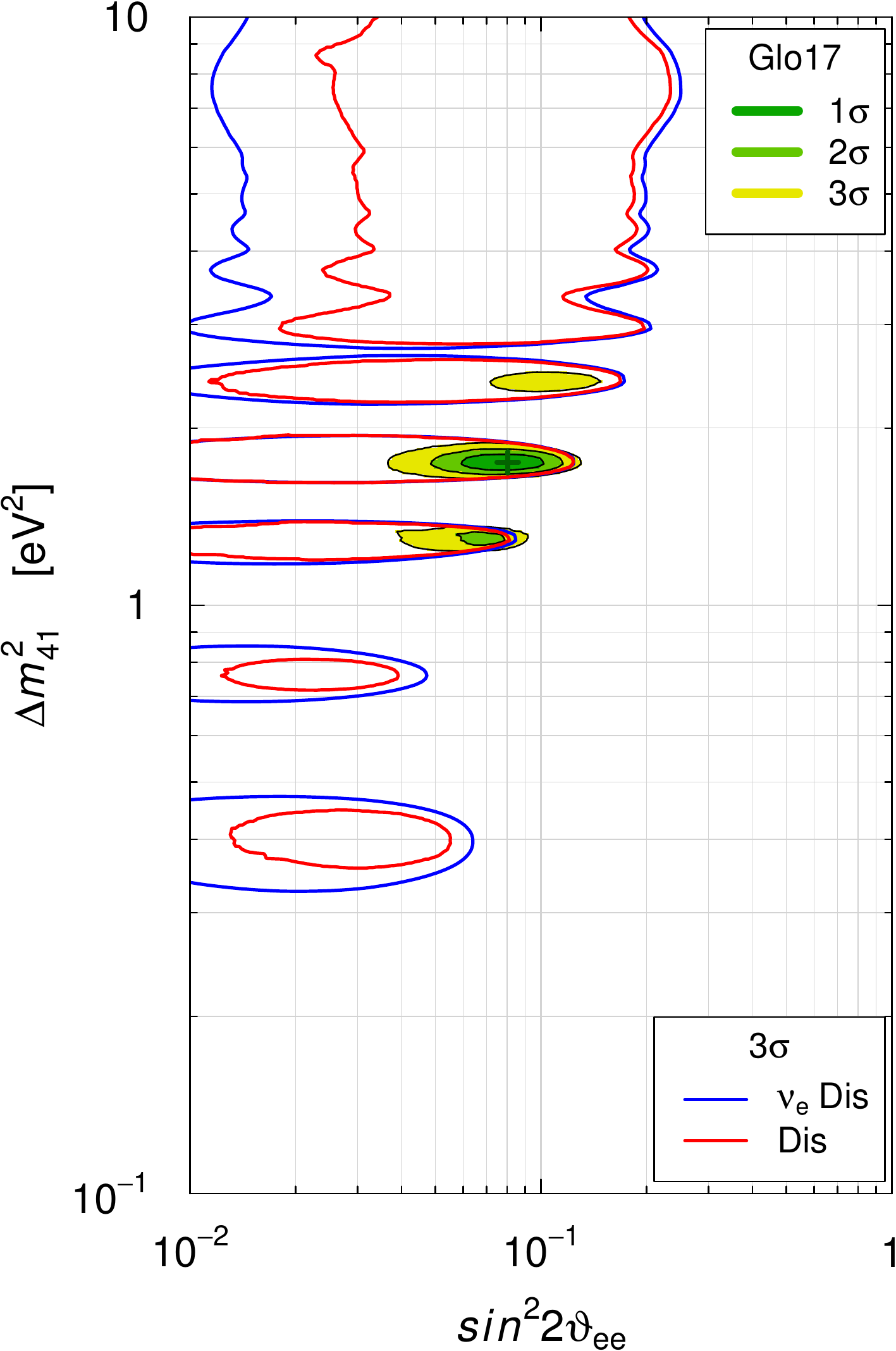}
}
&
\subfigure[]{\label{fig:all-smm}
\includegraphics*[width=0.3\linewidth]{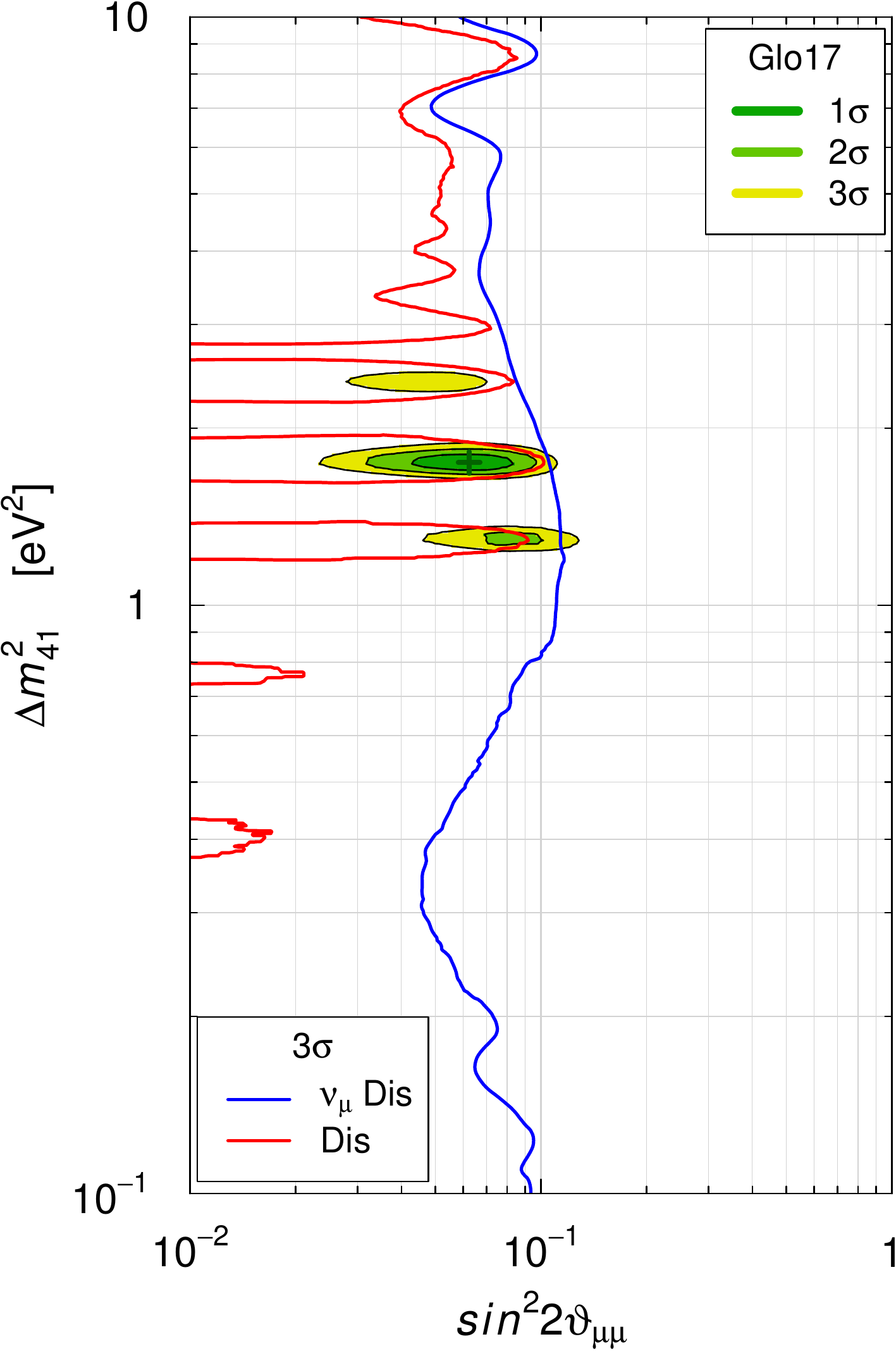}
}
\end{tabular}
\caption{ \label{fig:all}
Allowed regions in the
$\sin^{2}2\vartheta_{e\mu}$--$\Delta{m}^{2}_{41}$ \subref{fig:all-sem},
$\sin^{2}2\vartheta_{ee}$--$\Delta{m}^{2}_{41}$ \subref{fig:all-see}, and
$\sin^{2}2\vartheta_{\mu\mu}$--$\Delta{m}^{2}_{41}$ \subref{fig:all-smm},
planes
obtained in the 3+1 global fit
``Glo17''
of all SBL data.
There is a comparison with the $3\sigma$ allowed regions
obtained from
$\protect\nua{\mu}\to\protect\nua{e}$
SBL appearance data (App)
and the $3\sigma$ constraints obtained from
$\protect\nua{e}$
SBL disappearance data ($\nu_{e}$ Dis),
$\protect\nua{\mu}$
SBL disappearance data ($\nu_{\mu}$ Dis)
and the
combined $\protect\nua{e}$ and $\protect\nua{\mu}$ SBL disappearance data (Dis).
The best-fit points of the Glo17 and App fits are indicated by crosses.
}
\end{figure}

\begin{figure}[!t]
\centering
\setlength{\tabcolsep}{0pt}
\begin{tabular}{ccc}
\subfigure[]{\label{fig:prg-sem}
\includegraphics*[width=0.3\linewidth]{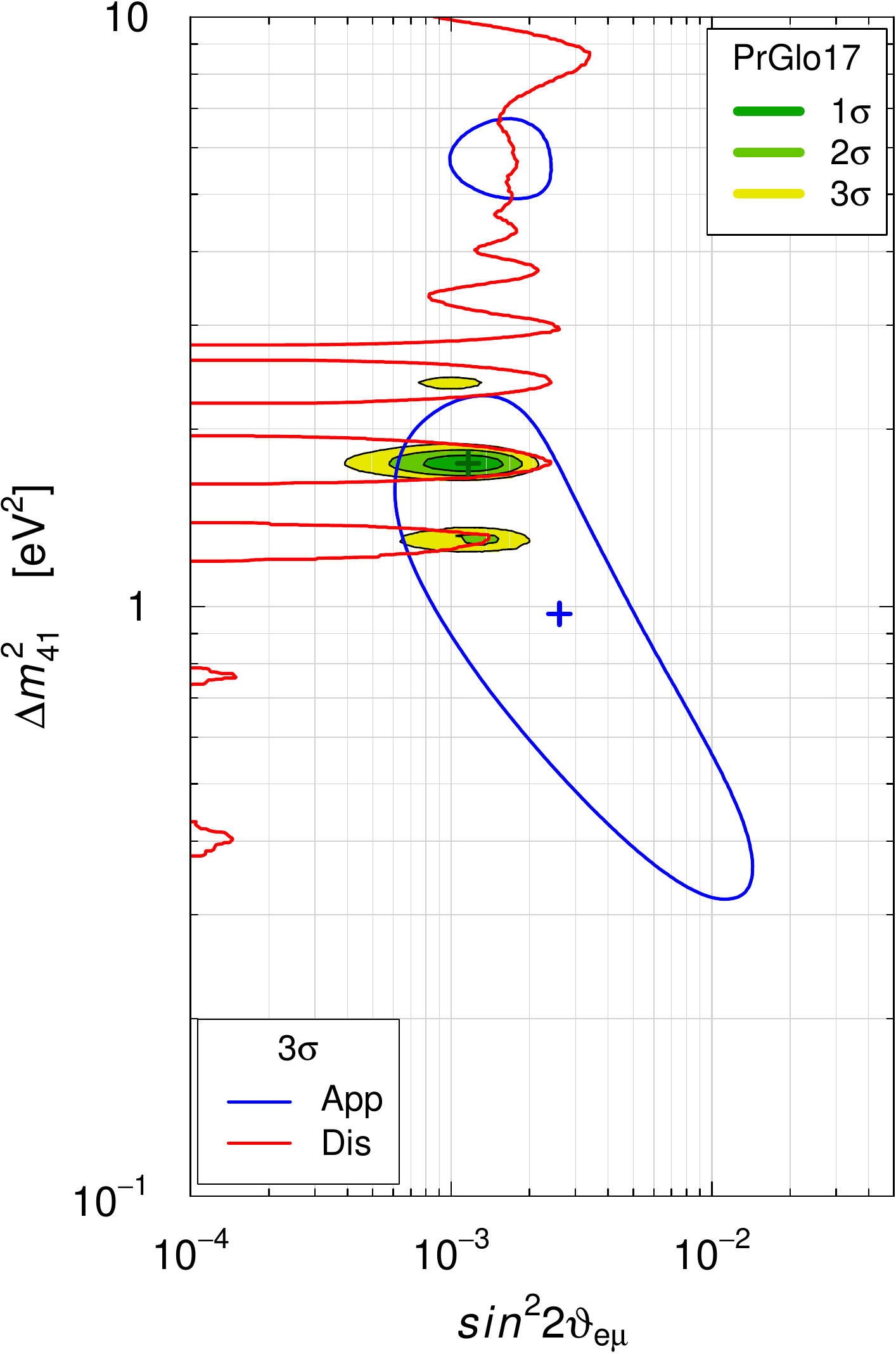}
}
&
\subfigure[]{\label{fig:prg-see}
\includegraphics*[width=0.3\linewidth]{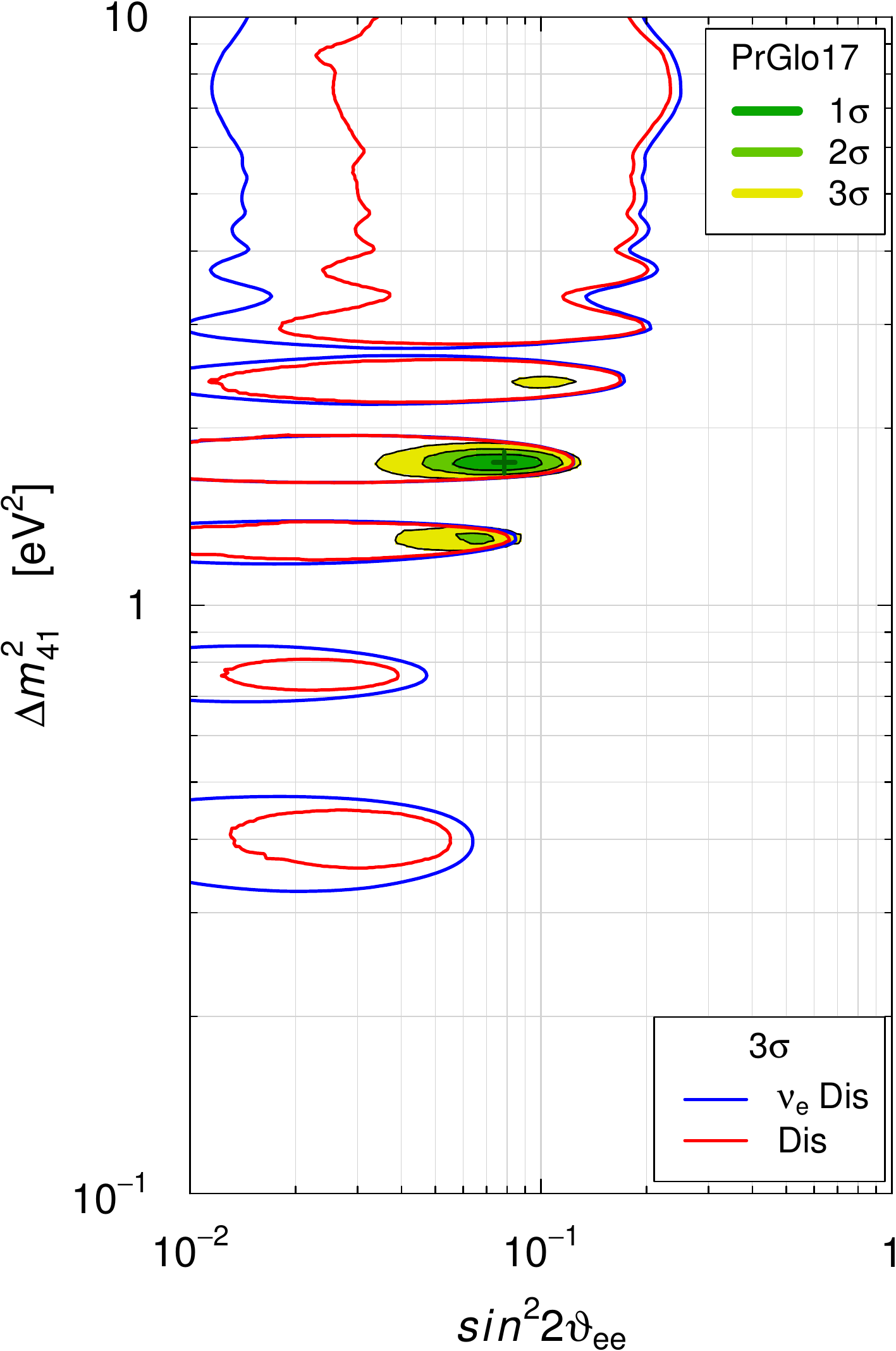}
}
&
\subfigure[]{\label{fig:prg-smm}
\includegraphics*[width=0.3\linewidth]{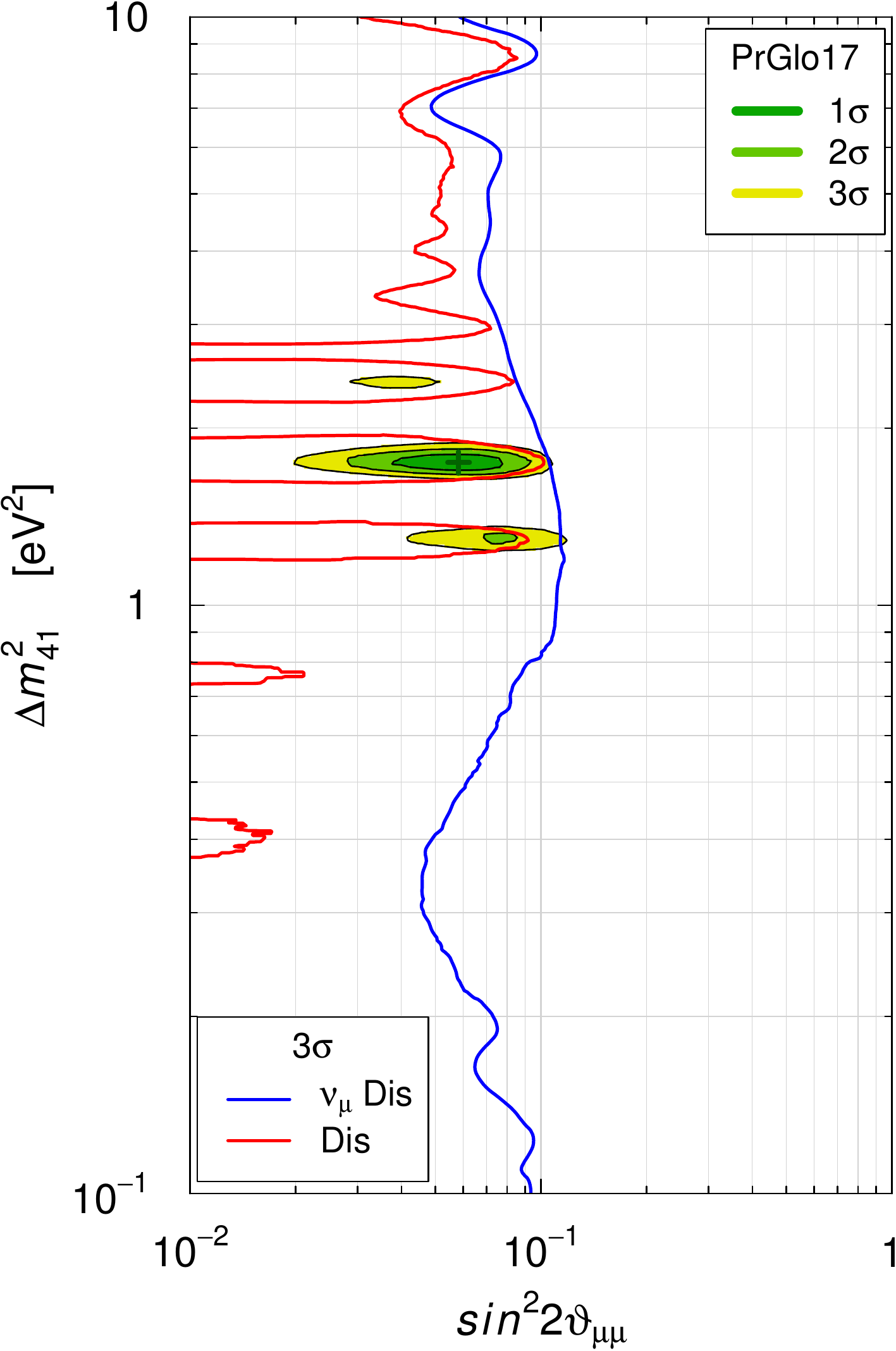}
}
\end{tabular}
\caption{ \label{fig:prg}
Allowed regions in the
$\sin^{2}2\vartheta_{e\mu}$--$\Delta{m}^{2}_{41}$ \subref{fig:prg-sem},
$\sin^{2}2\vartheta_{ee}$--$\Delta{m}^{2}_{41}$ \subref{fig:prg-see}, and
$\sin^{2}2\vartheta_{\mu\mu}$--$\Delta{m}^{2}_{41}$ \subref{fig:prg-smm},
planes
obtained in the pragmatic 3+1 global fit
``PrGlo17''
of SBL data.
There is a comparison with the $3\sigma$ allowed regions
obtained from
$\protect\nua{\mu}\to\protect\nua{e}$
SBL appearance data (App)
and the $3\sigma$ constraints obtained from
$\protect\nua{e}$
SBL disappearance data ($\nu_{e}$ Dis),
$\protect\nua{\mu}$
SBL disappearance data ($\nu_{\mu}$ Dis)
and the
combined $\protect\nua{e}$ and $\protect\nua{\mu}$ SBL disappearance data (Dis).
The best-fit points of the PrGlo17 and App fits are indicated by crosses.
}
\end{figure}

The results of the pragmatic 3+1 global fit ``PrGlo17'',
which includes
the MINOS, IceCube and NEOS data,
are shown by
the fourth column of Tab.~\ref{tab:all},
by Fig.~\ref{fig:prg}, and
by the dashed red curves in Fig.~\ref{fig:mar}.

From Tab.~\ref{tab:all}
one can see that, as expected,
the exclusion from the fit of the MiniBooNE low-energy data
leads to an increase of the
parameter goodness of fit from the unacceptable
$0.019\%$
of the Glo17 fit to the acceptable
$2.7\%$
of the PrGlo17 fit.
There is still a mild appearance-disappearance tension,
but the tolerable value of
parameter goodness of fit leads us to
consider the PrGlo17 fit as acceptable.

Comparing the allowed regions of the oscillation parameters
in Fig.~\ref{fig:prg} for the PrGlo17 fit
with
those in Fig.~\ref{fig:all} for the Glo17 fit
and the corresponding marginal $\Delta\chi^2$ curves in Fig.~\ref{fig:mar},
one can see that the differences are small.
As a consequence of the larger overlap
of the regions allowed by the fits of appearance and disappearance data,
the PrGlo17 fit has a minimum $\chi^2$ significantly smaller than the Glo17 fit,
which leads to an increased preference of the best-fit island at
$\Delta{m}^{2}_{41} \approx 1.7 \, \text{eV}^2$,
to a small shrink of the island at
$\Delta{m}^{2}_{41} \approx 1.3 \, \text{eV}^2$,
and at a significant reduction of the island at
$\Delta{m}^{2}_{41} \approx 2.4 \, \text{eV}^2$
(the corresponding $3\sigma$ interval for one degree of freedom allowed by the marginal $\Delta\chi^2$ in Fig.~\ref{fig:mar-d41}
disappears).

Table~\ref{tab:prgint} gives the marginal allowed intervals of the mixing parameters
$|U_{e4}|^2$,
$|U_{\mu4}|^2$, and
$|U_{\tau4}|^2$.
The stringent upper bounds on $|U_{\tau4}|^2$
slightly improve those found in the Glo16B fit
(see Eq.~(\ref{ut4glo16b}) and Fig.~\ref{fig:mar-ut4}).
At 90\% CL we have
$
|U_{\tau4}|^2
\lesssim
0.011
$
and
$
\vartheta_{34} \lesssim 6^{\circ}
$.

\begin{figure}[!t]
\centering
\setlength{\tabcolsep}{0pt}
\begin{tabular}{cc}
\subfigure[]{\label{fig:fut-sem}
\includegraphics*[width=0.49\linewidth]{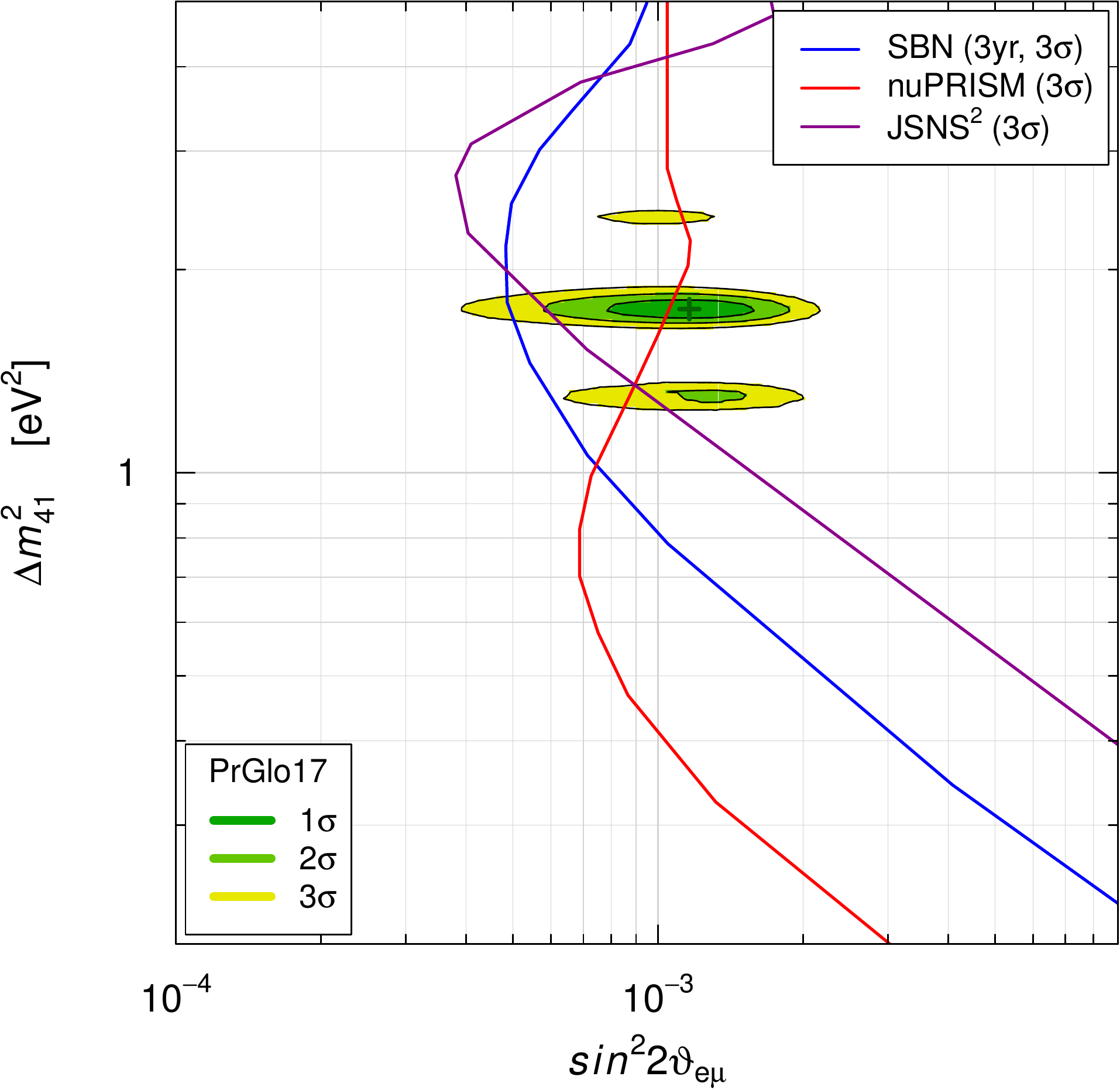}
}
&
\subfigure[]{\label{fig:fut-smm}
\includegraphics*[width=0.49\linewidth]{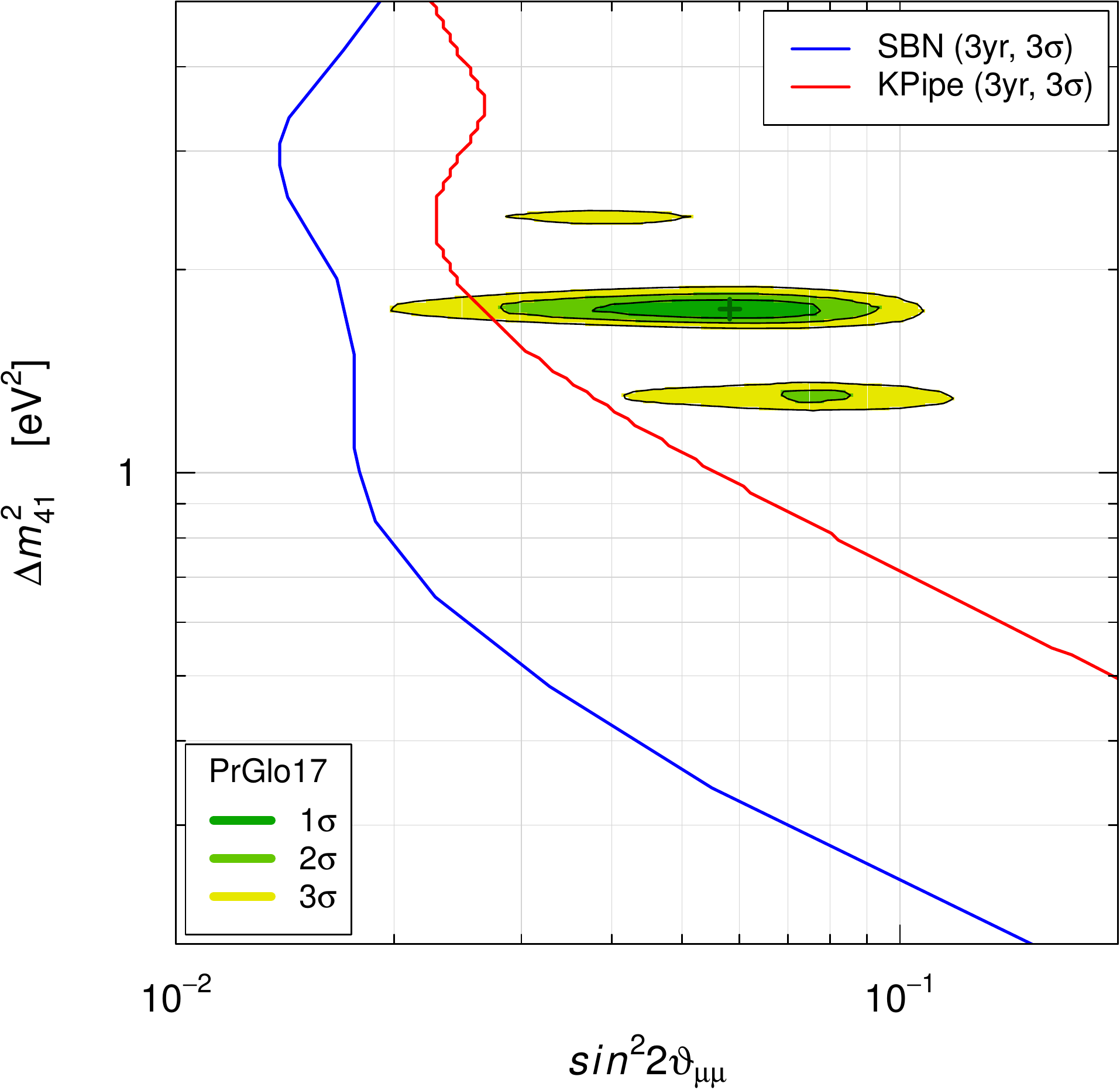}
}
\\
\subfigure[]{\label{fig:fut-see-rea}
\includegraphics*[width=0.49\linewidth]{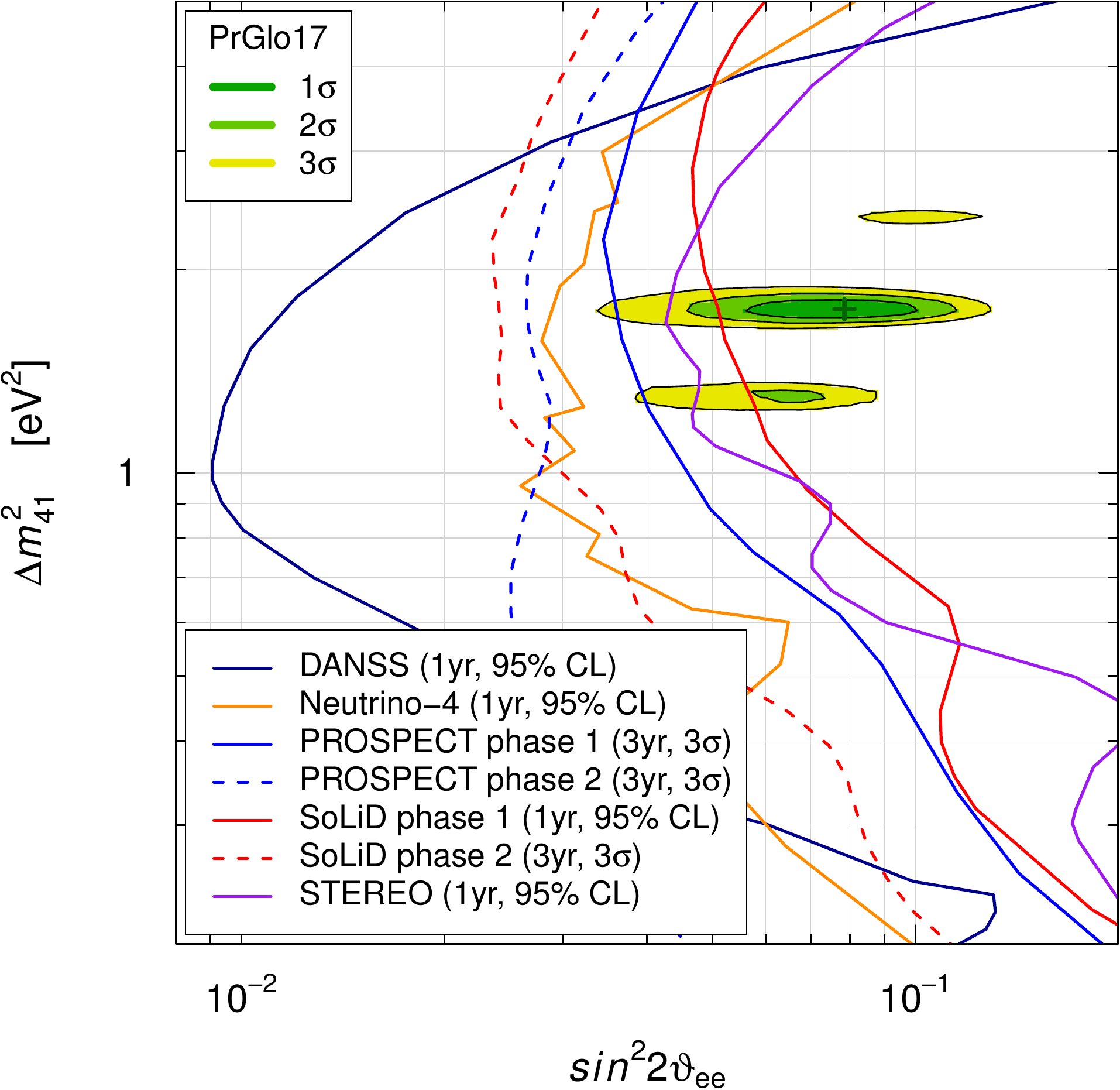}
}
&
\subfigure[]{\label{fig:fut-see-rad}
\includegraphics*[width=0.49\linewidth]{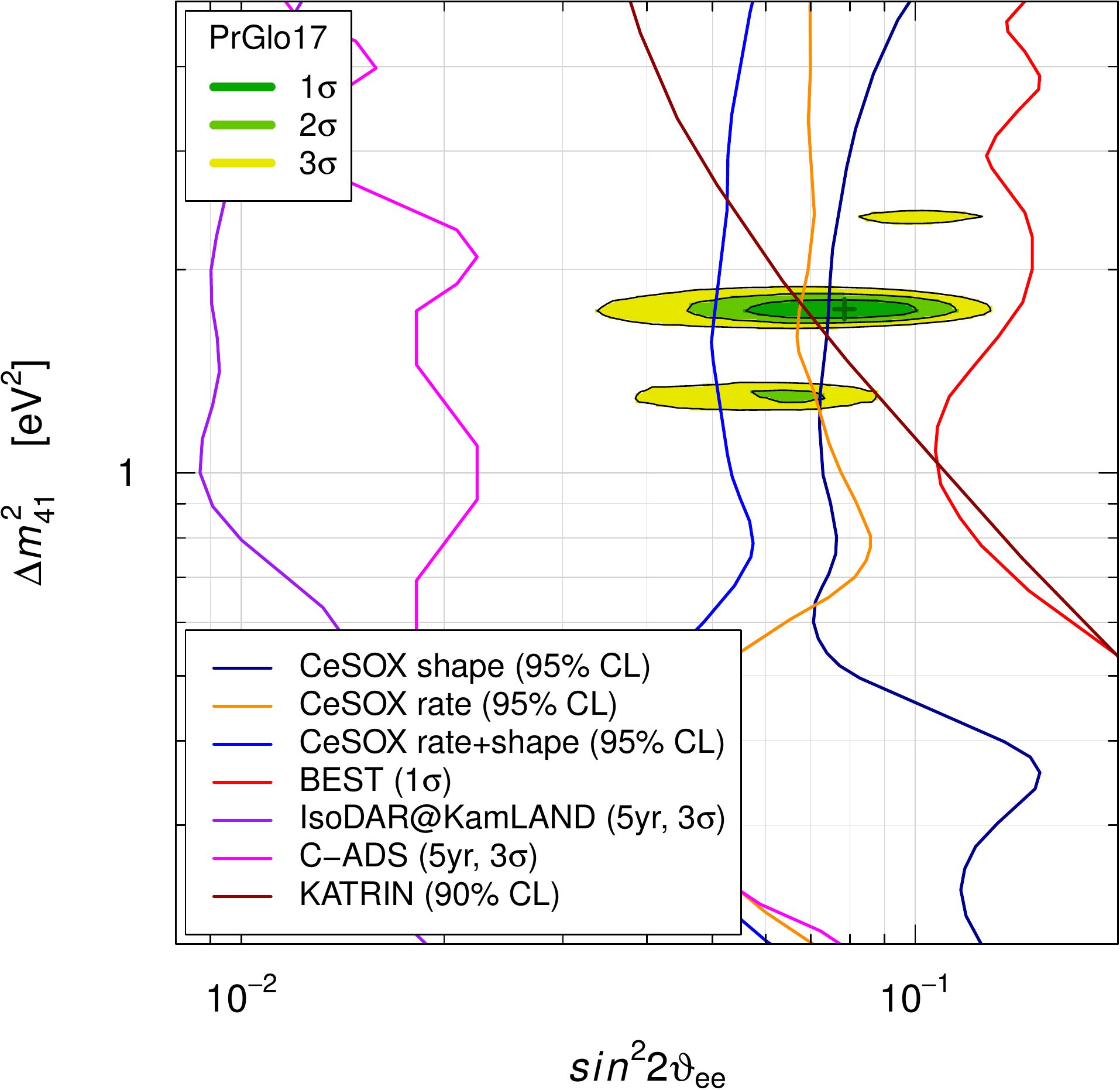}
}
\end{tabular}
\caption{ \label{fig:fut}
Sensitivities of future experiments
compared with the PrGlo17 allowed regions of Fig.~\ref{fig:prg}.
}
\end{figure}

We consider the results of the
PrGlo17 fit as the current status of our
3+1 analysis of short-baseline neutrino oscillation data.
Figure~\ref{fig:fut}
shows a comparison of the sensitivities of future experiments
with the PrGlo17 allowed regions of Fig.~\ref{fig:prg}
for:
\ref{fig:fut-sem}
$\nua{\mu}\to\nua{e}$
transitions
(SBN \cite{Antonello:2015lea},
nuPRISM \cite{Bhadra:2014oma},
JSNS$^2$ \cite{Harada:2013yaa});
\ref{fig:fut-smm}
$\nua{\mu}$
disappearance
(SBN \cite{Antonello:2015lea},
KPipe \cite{Axani:2015zxa});
\ref{fig:fut-see-rea},\subref{fig:fut-see-rad}
$\nua{e}$
disappearance
(DANSS \cite{Alekseev:2016llm},
Neutrino-4 \cite{Serebrov:2012sq},
PROSPECT \cite{Ashenfelter:2015uxt},
SoLid \cite{Ryder:2015sma},
STEREO \cite{Helaine:2016bmc},
CeSOX \cite{Borexino:2013xxa},
BEST \cite{Barinov:2016znv}
IsoDAR@KamLAND \cite{Abs:2015tbh},
C-ADS \cite{Ciuffoli:2015uta},
KATRIN \cite{Drexlin-NOW2016}).
It is clear that these experiments will
give definitive information on the existence of
active-sterile short-baseline oscillations
connected with the
LSND, Gallium and reactor anomalies.

\section{Conclusions}
\label{sec:conclusions}

In this paper we updated the global fit of short-baseline neutrino oscillation data
in the framework of 3+1 active-sterile neutrino mixing
\cite{Giunti:2012tn,Giunti:2012bc,Giunti:2013aea,Gariazzo:2015rra}.

We considered first, in section~\ref{sec:nuedis},
the data on $\nu_{e}$ and $\bar\nu_{e}$ disappearance
which include
the Gallium neutrino anomaly
data~\cite{Abdurashitov:2005tb,Laveder:2007zz,Giunti:2006bj,Giunti:2010zu,Giunti:2012tn}
and
the reactor antineutrino anomaly data~\cite{Mention:2011rk}.
The resulting allowed region
in the $\sin^{2}2\vartheta_{ee}$--$\Delta{m}^{2}_{41}$ plane
is rather wide,
as shown in Fig.~\ref{fig:nue-dis},
but it is smaller than that found in our previous analysis
\cite{Giunti:2012tn},
mainly as a result of the constraints given by the recent NEOS \cite{Ko:2016owz}
experiment.
The allowed region obtained with neutrino oscillation data alone has no upper bound for
$\Delta{m}^{2}_{41}$,
but it can be limited
\cite{Giunti:2012bc}
using the constraints found in the
Mainz
\cite{Kraus:2012he}
and
Troitsk
\cite{Belesev:2012hx,Belesev:2013cba}
$\beta$-decay experiments,
as shown in Fig.~\ref{fig:nue-mbt}.
We found the upper limit
$
\Delta{m}^{2}_{41}
\lesssim
148
\, \text{eV}^2
$
at
$3\sigma$.
Hence,
as shown in Fig.~\ref{fig:nuedisfut},
the ongoing
reactor, source and $\beta$-decay experiments
can clarify in a definitive way
the existence of short-baseline $\nua{e}$ disappearance
due to active-sterile neutrino mixing.

We presented also, in section~\ref{sec:fits},
the results of global fits of all the available
$\nua{\mu}\to\nua{e}$
appearance data,
$\nua{\mu}$ disappearance data,
in addition to the
$\nua{e}$ disappearance data
considered in section~\ref{sec:nuedis}.
We discussed the effects on the global fits of the recent data of the
MINOS \cite{MINOS:2016viw},
IceCube \cite{TheIceCube:2016oqi}, and
NEOS \cite{Ko:2016owz}
experiments.
As expected,
the MINOS, IceCube and NEOS data aggravate the appearance-disappearance tension,
which becomes tolerable only in the pragmatic PrGlo17 fit
discussed in subsection~\ref{sub:PrGlo17},
which is our recommended result.

We found that,
as expected
\cite{Collin:2016aqd,Giunti:2016oan},
the MINOS and IceCube constraints on
$\nua{\mu}$ disappearance
disfavor
the low-$\Delta{m}^{2}_{41}$--high-$\sin^{2}2\vartheta_{\mu\mu}$
and
the low-$\Delta{m}^{2}_{41}$--high-$\sin^{2}2\vartheta_{e\mu}$
parts of the allowed region.
The addition of the NEOS data has the more dramatic effect
of reducing the allowed region to three islands with narrow
$\Delta{m}^{2}_{41}$ widths and
$
0.00048
\lesssim
\sin^{2}2\vartheta_{e\mu}
\lesssim
0.0020
$
at $3\sigma$.
The best-fit island is at
$\Delta{m}^{2}_{41} \approx 1.7 \, \text{eV}^2$.
There is an island allowed at $2\sigma$ at
$\Delta{m}^{2}_{41} \approx 1.3 \, \text{eV}^2$,
and
an island allowed at $3\sigma$ at
$\Delta{m}^{2}_{41} \approx 2.4 \, \text{eV}^2$.
However,
as illustrated in Fig.~\ref{fig:fut},
the ongoing and planned experiments have the possibility to cover
all the allowed regions of the mixing parameters
and we expect that they will reach in a few years
a definitive conclusion on the existence
of the short-baseline oscillations
indicated by the LSND experiment and by the Gallium and reactor neutrino anomalies.

An interesting feature of the 3+1 analysis of the
MINOS and IceCube data
is that there is a dependence on $|U_{\tau4}|^2$
\cite{Nunokawa:2003ep,Choubey:2007ji,Razzaque:2011ab,Razzaque:2012tp,Esmaili:2012nz,Esmaili:2013vza,Esmaili:2013cja,Lindner:2015iaa}.
We obtained the stringent bounds on the value of $|U_{\tau4}|^2$
given in Tab.~\ref{tab:prgint},
which are comparable to those obtained in Ref.~\cite{Collin:2016aqd}.

The determination of active-sterile neutrino mixing
presented in this paper is of interest also for the phenomenology of long-baseline experiments
\cite{deGouvea:2014aoa,Klop:2014ima,Berryman:2015nua,Gandhi:2015xza,Palazzo:2015gja,Agarwalla:2016mrc,Agarwalla:2016xxa,Choubey:2016fpi,Agarwalla:2016xlg,Capozzi:2016vac},
and neutrinoless double-$\beta$ decay experiments
\cite{Barry:2011wb,Li:2011ss,Rodejohann:2012xd,Giunti:2012tn,Girardi:2013zra,Pascoli:2013fiz,Meroni:2014tba,Abada:2014nwa,Giunti:2015kza,Pas:2015eia}.

We did not consider the problem of the cosmological bounds on active-sterile neutrino mixing
\cite{Ade:2015xua},
which most likely must be solved with a non-standard effect as
a large lepton asymmetry
\cite{Hannestad:2012ky,Mirizzi:2012we,Saviano:2013ktj,Hannestad:2013pha}
or secret interactions of the sterile neutrino mediated by a massive vector or pseudoscalar boson
\cite{Hannestad:2013ana,Dasgupta:2013zpn,Mirizzi:2014ama,Saviano:2014esa,Forastieri:2015paa,Chu:2015ipa,Archidiacono:2016kkh},
which suppress the thermalization of the sterile neutrino
in the early Universe.

In conclusion,
this paper gives information on what are the regions
of the parameter space of 3+1 neutrino mixing which must be explored
by new experiments in order to check the indications
given by the LSND, Gallium and reactor anomalies.
Let us emphasize the importance of an experimental confirmation of these oscillations,
that would imply the existence of light sterile neutrinos.
These are new particles with properties outside the realm of the Standard Model
and their discovery
would open a prodigious window on new low-energy physics.

\section*{Acknowledgments}

We are very grateful to the NEOS Collaboration
for giving us the table of $\chi^2$ corresponding to
Fig.~4 of Ref.~\cite{Ko:2016owz}.
The work of S.G.~is supported by the Spanish grants FPA2014-58183-P, Multidark CSD2009-00064 and SEV-2014-0398 (MINECO), and PROMETEOII/2014/084 (Generalitat Valenciana).
The work of C.G. and M.L. was partially supported by the research grant {\sl Theoretical Astroparticle Physics} number 2012CPPYP7 under the program PRIN 2012 funded by the Italian Ministero dell'Istruzione, Universit\`a e della Ricerca (MIUR) and by the research project
{\em TAsP} funded by the Instituto Nazionale di Fisica Nucleare (INFN).
The work of Y.F.L. was supported in part by the National Natural Science Foundation of China under Grant Nos. 11305193 and 11135009, by the Strategic Priority Research Program of the Chinese Academy of Sciences under Grant No. XDA10010100, by the CAS Center for Excellence in Particle Physics (CCEPP).

\providecommand{\href}[2]{#2}\begingroup\raggedright\endgroup

\begin{thebibliography}{100}

\bibitem{Athanassopoulos:1995iw}
{\scshape LSND} collaboration, C.~Athanassopoulos et~al., \emph{{Candidate
events in a search for $\bar\nu_\mu \to \bar\nu_e$ oscillations}},
{\emph{Phys. Rev. Lett.} {\bfseries 75} (1995) 2650--2653},
[\href{https://arxiv.org/abs/nucl-ex/9504002}{{\ttfamily nucl-ex/9504002}}].

\bibitem{Aguilar:2001ty}
{\scshape LSND} collaboration, A.~Aguilar et~al., \emph{{Evidence for neutrino
oscillations from the observation of $\bar\nu_e$ appearance in a
$\bar\nu_\mu$ beam}}, {\emph{Phys. Rev.} {\bfseries D64} (2001) 112007},
[\href{https://arxiv.org/abs/hep-ex/0104049}{{\ttfamily hep-ex/0104049}}].

\bibitem{Abdurashitov:2005tb}
{\scshape SAGE} collaboration, J.~N. Abdurashitov et~al., \emph{{Measurement of
the response of a Ga solar neutrino experiment to neutrinos from an Ar-37
source}}, {\emph{Phys. Rev.} {\bfseries C73} (2006) 045805},
[\href{https://arxiv.org/abs/nucl-ex/0512041}{{\ttfamily nucl-ex/0512041}}].

\bibitem{Laveder:2007zz}
M.~Laveder, \emph{{Unbound neutrino roadmaps}},
\href{http://dx.doi.org/10.1016/j.nuclphysbps.2007.02.037}{\emph{Nucl. Phys.
Proc. Suppl.} {\bfseries 168} (2007) 344--346}.

\bibitem{Giunti:2006bj}
C.~Giunti and M.~Laveder, \emph{{Short-Baseline Active-Sterile Neutrino
Oscillations?}}, {\emph{Mod. Phys. Lett.} {\bfseries A22} (2007) 2499--2509},
[\href{https://arxiv.org/abs/hep-ph/0610352}{{\ttfamily hep-ph/0610352}}].

\bibitem{Giunti:2010zu}
C.~Giunti and M.~Laveder, \emph{{Statistical Significance of the Gallium
Anomaly}}, {\emph{Phys. Rev.} {\bfseries C83} (2011) 065504},
[\href{https://arxiv.org/abs/arXiv:1006.3244}{{\ttfamily arXiv:1006.3244}}].

\bibitem{Giunti:2012tn}
C.~Giunti, M.~Laveder, Y.~Li, Q.~Liu and H.~Long, \emph{{Update of
Short-Baseline Electron Neutrino and Antineutrino Disappearance}},
{\emph{Phys. Rev.} {\bfseries D86} (2012) 113014},
[\href{https://arxiv.org/abs/arXiv:1210.5715}{{\ttfamily arXiv:1210.5715}}].

\bibitem{Mention:2011rk}
G.~Mention et~al., \emph{{The Reactor Antineutrino Anomaly}}, {\emph{Phys.
Rev.} {\bfseries D83} (2011) 073006},
[\href{https://arxiv.org/abs/arXiv:1101.2755}{{\ttfamily arXiv:1101.2755}}].

\bibitem{Bilenky:1998dt}
S.~M. Bilenky, C.~Giunti and W.~Grimus, \emph{{Phenomenology of neutrino
oscillations}}, {\emph{Prog. Part. Nucl. Phys.} {\bfseries 43} (1999) 1},
[\href{https://arxiv.org/abs/hep-ph/9812360}{{\ttfamily hep-ph/9812360}}].

\bibitem{GonzalezGarcia:2007ib}
M.~C. Gonzalez-Garcia and M.~Maltoni, \emph{{Phenomenology with Massive
Neutrinos}}, {\emph{Phys. Rept.} {\bfseries 460} (2008) 1--129},
[\href{https://arxiv.org/abs/arXiv:0704.1800}{{\ttfamily arXiv:0704.1800}}].

\bibitem{Conrad:2012qt}
J.~Conrad, C.~Ignarra, G.~Karagiorgi, M.~Shaevitz and J.~Spitz, \emph{{Sterile
Neutrino Fits to Short Baseline Neutrino Oscillation Measurements}},
\href{http://dx.doi.org/10.1155/2013/163897}{\emph{Adv.High Energy Phys.}
{\bfseries 2013} (2013) 163897},
[\href{https://arxiv.org/abs/arXiv:1207.4765}{{\ttfamily arXiv:1207.4765}}].

\bibitem{Gariazzo:2015rra}
S.~Gariazzo, C.~Giunti, M.~Laveder, Y.~Li and E.~Zavanin, \emph{{Light sterile
neutrinos}}, {\emph{J. Phys.} {\bfseries G43} (2016) 033001},
[\href{https://arxiv.org/abs/arXiv:1507.08204}{{\ttfamily
arXiv:1507.08204}}].

\bibitem{Maltoni:2004ei}
M.~Maltoni, T.~Schwetz, M.~Tortola and J.~Valle, \emph{{Status of global fits
to neutrino oscillations}}, {\emph{New J. Phys.} {\bfseries 6} (2004) 122},
[\href{https://arxiv.org/abs/hep-ph/0405172}{{\ttfamily hep-ph/0405172}}].

\bibitem{Ade:2015xua}
{\scshape Planck} collaboration, P.~A.~R. Ade et~al., \emph{{Planck 2015
results. XIII. Cosmological parameters}}, {\emph{Astron.Astrophys.}
{\bfseries 594} (2016) A13},
[\href{https://arxiv.org/abs/arXiv:1502.01589}{{\ttfamily
arXiv:1502.01589}}].

\bibitem{Bilenky:2014uka}
S.~Bilenky and C.~Giunti, \emph{{Neutrinoless Double-Beta Decay: a Probe of
Physics Beyond the Standard Model}}, {\emph{Int.J.Mod.Phys.} {\bfseries A30}
(2015) 0001}, [\href{https://arxiv.org/abs/arXiv:1411.4791}{{\ttfamily
arXiv:1411.4791}}].

\bibitem{DellOro:2016tmg}
S.~Dell'Oro, S.~Marcocci, M.~Viel and F.~Vissani, \emph{{Neutrinoless double
beta decay: 2015 review}}, {\emph{Adv.High Energy Phys.} {\bfseries 2016}
(2016) 2162659}, [\href{https://arxiv.org/abs/arXiv:1601.07512}{{\ttfamily
arXiv:1601.07512}}].

\bibitem{Giunti:2015mwa}
C.~Giunti and E.~M. Zavanin, \emph{{Appearance-Disappearance Relation in
3+$N_{s}$ Short-Baseline Neutrino Oscillations}}, {\emph{Mod. Phys. Lett.}
{\bfseries A31} (2016) 1650003},
[\href{https://arxiv.org/abs/arXiv:1508.03172}{{\ttfamily
arXiv:1508.03172}}].

\bibitem{Capozzi:2016rtj}
F.~Capozzi, E.~Lisi, A.~Marrone, D.~Montanino and A.~Palazzo, \emph{{Neutrino
masses and mixings: Status of known and unknown $3\nu$ parameters}},
{\emph{Nucl. Phys.} {\bfseries B908} (2016) 218--234},
[\href{https://arxiv.org/abs/arXiv:1601.07777}{{\ttfamily
arXiv:1601.07777}}].

\bibitem{Esteban:2016qun}
I.~Esteban, M.~C. Gonzalez-Garcia, M.~Maltoni, I.~Martinez-Soler and
T.~Schwetz, \emph{{Updated fit to three neutrino mixing: exploring the
accelerator-reactor complementarity}}, {\emph{JHEP} {\bfseries 1701} (2017)
087}, [\href{https://arxiv.org/abs/arXiv:1611.01514}{{\ttfamily
arXiv:1611.01514}}].

\bibitem{Bilenky:1996rw}
S.~M. Bilenky, C.~Giunti and W.~Grimus, \emph{{Neutrino mass spectrum from the
results of neutrino oscillation experiments}}, {\emph{Eur. Phys. J.}
{\bfseries C1} (1998) 247--253},
[\href{https://arxiv.org/abs/hep-ph/9607372}{{\ttfamily hep-ph/9607372}}].

\bibitem{deGouvea:2014aoa}
A.~de~Gouvea, K.~J. Kelly and A.~Kobach, \emph{{CP-Invariance Violation at
Short-Baseline Experiments in 3+1 Scenarios}}, {\emph{Phys. Rev.} {\bfseries
D91} (2015) 053005}, [\href{https://arxiv.org/abs/arXiv:1412.1479}{{\ttfamily
arXiv:1412.1479}}].

\bibitem{Klop:2014ima}
N.~Klop and A.~Palazzo, \emph{{Imprints of CP-violating phases induced by
sterile neutrinos in T2K}}, {\emph{Phys. Rev.} {\bfseries D91} (2015)
073017}, [\href{https://arxiv.org/abs/arXiv:1412.7524}{{\ttfamily
arXiv:1412.7524}}].

\bibitem{Berryman:2015nua}
J.~M. Berryman, A.~de~Gouvea, K.~J. Kelly and A.~Kobach, \emph{{A Sterile
Neutrino at DUNE}}, {\emph{Phys. Rev.} {\bfseries D92} (2015) 073012},
[\href{https://arxiv.org/abs/arXiv:1507.03986}{{\ttfamily
arXiv:1507.03986}}].

\bibitem{Gandhi:2015xza}
R.~Gandhi, B.~Kayser, M.~Masud and S.~Prakash, \emph{{The impact of sterile
neutrinos on CP measurements at long baselines}}, {\emph{JHEP} {\bfseries 11}
(2015) 039}, [\href{https://arxiv.org/abs/arXiv:1508.06275}{{\ttfamily
arXiv:1508.06275}}].

\bibitem{Palazzo:2015gja}
A.~Palazzo, \emph{{3-flavor and 4-flavor implications of the latest T2K and
NO$\nu$A electron (anti-)neutrino appearance results}}, {\emph{Phys.Lett.}
{\bfseries B757} (2016) 142--147},
[\href{https://arxiv.org/abs/arXiv:1509.03148}{{\ttfamily
arXiv:1509.03148}}].

\bibitem{Agarwalla:2016mrc}
S.~K. Agarwalla, S.~S. Chatterjee, A.~Dasgupta and A.~Palazzo, \emph{{Discovery
Potential of T2K and NOvA in the Presence of a Light Sterile Neutrino}},
{\emph{JHEP} {\bfseries 02} (2016) 111},
[\href{https://arxiv.org/abs/arXiv:1601.05995}{{\ttfamily
arXiv:1601.05995}}].

\bibitem{Agarwalla:2016xxa}
S.~K. Agarwalla, S.~S. Chatterjee and A.~Palazzo, \emph{{Physics Reach of DUNE
with a Light Sterile Neutrino}}, {\emph{JHEP} {\bfseries 1609} (2016) 016},
[\href{https://arxiv.org/abs/arXiv:1603.03759}{{\ttfamily
arXiv:1603.03759}}].

\bibitem{Choubey:2016fpi}
S.~Choubey and D.~Pramanik, \emph{{Constraints on Sterile Neutrino Oscillations
using DUNE Near Detector}}, {\emph{Phys.Lett.} {\bfseries B764} (2017)
135--141}, [\href{https://arxiv.org/abs/arXiv:1604.04731}{{\ttfamily
arXiv:1604.04731}}].

\bibitem{Agarwalla:2016xlg}
S.~K. Agarwalla, S.~S. Chatterjee and A.~Palazzo, \emph{{Octant of
$\theta_{23}$ in danger with a light sterile neutrino}}, {\emph{Phys. Rev.
Lett.} {\bfseries 118} (2017) 031804},
[\href{https://arxiv.org/abs/arXiv:1605.04299}{{\ttfamily
arXiv:1605.04299}}].

\bibitem{Capozzi:2016vac}
F.~Capozzi, C.~Giunti, M.~Laveder and A.~Palazzo, \emph{{Joint short- and
long-baseline constraints on light sterile neutrinos}}, {\emph{Phys.Rev.}
{\bfseries D95} (2017) 033006},
[\href{https://arxiv.org/abs/arXiv:1612.07764}{{\ttfamily
arXiv:1612.07764}}].

\bibitem{Long:2013hwa}
H.~Long, Y.~Li and C.~Giunti, \emph{{CP-violating Phases in Active-Sterile
Solar Neutrino Oscillations}}, {\emph{Phys. Rev. D 87,} {\bfseries 113004}
(2013) 113004}, [\href{https://arxiv.org/abs/arXiv:1304.2207}{{\ttfamily
arXiv:1304.2207}}].

\bibitem{Giunti:2012bc}
C.~Giunti, M.~Laveder, Y.~Li and H.~Long, \emph{{Short-Baseline Electron
Neutrino Oscillation Length After Troitsk}}, {\emph{Phys. Rev.} {\bfseries
D87} (2013) 013004}, [\href{https://arxiv.org/abs/arXiv:1212.3805}{{\ttfamily
arXiv:1212.3805}}].

\bibitem{Giunti:2013aea}
C.~Giunti, M.~Laveder, Y.~Li and H.~Long, \emph{{A Pragmatic View of
Short-Baseline Neutrino Oscillations}}, {\emph{Phys. Rev.} {\bfseries D88}
(2013) 073008}, [\href{https://arxiv.org/abs/arXiv:1308.5288}{{\ttfamily
arXiv:1308.5288}}].

\bibitem{Giunti:2016elf}
C.~Giunti, \emph{{Precise Determination of the ${}^{235}\text{U}$ Reactor
Antineutrino Cross Section per Fission}}, {\emph{Phys.Lett.} {\bfseries B764}
(2017) 145--149}, [\href{https://arxiv.org/abs/arXiv:1608.04096}{{\ttfamily
arXiv:1608.04096}}].

\bibitem{MINOS:2016viw}
{\scshape MINOS} collaboration, P.~Adamson et~al., \emph{{A search for sterile
neutrinos mixing with muon neutrinos in MINOS}}, {\emph{Phys. Rev. Lett.}
{\bfseries 117} (2016) 151803},
[\href{https://arxiv.org/abs/arXiv:1607.01176}{{\ttfamily
arXiv:1607.01176}}].

\bibitem{TheIceCube:2016oqi}
{\scshape IceCube} collaboration, M.~G. Aartsen et~al., \emph{{Searches for
Sterile Neutrinos with the IceCube Detector}}, {\emph{Phys. Rev. Lett.}
{\bfseries 117} (2016) 071801},
[\href{https://arxiv.org/abs/arXiv:1605.01990}{{\ttfamily
arXiv:1605.01990}}].

\bibitem{Ko:2016owz}
{\scshape NEOS} collaboration, Y.~Ko et~al., \emph{{A sterile neutrino search
at NEOS Experiment}}, {\emph{Phys.Rev.Lett.} {\bfseries 118} (2017) 121802},
[\href{https://arxiv.org/abs/arXiv:1610.05134}{{\ttfamily
arXiv:1610.05134}}].

\bibitem{Giunti:2016oan}
C.~Giunti, \emph{{Oscillations Beyond Three-Neutrino Mixing}},
\href{https://arxiv.org/abs/arXiv:1609.04688}{{\ttfamily arXiv:1609.04688}}.

\bibitem{Collin:2016aqd}
G.~Collin, C.~Arguelles, J.~Conrad and M.~Shaevitz, \emph{{First Constraints on
the Complete Neutrino Mixing Matrix with a Sterile Neutrino}}, {\emph{Phys.
Rev. Lett.} {\bfseries 117} (2016) 221801},
[\href{https://arxiv.org/abs/arXiv:1607.00011}{{\ttfamily
arXiv:1607.00011}}].

\bibitem{Collin:2016rao}
G.~H. Collin, C.~A. Arguelles, J.~M. Conrad and M.~H. Shaevitz, \emph{{Sterile
Neutrino Fits to Short Baseline Data}}, {\emph{Nucl. Phys.} {\bfseries B908}
(2016) 354--365}, [\href{https://arxiv.org/abs/arXiv:1602.00671}{{\ttfamily
arXiv:1602.00671}}].

\bibitem{An:2016srz}
{\scshape Daya Bay} collaboration, F.~An et~al., \emph{{Improved Measurement of
the Reactor Antineutrino Flux and Spectrum at Daya Bay}}, {\emph{Chin.Phys.}
{\bfseries C41} (2017) 013002},
[\href{https://arxiv.org/abs/arXiv:1607.05378}{{\ttfamily
arXiv:1607.05378}}].

\bibitem{Okada:1996kw}
N.~Okada and O.~Yasuda, \emph{{A sterile neutrino scenario constrained by
experiments and cosmology}}, {\emph{Int. J. Mod. Phys.} {\bfseries A12}
(1997) 3669--3694}, [\href{https://arxiv.org/abs/hep-ph/9606411}{{\ttfamily
hep-ph/9606411}}].

\bibitem{Kopp:2011qd}
J.~Kopp, M.~Maltoni and T.~Schwetz, \emph{{Are there sterile neutrinos at the
eV scale?}}, {\emph{Phys. Rev. Lett.} {\bfseries 107} (2011) 091801},
[\href{https://arxiv.org/abs/arXiv:1103.4570}{{\ttfamily arXiv:1103.4570}}].

\bibitem{Giunti:2011gz}
C.~Giunti and M.~Laveder, \emph{{3+1 and 3+2 Sterile Neutrino Fits}},
\href{http://dx.doi.org/10.1103/PhysRevD.84.073008}{\emph{Phys. Rev.}
{\bfseries D84} (2011) 073008},
[\href{https://arxiv.org/abs/arXiv:1107.1452}{{\ttfamily arXiv:1107.1452}}].

\bibitem{Giunti:2011hn}
C.~Giunti and M.~Laveder, \emph{{Status of 3+1 Neutrino Mixing}},
\href{http://dx.doi.org/10.1103/PhysRevD.84.093006}{\emph{Phys. Rev.}
{\bfseries D84} (2011) 093006},
[\href{https://arxiv.org/abs/arXiv:1109.4033}{{\ttfamily arXiv:1109.4033}}].

\bibitem{Giunti:2011cp}
C.~Giunti and M.~Laveder, \emph{{Implications of 3+1 Short-Baseline Neutrino
Oscillations}}, {\emph{Phys. Lett.} {\bfseries B706} (2011) 200--207},
[\href{https://arxiv.org/abs/arXiv:1111.1069}{{\ttfamily arXiv:1111.1069}}].

\bibitem{Archidiacono:2012ri}
M.~Archidiacono, N.~Fornengo, C.~Giunti and A.~Melchiorri, \emph{{Testing 3+1
and 3+2 neutrino mass models with cosmology and short baseline experiments}},
{\emph{Phys. Rev.} {\bfseries D86} (2012) 065028},
[\href{https://arxiv.org/abs/arXiv:1207.6515}{{\ttfamily arXiv:1207.6515}}].

\bibitem{Archidiacono:2013xxa}
M.~Archidiacono, N.~Fornengo, C.~Giunti, S.~Hannestad and A.~Melchiorri,
\emph{{Sterile Neutrinos: Cosmology vs Short-BaseLine Experiments}},
\href{http://dx.doi.org/10.1103/PhysRevD.87.125034}{\emph{Phys. Rev.}
{\bfseries D87} (2013) 125034},
[\href{https://arxiv.org/abs/arXiv:1302.6720}{{\ttfamily arXiv:1302.6720}}].

\bibitem{Kopp:2013vaa}
J.~Kopp, P.~A.~N. Machado, M.~Maltoni and T.~Schwetz, \emph{{Sterile Neutrino
Oscillations: The Global Picture}},
\href{http://dx.doi.org/10.1007/JHEP05(2013)050}{\emph{JHEP} {\bfseries 1305}
(2013) 050}, [\href{https://arxiv.org/abs/arXiv:1303.3011}{{\ttfamily
arXiv:1303.3011}}].

\bibitem{AguilarArevalo:2008rc}
{\scshape MiniBooNE} collaboration, A.~A. Aguilar-Arevalo et~al.,
\emph{{Unexplained Excess of Electron-Like Events From a 1-GeV Neutrino
Beam}}, {\emph{Phys. Rev. Lett.} {\bfseries 102} (2009) 101802},
[\href{https://arxiv.org/abs/arXiv:0812.2243}{{\ttfamily arXiv:0812.2243}}].

\bibitem{Aguilar-Arevalo:2013pmq}
{\scshape MiniBooNE} collaboration, A.~Aguilar-Arevalo et~al., \emph{{Improved
Search for $\bar\nu_\mu \to \bar\nu_e$ Oscillations in the MiniBooNE
Experiment}},
\href{http://dx.doi.org/10.1103/PhysRevLett.110.161801}{\emph{Phys. Rev.
Lett.} {\bfseries 110} (2013) 161801},
[\href{https://arxiv.org/abs/arXiv:1303.2588}{{\ttfamily arXiv:1303.2588}}].

\bibitem{Gollapinni:2015lca}
{\scshape MicroBooNE} collaboration, S.~Gollapinni, \emph{{Accelerator-based
Short-baseline Neutrino Oscillation Experiments}},
\href{https://arxiv.org/abs/arXiv:1510.04412}{{\ttfamily arXiv:1510.04412}}.

\bibitem{Declais:1995su}
{\scshape Bugey} collaboration, B.~Achkar et~al., \emph{{Search for neutrino
oscillations at 15-meters, 40-meters, and 95-meters from a nuclear power
reactor at Bugey}}, {\emph{Nucl. Phys.} {\bfseries B434} (1995) 503--534}.

\bibitem{Serebrov:2017nxa}
{\scshape Neutrino-4} collaboration, A.~P. Serebrov et~al., \emph{{Experiment
NEUTRINO-4 Search for Sterile Neutrino}}, {\emph{PoS} {\bfseries INPC2016}
(2017) 255}, [\href{https://arxiv.org/abs/arXiv:1702.00941}{{\ttfamily
arXiv:1702.00941}}].

\bibitem{Declais:1994ma}
{\scshape Bugey} collaboration, Y.~Declais et~al., \emph{{Study of reactor
anti-neutrino interaction with proton at Bugey nuclear power plant}},
{\emph{Phys. Lett.} {\bfseries B338} (1994) 383--389}.

\bibitem{Kuvshinnikov:1990ry}
A.~Kuvshinnikov, L.~Mikaelyan, S.~Nikolaev, M.~Skorokhvatov and A.~Etenko,
\emph{{Measuring the $\bar\nu_{e} + p \to n + e^{+}$ cross-section and beta
decay axial constant in a new experiment at Rovno NPP reactor}}, {\emph{JETP
Lett.} {\bfseries 54} (1991) 253--257}.

\bibitem{Zacek:1986cu}
{\scshape CalTech-SIN-TUM} collaboration, G.~Zacek et~al., \emph{{Neutrino
oscillation experiments at the Gosgen nuclear power reactor}}, {\emph{Phys.
Rev.} {\bfseries D34} (1986) 2621--2636}.

\bibitem{Kwon:1981ua}
H.~Kwon et~al., \emph{{Search for neutrino oscillations at a fission reactor}},
{\emph{Phys. Rev.} {\bfseries D24} (1981) 1097--1111}.

\bibitem{Hoummada:1995zz}
A.~Hoummada, S.~Lazrak~Mikou, G.~Bagieu, J.~Cavaignac and D.~Holm~Koang,
\emph{{Neutrino oscillations I.L.L. experiment reanalysis}}, {\emph{Applied
Radiation and Isotopes} {\bfseries 46} (1995) 449--450}.

\bibitem{Vidyakin:1987ue}
{\scshape Krasnoyarsk} collaboration, G.~S. Vidyakin et~al., \emph{{Detection
of anti-neutrinos in the flux from two reactors}}, {\emph{Sov. Phys. JETP}
{\bfseries 66} (1987) 243--247}.

\bibitem{Vidyakin:1990iz}
{\scshape Krasnoyarsk} collaboration, G.~S. Vidyakin et~al., \emph{{Bounds on
the neutrino oscillation parameters for reactor anti-neutrinos}}, {\emph{Sov.
Phys. JETP} {\bfseries 71} (1990) 424--426}.

\bibitem{Vidyakin:1994ut}
{\scshape Krasnoyarsk} collaboration, G.~S. Vidyakin et~al., \emph{{Limitations
on the characteristics of neutrino oscillations}}, {\emph{JETP Lett.}
{\bfseries 59} (1994) 390--393}.

\bibitem{Afonin:1988gx}
A.~I. Afonin et~al., \emph{{A study of the reaction $ \bar\nu_e + p \to e^+ + n
$ on a nuclear reactor}}, {\emph{Sov. Phys. JETP} {\bfseries 67} (1988)
213--221}.

\bibitem{Greenwood:1996pb}
Z.~D. Greenwood et~al., \emph{{Results of a two position reactor neutrino
oscillation experiment}}, {\emph{Phys. Rev.} {\bfseries D53} (1996)
6054--6064}.

\bibitem{Huber:2011wv}
P.~Huber, \emph{{On the determination of anti-neutrino spectra from nuclear
reactors}}, {\emph{Phys. Rev.} {\bfseries C84} (2011) 024617},
[\href{https://arxiv.org/abs/arXiv:1106.0687}{{\ttfamily arXiv:1106.0687}}].

\bibitem{Kozlov:1999ct}
Y.~V. Kozlov, S.~V. Khalturtsev, I.~N. Machulin, A.~V. Martemyanov, V.~P.
Martemyanov, S.~V. Sukhotin et~al., \emph{{Anti-neutrino deuteron experiment
at Krasnoyarsk}}, \href{http://dx.doi.org/10.1134/1.855742}{\emph{Phys. Atom.
Nucl.} {\bfseries 63} (2000) 1016--1019},
[\href{https://arxiv.org/abs/hep-ex/9912047}{{\ttfamily hep-ex/9912047}}].

\bibitem{Zhang:2013ela}
C.~Zhang, X.~Qian and P.~Vogel, \emph{{Reactor Antineutrino Anomaly with known
$\theta_{13}$}}, {\emph{Phys. Rev.} {\bfseries D87} (2013) 073018},
[\href{https://arxiv.org/abs/arXiv:1303.0900}{{\ttfamily arXiv:1303.0900}}].

\bibitem{Apollonio:2002gd}
{\scshape CHOOZ} collaboration, M.~Apollonio et~al., \emph{{Search for neutrino
oscillations on a long base-line at the CHOOZ nuclear power station}},
{\emph{Eur. Phys. J.} {\bfseries C27} (2003) 331},
[\href{https://arxiv.org/abs/hep-ex/0301017}{{\ttfamily hep-ex/0301017}}].

\bibitem{Boehm:2001ik}
{\scshape Palo Verde} collaboration, F.~Boehm et~al., \emph{{Final results from
the Palo Verde neutrino oscillation experiment}}, {\emph{Phys. Rev.}
{\bfseries D64} (2001) 112001},
[\href{https://arxiv.org/abs/hep-ex/0107009}{{\ttfamily hep-ex/0107009}}].

\bibitem{Boireau:2015dda}
{\scshape NUCIFER} collaboration, G.~Boireau et~al., \emph{{Online Monitoring
of the Osiris Reactor with the Nucifer Neutrino Detector}}, {\emph{Phys.
Rev.} {\bfseries D93} (2016) 112006},
[\href{https://arxiv.org/abs/arXiv:1509.05610}{{\ttfamily
arXiv:1509.05610}}].

\bibitem{RENO-AAP2016}
H.~Seo, \emph{{Recent Results from RENO}}, Talk presented at {AAP 2016,
Applied Antineutrino Physics, 1-2 December 2016, Liverpool, UK}.

\bibitem{Kozlov:1999cs}
Y.~V. Kozlov et~al., \emph{{Today and future neutrino experiments at
Krasnoyarsk nuclear reactor}},
\href{http://dx.doi.org/10.1016/S0920-5632(00)00738-6}{\emph{Nucl. Phys.
Proc. Suppl.} {\bfseries 87} (2000) 514--516},
[\href{https://arxiv.org/abs/hep-ex/9912046}{{\ttfamily hep-ex/9912046}}].

\bibitem{Schreckenbach:1985ep}
K.~Schreckenbach, G.~Colvin, W.~Gelletly and F.~Von~Feilitzsch,
\emph{{Determination of the anti-neutrino spectrum from U-235 thermal neutron
fission products up to 9.5-MeV}}, {\emph{Phys. Lett.} {\bfseries B160} (1985)
325--330}.

\bibitem{Hahn:1989zr}
A.~A. Hahn et~al., \emph{{Anti-neutrino spectra from Pu-241 and Pu-239 thermal
neutron fission products}}, {\emph{Phys. Lett.} {\bfseries B218} (1989)
365--368}.

\bibitem{Haag:2014kia}
N.~Haag, F.~von Feilitzsch, L.~Oberauer, W.~Potzel, K.~Schreckenbach et~al.,
\emph{{Re-publication of the data from the BILL magnetic spectrometer: The
cumulative $\beta$ spectra of the fission products of $^{235}$U, $^{239}$Pu,
and $^{241}$Pu}}, \href{https://arxiv.org/abs/arXiv:1405.3501}{{\ttfamily
arXiv:1405.3501}}.

\bibitem{RENO:2015ksa}
{\scshape RENO} collaboration, J.~Choi et~al., \emph{{Observation of Energy and
Baseline Dependent Reactor Antineutrino Disappearance in the RENO
Experiment}}, {\emph{Phys. Rev. Lett.} {\bfseries 116} (2016) 211801},
[\href{https://arxiv.org/abs/arXiv:1511.05849}{{\ttfamily
arXiv:1511.05849}}].

\bibitem{Abe:2014bwa}
{\scshape Double Chooz} collaboration, Y.~Abe et~al., \emph{{Improved
measurements of the neutrino mixing angle $\theta_{13}$ with the Double Chooz
detector}}, {\emph{JHEP} {\bfseries 1410} (2014) 86},
[\href{https://arxiv.org/abs/arXiv:1406.7763}{{\ttfamily arXiv:1406.7763}}].

\bibitem{An:2015nua}
{\scshape Daya Bay} collaboration, F.~P. An et~al., \emph{{Measurement of the
Reactor Antineutrino Flux and Spectrum at Daya Bay}}, {\emph{Phys. Rev.
Lett.} {\bfseries 116} (2016) 061801},
[\href{https://arxiv.org/abs/arXiv:1508.04233}{{\ttfamily
arXiv:1508.04233}}].

\bibitem{Huber:2016fkt}
P.~Huber, \emph{{Reactor antineutrino fluxes - status and challenges}},
{\emph{Nucl. Phys.} {\bfseries B908} (2016) 268--278},
[\href{https://arxiv.org/abs/arXiv:1602.01499}{{\ttfamily
arXiv:1602.01499}}].

\bibitem{Hayes:2016qnu}
A.~C. Hayes and P.~Vogel, \emph{{Reactor Neutrino Spectra}},
{\emph{Ann.Rev.Nucl.Part.Sci.} {\bfseries 66} (2016) 219--244},
[\href{https://arxiv.org/abs/arXiv:1605.02047}{{\ttfamily
arXiv:1605.02047}}].

\bibitem{Mueller:2011nm}
T.~A. Mueller et~al., \emph{{Improved Predictions of Reactor Antineutrino
Spectra}}, {\emph{Phys. Rev.} {\bfseries C83} (2011) 054615},
[\href{https://arxiv.org/abs/arXiv:1101.2663}{{\ttfamily arXiv:1101.2663}}].

\bibitem{Huber:2016xis}
P.~Huber, \emph{{The 5 MeV bump - a nuclear whodunit mystery}}, {\emph{Phys.
Rev. Lett.} {\bfseries 118} (2017) 042502},
[\href{https://arxiv.org/abs/arXiv:1609.03910}{{\ttfamily
arXiv:1609.03910}}].

\bibitem{Kraus:2012he}
C.~Kraus, A.~Singer, K.~Valerius and C.~Weinheimer, \emph{{Limit on sterile
neutrino contribution from the Mainz Neutrino Mass Experiment}},
\href{http://dx.doi.org/10.1140/epjc/s10052-013-2323-z}{\emph{Eur.Phys.J.}
{\bfseries C73} (2013) 2323},
[\href{https://arxiv.org/abs/arXiv:1210.4194}{{\ttfamily arXiv:1210.4194}}].

\bibitem{Belesev:2012hx}
A.~Belesev, A.~Berlev, E.~Geraskin, A.~Golubev, N.~Likhovid et~al., \emph{{An
upper limit on additional neutrino mass eigenstate in 2 to 100 eV region from
'Troitsk nu-mass' data}},
\href{http://dx.doi.org/10.1134/S0021364013020033}{\emph{JETP Lett.}
{\bfseries 97} (2013) 67--69},
[\href{https://arxiv.org/abs/arXiv:1211.7193}{{\ttfamily arXiv:1211.7193}}].

\bibitem{Belesev:2013cba}
A.~Belesev et~al., \emph{{A search for an additional neutrino mass eigenstate
in 2 to 100 eV region from 'Troitsk nu-mass' data - detailed analysis}},
{\emph{J. Phys.} {\bfseries G41} (2014) 015001},
[\href{https://arxiv.org/abs/arXiv:1307.5687}{{\ttfamily arXiv:1307.5687}}].

\bibitem{Giunti:2007ry}
C.~Giunti and C.~W. Kim, \emph{{Fundamentals of Neutrino Physics and
Astrophysics}}.
Oxford University Press, Oxford, UK, 2007.

\bibitem{Giunti:2009xz}
C.~Giunti and Y.~Li, \emph{{Matter Effects in Active-Sterile Solar Neutrino
Oscillations}}, {\emph{Phys. Rev.} {\bfseries D80} (2009) 113007},
[\href{https://arxiv.org/abs/arXiv:0910.5856}{{\ttfamily arXiv:0910.5856}}].

\bibitem{Palazzo:2011rj}
A.~Palazzo, \emph{{Testing the very-short-baseline neutrino anomalies at the
solar sector}}, {\emph{Phys. Rev.} {\bfseries D83} (2011) 113013},
[\href{https://arxiv.org/abs/arXiv:1105.1705}{{\ttfamily arXiv:1105.1705}}].

\bibitem{Palazzo:2012yf}
A.~Palazzo, \emph{{An estimate of $\vartheta_{14}$ independent of reactor
antineutrino fluxes}}, {\emph{Phys. Rev.} {\bfseries D85} (2012) 077301},
[\href{https://arxiv.org/abs/arXiv:1201.4280}{{\ttfamily arXiv:1201.4280}}].

\bibitem{Palazzo:2013me}
A.~Palazzo, \emph{{Phenomenology of light sterile neutrinos: a brief review}},
{\emph{Mod.Phys.Lett.} {\bfseries A28} (2013) 1330004},
[\href{https://arxiv.org/abs/arXiv:1302.1102}{{\ttfamily arXiv:1302.1102}}].

\bibitem{Armbruster:1998uk}
{\scshape KARMEN} collaboration, B.~Armbruster et~al., \emph{{New experimental
limits on $\nu_e \to \nu_\tau$ oscillations in 2-$\nu$ and 3-$\nu$ mixing
schemes}}, {\emph{Phys. Rev.} {\bfseries C57} (1998) 3414--3424},
[\href{https://arxiv.org/abs/hep-ex/9801007}{{\ttfamily hep-ex/9801007}}].

\bibitem{Auerbach:2001hz}
{\scshape LSND} collaboration, L.~B. Auerbach et~al., \emph{{Measurements of
charged current reactions of nu/e on C-12}}, {\emph{Phys. Rev.} {\bfseries
C64} (2001) 065501}, [\href{https://arxiv.org/abs/hep-ex/0105068}{{\ttfamily
hep-ex/0105068}}].

\bibitem{Conrad:2011ce}
J.~Conrad and M.~Shaevitz, \emph{{Limits on Electron Neutrino Disappearance
from the KARMEN and LSND electron neutrino - Carbon Cross Section Data}},
{\emph{Phys. Rev.} {\bfseries D85} (2012) 013017},
[\href{https://arxiv.org/abs/arXiv:1106.5552}{{\ttfamily arXiv:1106.5552}}].

\bibitem{Abe:2014nuo}
{\scshape T2K} collaboration, K.~Abe et~al., \emph{{Search for short baseline
$\nu_e$ disappearance with the T2K near detector}}, {\emph{Phys. Rev.}
{\bfseries D91} (2015) 051102},
[\href{https://arxiv.org/abs/arXiv:1410.8811}{{\ttfamily arXiv:1410.8811}}].

\bibitem{Abe:2016nxk}
{\scshape Super-Kamiokande} collaboration, K.~Abe et~al., \emph{{Solar Neutrino
Measurements in Super-Kamiokande-IV}}, {\emph{Phys. Rev.} {\bfseries D94}
(2016) 052010}, [\href{https://arxiv.org/abs/arXiv:1606.07538}{{\ttfamily
arXiv:1606.07538}}].

\bibitem{Bellini:2013lnn}
{\scshape Borexino} collaboration, G.~Bellini et~al., \emph{{Final results of
Borexino Phase-I on low energy solar neutrino spectroscopy}}, {\emph{Phys.
Rev.} {\bfseries D89} (2014) 112007},
[\href{https://arxiv.org/abs/arXiv:1308.0443}{{\ttfamily arXiv:1308.0443}}].

\bibitem{Olive:2016xmw}
{\scshape Particle Data Group} collaboration, C.~Patrignani et~al.,
\emph{{Review of Particle Physics}},
\href{http://dx.doi.org/10.1088/1674-1137/40/10/100001}{\emph{Chin. Phys.}
{\bfseries C40} (2016) 100001}.

\bibitem{Alekseev:2016llm}
{\scshape DANSS} collaboration, I.~Alekseev et~al., \emph{{DANSS: Detector of
the reactor AntiNeutrino based on Solid Scintillator}}, {\emph{JINST}
{\bfseries 11} (2016) P11011},
[\href{https://arxiv.org/abs/arXiv:1606.02896}{{\ttfamily
arXiv:1606.02896}}].

\bibitem{Ashenfelter:2015uxt}
{\scshape PROSPECT} collaboration, J.~Ashenfelter et~al., \emph{{The PROSPECT
Physics Program}}, {\emph{J. Phys.} {\bfseries G43} (2016) 113001},
[\href{https://arxiv.org/abs/arXiv:1512.02202}{{\ttfamily
arXiv:1512.02202}}].

\bibitem{Michiels:2016qui}
{\scshape SoLid} collaboration, I.~Michiels, \emph{{SoLid: Search for
Oscillation with a 6Li Detector at the BR2 research reactor}},
\href{https://arxiv.org/abs/arXiv:1605.00215}{{\ttfamily arXiv:1605.00215}}.

\bibitem{Manzanillas:2017rta}
{\scshape STEREO} collaboration, L.~Manzanillas, \emph{{STEREO: Search for
sterile neutrinos at the ILL}},
\href{https://arxiv.org/abs/arXiv:1702.02498}{{\ttfamily arXiv:1702.02498}}.

\bibitem{Borexino:2013xxa}
{\scshape Borexino} collaboration, G.~Bellini et~al., \emph{{SOX: Short
distance neutrino Oscillations with BoreXino}}, {\emph{JHEP} {\bfseries 1308}
(2013) 038}, [\href{https://arxiv.org/abs/arXiv:1304.7721}{{\ttfamily
arXiv:1304.7721}}].

\bibitem{Barinov:2016znv}
V.~Barinov, V.~Gavrin, D.~Gorbunov and T.~Ibragimova, \emph{{BEST sensitivity
to O(1) eV sterile neutrino}}, {\emph{Phys. Rev.} {\bfseries D93} (2016)
073002}, [\href{https://arxiv.org/abs/arXiv:1602.03826}{{\ttfamily
arXiv:1602.03826}}].

\bibitem{Serebrov:2012sq}
{\scshape Neutrino-4} collaboration, A.~P. Serebrov et~al., \emph{{NEUTRINO-4
experiment: preparations for search for sterile neutrino at 100 MW reactor
SM-3 at 6-12 meters}},
\href{https://arxiv.org/abs/arXiv:1205.2955}{{\ttfamily arXiv:1205.2955}}.

\bibitem{Ryder:2015sma}
{\scshape SoLid} collaboration, N.~Ryder, \emph{{First results of the
deployment of a SoLid detector module at the SCK-CEN BR2 reactor}},
{\emph{PoS} {\bfseries EPS-HEP2015} (2015) 071},
[\href{https://arxiv.org/abs/arXiv:1510.07835}{{\ttfamily
arXiv:1510.07835}}].

\bibitem{Helaine:2016bmc}
{\scshape STEREO} collaboration, V.~Helaine, \emph{{Sterile neutrino search at
the ILL nuclear reactor: the STEREO experiment}},
\href{https://arxiv.org/abs/arXiv:1604.08877}{{\ttfamily arXiv:1604.08877}}.

\bibitem{Abs:2015tbh}
M.~Abs et~al., \emph{{IsoDAR@KamLAND: A Conceptual Design Report for the
Technical Facility}},
\href{https://arxiv.org/abs/arXiv:1511.05130}{{\ttfamily arXiv:1511.05130}}.

\bibitem{Ciuffoli:2015uta}
E.~Ciuffoli, J.~Evslin and F.~Zhao, \emph{{Neutrino Physics with Accelerator
Driven Subcritical Reactors}}, {\emph{JHEP} {\bfseries 01} (2016) 004},
[\href{https://arxiv.org/abs/arXiv:1509.03494}{{\ttfamily
arXiv:1509.03494}}].

\bibitem{Drexlin-NOW2016}
G.~Drexlin, \emph{{KATRIN}}, Talk presented at {NOW 2016, 4-11 September 2016,
Otranto, Lecce, Italy}.

\bibitem{Riis:2010zm}
A.~S. Riis and S.~Hannestad, \emph{{Detecting sterile neutrinos with KATRIN
like experiments}}, {\emph{JCAP} {\bfseries 1102} (2011) 011},
[\href{https://arxiv.org/abs/arXiv:1008.1495}{{\ttfamily arXiv:1008.1495}}].

\bibitem{Formaggio:2011jg}
J.~A. Formaggio and J.~Barrett, \emph{{Resolving the Reactor Neutrino Anomaly
with the KATRIN Neutrino Experiment}}, {\emph{Phys. Lett.} {\bfseries B706}
(2011) 68--71}, [\href{https://arxiv.org/abs/arXiv:1105.1326}{{\ttfamily
arXiv:1105.1326}}].

\bibitem{SejersenRiis:2011sj}
A.~S. Riis, S.~Hannestad and C.~Weinheimer, \emph{{Analysis of KATRIN data
using Bayesian inference}}, {\emph{Phys. Rev.} {\bfseries C84} (2011)
045503}, [\href{https://arxiv.org/abs/arXiv:1105.6005}{{\ttfamily
arXiv:1105.6005}}].

\bibitem{Esmaili:2012vg}
A.~Esmaili and O.~L.~G. Peres, \emph{{KATRIN Sensitivity to Sterile Neutrino
Mass in the Shadow of Lightest Neutrino Mass}}, {\emph{Phys. Rev.} {\bfseries
D85} (2012) 117301}, [\href{https://arxiv.org/abs/arXiv:1203.2632}{{\ttfamily
arXiv:1203.2632}}].

\bibitem{Gastaldo:2016kak}
L.~Gastaldo, C.~Giunti and E.~M. Zavanin, \emph{{Light sterile neutrino
sensitivity of 163Ho experiments}}, {\emph{JHEP} {\bfseries 1606} (2016)
061}, [\href{https://arxiv.org/abs/arXiv:1605.05497}{{\ttfamily
arXiv:1605.05497}}].

\bibitem{Borodovsky:1992pn}
{\scshape BNL-E776} collaboration, L.~Borodovsky et~al., \emph{{Search for
muon-neutrino oscillations $ \nu_\mu \to \nu_e $ ($ \bar\nu_\mu \to \bar\mu_e
$) in a wide band neutrino beam}}, {\emph{Phys. Rev. Lett.} {\bfseries 68}
(1992) 274--277}.

\bibitem{Armbruster:2002mp}
{\scshape KARMEN} collaboration, B.~Armbruster et~al., \emph{{Upper limits for
neutrino oscillations $\bar\nu_\mu\to\bar\nu_e$ from muon decay at rest}},
{\emph{Phys. Rev.} {\bfseries D65} (2002) 112001},
[\href{https://arxiv.org/abs/hep-ex/0203021}{{\ttfamily hep-ex/0203021}}].

\bibitem{Astier:2003gs}
{\scshape NOMAD} collaboration, P.~Astier et~al., \emph{{Search for $\nu_\mu
\to \nu_e$ Oscillations in the NOMAD Experiment}}, {\emph{Phys. Lett.}
{\bfseries B570} (2003) 19--31},
[\href{https://arxiv.org/abs/hep-ex/0306037}{{\ttfamily hep-ex/0306037}}].

\bibitem{Antonello:2013gut}
{\scshape ICARUS} collaboration, M.~Antonello et~al., \emph{{Search for
anomalies in the $\nu_e$ appearance from a $\nu_\mu$ beam}},
{\emph{Eur.Phys.J.} {\bfseries C73} (2013) 2599},
[\href{https://arxiv.org/abs/arXiv:1307.4699}{{\ttfamily arXiv:1307.4699}}].

\bibitem{Agafonova:2013xsk}
{\scshape OPERA} collaboration, N.~Agafonova et~al., \emph{{Search for
$\nu_\mu\to\nu_e$ oscillations with the OPERA experiment in the CNGS beam}},
{\emph{JHEP} {\bfseries 1307} (2013) 004},
[\href{https://arxiv.org/abs/arXiv:1303.3953}{{\ttfamily arXiv:1303.3953}}].

\bibitem{Dydak:1983zq}
{\scshape CDHSW} collaboration, F.~Dydak et~al., \emph{{A search for $\nu_\mu$
oscillations in the $\Delta m^2$ range $0.3-90 \, \mathrm{eV}^2$}},
{\emph{Phys. Lett.} {\bfseries B134} (1984) 281}.

\bibitem{Maltoni:2007zf}
M.~Maltoni and T.~Schwetz, \emph{{Sterile neutrino oscillations after first
MiniBooNE results}}, {\emph{Phys. Rev.} {\bfseries D76} (2007) 093005},
[\href{https://arxiv.org/abs/arXiv:0705.0107}{{\ttfamily arXiv:0705.0107}}].

\bibitem{Mahn:2011ea}
{\scshape SciBooNE-MiniBooNE} collaboration, K.~B.~M. Mahn et~al., \emph{{Dual
baseline search for muon neutrino disappearance at $0.5 < \Delta{m}^2 < 40 \,
\text{eV}^2$}}, {\emph{Phys. Rev.} {\bfseries D85} (2012) 032007},
[\href{https://arxiv.org/abs/arXiv:1106.5685}{{\ttfamily arXiv:1106.5685}}].

\bibitem{Cheng:2012yy}
{\scshape SciBooNE-MiniBooNE} collaboration, G.~Cheng et~al., \emph{{Dual
baseline search for muon antineutrino disappearance at $0.1 \text{eV}^2 <
\Delta{m}^2 < 100 \text{eV}^2$}}, {\emph{Phys. Rev.} {\bfseries D86} (2012)
052009}, [\href{https://arxiv.org/abs/arXiv:1208.0322}{{\ttfamily
arXiv:1208.0322}}].

\bibitem{GonzalezGarcia:2005xw}
M.~Gonzalez-Garcia, F.~Halzen and M.~Maltoni, \emph{{Physics Reach of
High-Energy and High-Statistics IceCube Atmospheric Neutrino Data}},
{\emph{Phys. Rev.} {\bfseries D71} (2005) 093010},
[\href{https://arxiv.org/abs/hep-ph/0502223}{{\ttfamily hep-ph/0502223}}].

\bibitem{Aartsen:2015rwa}
{\scshape IceCube} collaboration, M.~G. Aartsen et~al., \emph{{Evidence for
Astrophysical Muon Neutrinos from the Northern Sky with IceCube}},
{\emph{Phys. Rev. Lett.} {\bfseries 115} (2015) 081102},
[\href{https://arxiv.org/abs/arXiv:1507.04005}{{\ttfamily
arXiv:1507.04005}}].

\bibitem{Aartsen:2014muf}
{\scshape IceCube} collaboration, M.~G. Aartsen et~al., \emph{{Atmospheric and
Astrophysical Neutrinos above 1 TeV Interacting in IceCube}}, {\emph{Phys.
Rev.} {\bfseries D91} (2015) 022001},
[\href{https://arxiv.org/abs/arXiv:1410.1749}{{\ttfamily arXiv:1410.1749}}].

\bibitem{Delgado:2014kpa}
C.~A.~A. Delgado, J.~Salvado and C.~N. Weaver, \emph{{A Simple Quantum
Integro-Differential Solver (SQuIDS)}}, {\emph{Comput. Phys. Commun.}
{\bfseries 196} (2015) 569--591},
[\href{https://arxiv.org/abs/arXiv:1412.3832}{{\ttfamily arXiv:1412.3832}}].

\bibitem{Honda:1995hz}
M.~Honda, T.~Kajita, K.~Kasahara and S.~Midorikawa, \emph{{Calculation of the
flux of atmospheric neutrinos}}, {\emph{Phys. Rev.} {\bfseries D52} (1995)
4985--5005}, [\href{https://arxiv.org/abs/hep-ph/9503439}{{\ttfamily
hep-ph/9503439}}].

\bibitem{Honda:2004yz}
M.~Honda, T.~Kajita, K.~Kasahara and S.~Midorikawa, \emph{{A New calculation of
the atmospheric neutrino flux in a 3-dimensional scheme}}, {\emph{Phys. Rev.}
{\bfseries D70} (2004) 043008},
[\href{https://arxiv.org/abs/astro-ph/0404457}{{\ttfamily
astro-ph/0404457}}].

\bibitem{Sanuki:2006yd}
T.~Sanuki et~al., \emph{{Study of cosmic ray interaction model based on
atmospheric muons for the neutrino flux calculation}}, {\emph{Phys. Rev.}
{\bfseries D75} (2007) 043005},
[\href{https://arxiv.org/abs/astro-ph/0611201}{{\ttfamily
astro-ph/0611201}}].

\bibitem{Honda:2006qj}
M.~Honda et~al., \emph{{Calculation of atmospheric neutrino flux using the
interaction model calibrated with atmospheric muon data}}, {\emph{Phys. Rev.}
{\bfseries D75} (2007) 043006},
[\href{https://arxiv.org/abs/astro-ph/0611418}{{\ttfamily
astro-ph/0611418}}].

\bibitem{Gaisser:2013bla}
T.~K. Gaisser, T.~Stanev and S.~Tilav, \emph{{Cosmic Ray Energy Spectrum from
Measurements of Air Showers}},
\href{http://dx.doi.org/10.1007/s11467-013-0319-7}{\emph{Front.
Phys.(Beijing)} {\bfseries 8} (2013) 748--758},
[\href{https://arxiv.org/abs/arXiv:1303.3565}{{\ttfamily arXiv:1303.3565}}].

\bibitem{Dziewonski:1981xy}
A.~M. Dziewonski and D.~L. Anderson, \emph{{Preliminary reference earth
model}}, {\emph{Phys. Earth Planet. Interiors} {\bfseries 25} (1981)
297--356}.

\bibitem{Arguelles:2015wba}
C.~A. Arguelles, F.~Halzen, L.~Will, M.~Kroll and M.~H. Reno, \emph{{The
High-Energy Behavior of Photon, Neutrino and Proton Cross Sections}},
{\emph{Phys. Rev.} {\bfseries D92} (2015) 074040},
[\href{https://arxiv.org/abs/arXiv:1504.06639}{{\ttfamily
arXiv:1504.06639}}].

\bibitem{Arguelles:2015a}
C.~A. Arguelles, \emph{{New physics with atmospheric Neutrinos}}, {PhD thesis,
ISBN 978-1-339-06088-0}.

\bibitem{Jones:2015bya}
B.~J.~P. Jones, \emph{{Sterile Neutrinos in Cold Climates}}, {PhD thesis,
FERMILAB-THESIS-2015-17}.

\bibitem{Nelder:1965a}
J.~A. Nelder and R.~Mead, \emph{{A Simplex Method for Function Minimization}},
\href{http://dx.doi.org/10.1093/comjnl/7.4.308}{\emph{The Computer Journal}
{\bfseries 7} (1965) 308--313}.

\bibitem{Nunokawa:2003ep}
H.~Nunokawa, O.~L.~G. Peres and R.~Z. Funchal, \emph{{Probing the LSND scale
and four neutrino scenarios with a neutrino telescope}}, {\emph{Phys. Lett.}
{\bfseries B562} (2003) 279},
[\href{https://arxiv.org/abs/hep-ph/0302039}{{\ttfamily hep-ph/0302039}}].

\bibitem{Choubey:2007ji}
S.~Choubey, \emph{{Signature of sterile species in atmospheric neutrino data at
neutrino telescopes}}, {\emph{JHEP} {\bfseries 0712} (2007) 014},
[\href{https://arxiv.org/abs/arXiv:0709.1937}{{\ttfamily arXiv:0709.1937}}].

\bibitem{Razzaque:2011ab}
S.~Razzaque and A.~Y. Smirnov, \emph{{Searching for sterile neutrinos in ice}},
{\emph{JHEP} {\bfseries 1107} (2011) 084},
[\href{https://arxiv.org/abs/arXiv:1104.1390}{{\ttfamily arXiv:1104.1390}}].

\bibitem{Razzaque:2012tp}
S.~Razzaque and A.~Y. Smirnov, \emph{{Searches for sterile neutrinos with
IceCube DeepCore}}, {\emph{Phys. Rev.} {\bfseries D85} (2012) 093010},
[\href{https://arxiv.org/abs/arXiv:1203.5406}{{\ttfamily arXiv:1203.5406}}].

\bibitem{Esmaili:2012nz}
A.~Esmaili, F.~Halzen and O.~L.~G. Peres, \emph{{Constraining Sterile Neutrinos
with AMANDA and IceCube Atmospheric Neutrino Data}}, {\emph{JCAP} {\bfseries
1211} (2012) 041}, [\href{https://arxiv.org/abs/arXiv:1206.6903}{{\ttfamily
arXiv:1206.6903}}].

\bibitem{Esmaili:2013vza}
A.~Esmaili and A.~Y. Smirnov, \emph{{Restricting the LSND and MiniBooNE sterile
neutrinos with the IceCube atmospheric neutrino data}}, {\emph{JHEP}
{\bfseries 1312} (2013) 014},
[\href{https://arxiv.org/abs/arXiv:1307.6824}{{\ttfamily arXiv:1307.6824}}].

\bibitem{Esmaili:2013cja}
A.~Esmaili, F.~Halzen and O.~L.~G. Peres, \emph{{Exploring $\nu_{\tau}-\nu_{s}$
mixing with cascade events in DeepCore}}, {\emph{JCAP} {\bfseries 1307}
(2013) 048}, [\href{https://arxiv.org/abs/arXiv:1303.3294}{{\ttfamily
arXiv:1303.3294}}].

\bibitem{Lindner:2015iaa}
M.~Lindner, W.~Rodejohann and X.-J. Xu, \emph{{Sterile neutrinos in the light
of IceCube}}, {\emph{JHEP} {\bfseries 1601} (2016) 124},
[\href{https://arxiv.org/abs/arXiv:1510.00666}{{\ttfamily
arXiv:1510.00666}}].

\bibitem{Abe:2014gda}
{\scshape Super-Kamiokande} collaboration, K.~Abe et~al., \emph{{Limits on
Sterile Neutrino Mixing using Atmospheric Neutrinos in Super-Kamiokande}},
{\emph{Phys. Rev.} {\bfseries D91} (2015) 052019},
[\href{https://arxiv.org/abs/arXiv:1410.2008}{{\ttfamily arXiv:1410.2008}}].

\bibitem{Aartsen:2017bap}
{\scshape IceCube} collaboration, M.~G. Aartsen et~al., \emph{{Search for
sterile neutrino mixing using three years of IceCube DeepCore data}},
\href{https://arxiv.org/abs/arXiv:1702.05160}{{\ttfamily arXiv:1702.05160}}.

\bibitem{Adamson:2010wi}
{\scshape The MINOS} collaboration, P.~Adamson et~al., \emph{{Search for
sterile neutrino mixing in the MINOS long- baseline experiment}},
\href{http://dx.doi.org/10.1103/PhysRevD.81.052004}{\emph{Phys. Rev.}
{\bfseries D81} (2010) 052004},
[\href{https://arxiv.org/abs/arXiv:1001.0336}{{\ttfamily arXiv:1001.0336}}].

\bibitem{Adamson:2011ku}
{\scshape MINOS} collaboration, P.~Adamson et~al., \emph{{Active to sterile
neutrino mixing limits from neutral-current interactions in MINOS}},
{\emph{Phys. Rev. Lett.} {\bfseries 107} (2011) 011802},
[\href{https://arxiv.org/abs/arXiv:1104.3922}{{\ttfamily arXiv:1104.3922}}].

\bibitem{Agafonova:2015neo}
{\scshape OPERA} collaboration, N.~Agafonova et~al., \emph{{Limits on
muon-neutrino to tau-neutrino oscillations induced by a sterile neutrino
state obtained by OPERA at the CNGS beam}}, {\emph{JHEP} {\bfseries 1506}
(2015) 069}, [\href{https://arxiv.org/abs/arXiv:1503.01876}{{\ttfamily
arXiv:1503.01876}}].

\bibitem{Martini:2012fa}
M.~Martini, M.~Ericson and G.~Chanfray, \emph{{Neutrino energy reconstruction
problems and neutrino oscillations}}, {\emph{Phys. Rev.} {\bfseries D85}
(2012) 093012}, [\href{https://arxiv.org/abs/arXiv:1202.4745}{{\ttfamily
arXiv:1202.4745}}].

\bibitem{Martini:2012uc}
M.~Martini, M.~Ericson and G.~Chanfray, \emph{{Energy reconstruction effects in
neutrino oscillation experiments and implications for the analysis}},
{\emph{Phys. Rev.} {\bfseries D87} (2013) 013009},
[\href{https://arxiv.org/abs/arXiv:1211.1523}{{\ttfamily arXiv:1211.1523}}].

\bibitem{Ericson:2016yjn}
M.~Ericson, M.~V. Garzelli, C.~Giunti and M.~Martini, \emph{{Assessing the role
of nuclear effects in the interpretation of the MiniBooNE low-energy
anomaly}}, {\emph{Phys. Rev.} {\bfseries D93} (2016) 073008},
[\href{https://arxiv.org/abs/arXiv:1602.01390}{{\ttfamily
arXiv:1602.01390}}].

\bibitem{Antonello:2015lea}
{\scshape MicroBooNE, LAr1-ND, ICARUS-WA104} collaboration, R.~Acciarri et~al.,
\emph{{A Proposal for a Three Detector Short-Baseline Neutrino Oscillation
Program in the Fermilab Booster Neutrino Beam}},
\href{https://arxiv.org/abs/arXiv:1503.01520}{{\ttfamily arXiv:1503.01520}}.

\bibitem{Bhadra:2014oma}
{\scshape nuPRISM} collaboration, S.~Bhadra et~al., \emph{{Letter of Intent to
Construct a nuPRISM Detector in the J-PARC Neutrino Beamline}},
\href{https://arxiv.org/abs/arXiv:1412.3086}{{\ttfamily arXiv:1412.3086}}.

\bibitem{Harada:2013yaa}
{\scshape JSNS2} collaboration, M.~Harada et~al., \emph{{Proposal: A Search for
Sterile Neutrino at J-PARC Materials and Life Science Experimental
Facility}}, \href{https://arxiv.org/abs/arXiv:1310.1437}{{\ttfamily
arXiv:1310.1437}}.

\bibitem{Axani:2015zxa}
S.~N. Axani et~al., \emph{{KPipe: a decisive test for muon neutrino
disappearance}}, \href{https://arxiv.org/abs/arXiv:1510.06994}{{\ttfamily
arXiv:1510.06994}}.

\bibitem{Barry:2011wb}
J.~Barry, W.~Rodejohann and H.~Zhang, \emph{{Light Sterile Neutrinos: Models
and Phenomenology}}, {\emph{JHEP} {\bfseries 07} (2011) 091},
[\href{https://arxiv.org/abs/arXiv:1105.3911}{{\ttfamily arXiv:1105.3911}}].

\bibitem{Li:2011ss}
Y.~Li and S.~Liu, \emph{{Vanishing effective mass of the neutrinoless double
beta decay including light sterile neutrinos}}, {\emph{Phys. Lett.}
{\bfseries B706} (2012) 406--411},
[\href{https://arxiv.org/abs/arXiv:1110.5795}{{\ttfamily arXiv:1110.5795}}].

\bibitem{Rodejohann:2012xd}
W.~Rodejohann, \emph{{Neutrinoless double beta decay and neutrino physics}},
{\emph{J. Phys.} {\bfseries G39} (2012) 124008},
[\href{https://arxiv.org/abs/arXiv:1206.2560}{{\ttfamily arXiv:1206.2560}}].

\bibitem{Girardi:2013zra}
I.~Girardi, A.~Meroni and S.~T. Petcov, \emph{{Neutrinoless Double Beta Decay
in the Presence of Light Sterile Neutrinos}}, {\emph{JHEP} {\bfseries 1311}
(2013) 146}, [\href{https://arxiv.org/abs/arXiv:1308.5802}{{\ttfamily
arXiv:1308.5802}}].

\bibitem{Pascoli:2013fiz}
S.~Pascoli, M.~Mitra and S.~Wong, \emph{{The Effect of Cancellation in
Neutrinoless Double Beta Decay}}, {\emph{Phys. Rev.} {\bfseries D90} (2014)
093005}, [\href{https://arxiv.org/abs/arXiv:1310.6218}{{\ttfamily
arXiv:1310.6218}}].

\bibitem{Meroni:2014tba}
A.~Meroni and E.~Peinado, \emph{{The quest for neutrinoless double beta decay:
Pseudo-Dirac, Majorana and sterile neutrinos}}, {\emph{Phys. Rev.} {\bfseries
D90} (2014) 053002}, [\href{https://arxiv.org/abs/arXiv:1406.3990}{{\ttfamily
arXiv:1406.3990}}].

\bibitem{Abada:2014nwa}
A.~Abada, V.~D. Romeri and A.~Teixeira, \emph{{Effect of steriles states on
lepton magnetic moments and neutrinoless double beta decay}}, {\emph{JHEP}
{\bfseries 1409} (2014) 074},
[\href{https://arxiv.org/abs/arXiv:1406.6978}{{\ttfamily arXiv:1406.6978}}].

\bibitem{Giunti:2015kza}
C.~Giunti and E.~M. Zavanin, \emph{{Predictions for Neutrinoless Double-Beta
Decay in the 3+1 Sterile Neutrino Scenario}}, {\emph{JHEP} {\bfseries 07}
(2015) 171}, [\href{https://arxiv.org/abs/arXiv:1505.00978}{{\ttfamily
arXiv:1505.00978}}].

\bibitem{Pas:2015eia}
H.~Pas and W.~Rodejohann, \emph{{Neutrinoless Double Beta Decay}}, {\emph{New
J. Phys.} {\bfseries 17} (2015) 115010},
[\href{https://arxiv.org/abs/arXiv:1507.00170}{{\ttfamily
arXiv:1507.00170}}].

\bibitem{Hannestad:2012ky}
S.~Hannestad, I.~Tamborra and T.~Tram, \emph{{Thermalisation of light sterile
neutrinos in the early universe}}, {\emph{JCAP} {\bfseries 1207} (2012) 025},
[\href{https://arxiv.org/abs/arXiv:1204.5861}{{\ttfamily arXiv:1204.5861}}].

\bibitem{Mirizzi:2012we}
A.~Mirizzi, N.~Saviano, G.~Miele and P.~D. Serpico, \emph{{Light sterile
neutrino production in the early universe with dynamical neutrino
asymmetries}},
\href{http://dx.doi.org/10.1103/PhysRevD.86.053009}{\emph{Phys. Rev.}
{\bfseries D86} (2012) 053009},
[\href{https://arxiv.org/abs/arXiv:1206.1046}{{\ttfamily arXiv:1206.1046}}].

\bibitem{Saviano:2013ktj}
N.~Saviano et~al., \emph{{Multi-momentum and multi-flavour active-sterile
neutrino oscillations in the early universe: role of neutrino asymmetries and
effects on nucleosynthesis}}, {\emph{Phys. Rev.} {\bfseries D87} (2013)
073006}, [\href{https://arxiv.org/abs/arXiv:1302.1200}{{\ttfamily
arXiv:1302.1200}}].

\bibitem{Hannestad:2013pha}
S.~Hannestad, R.~S. Hansen and T.~Tram, \emph{{Can active-sterile neutrino
oscillations lead to chaotic behavior of the cosmological lepton
asymmetry?}}, {\emph{JCAP} {\bfseries 1304} (2013) 032},
[\href{https://arxiv.org/abs/arXiv:1302.7279}{{\ttfamily arXiv:1302.7279}}].

\bibitem{Hannestad:2013ana}
S.~Hannestad, R.~S. Hansen and T.~Tram, \emph{{How secret interactions can
reconcile sterile neutrinos with cosmology}}, {\emph{Phys. Rev. Lett.}
{\bfseries 112} (2014) 031802},
[\href{https://arxiv.org/abs/arXiv:1310.5926}{{\ttfamily arXiv:1310.5926}}].

\bibitem{Dasgupta:2013zpn}
B.~Dasgupta and J.~Kopp, \emph{{A menage a trois of eV-scale sterile neutrinos,
cosmology, and structure formation}}, {\emph{Phys. Rev. Lett.} {\bfseries
112} (2014) 031803}, [\href{https://arxiv.org/abs/arXiv:1310.6337}{{\ttfamily
arXiv:1310.6337}}].

\bibitem{Mirizzi:2014ama}
A.~Mirizzi, G.~Mangano, O.~Pisanti and N.~Saviano, \emph{{Tension between
secret sterile neutrino interactions and cosmological neutrino mass bounds}},
{\emph{Phys. Rev.} {\bfseries D91} (2015) 025019},
[\href{https://arxiv.org/abs/arXiv:1410.1385}{{\ttfamily arXiv:1410.1385}}].

\bibitem{Saviano:2014esa}
N.~Saviano, O.~Pisanti, G.~Mangano and A.~Mirizzi, \emph{{Unveiling secret
interactions among sterile neutrinos with big-bang nucleosynthesis}},
{\emph{Phys. Rev.} {\bfseries D90} (2014) 113009},
[\href{https://arxiv.org/abs/arXiv:1409.1680}{{\ttfamily arXiv:1409.1680}}].

\bibitem{Forastieri:2015paa}
F.~Forastieri, M.~Lattanzi and P.~Natoli, \emph{{Constraints on secret neutrino
interactions after Planck}}, {\emph{JCAP} {\bfseries 1507} (2015) 014},
[\href{https://arxiv.org/abs/arXiv:1504.04999}{{\ttfamily
arXiv:1504.04999}}].

\bibitem{Chu:2015ipa}
X.~Chu, B.~Dasgupta and J.~Kopp, \emph{{Sterile Neutrinos with Secret
Interactions - Lasting Friendship with Cosmology}}, {\emph{JCAP} {\bfseries
1510} (2015) 011}, [\href{https://arxiv.org/abs/arXiv:1505.02795}{{\ttfamily
arXiv:1505.02795}}].

\bibitem{Archidiacono:2016kkh}
M.~Archidiacono et~al., \emph{{Pseudoscalar - sterile neutrino interactions:
reconciling the cosmos with neutrino oscillations}}, {\emph{JCAP} {\bfseries
1608} (2016) 067}, [\href{https://arxiv.org/abs/arXiv:1606.07673}{{\ttfamily
arXiv:1606.07673}}].

\bibitem{Maltoni:2003cu}
M.~Maltoni and T.~Schwetz, \emph{{Testing the statistical compatibility of
independent data sets}}, {\emph{Phys. Rev.} {\bfseries D68} (2003) 033020},
[\href{https://arxiv.org/abs/hep-ph/0304176}{{\ttfamily hep-ph/0304176}}].

\end{thebibliography}

\clearpage

\begin{table}[!t]
\centering
\resizebox{\textwidth}{!}{
\begin{tabular}{D{.}{.}{2.0}*{6}{c}D{.}{.}{2.1}*{3}{c}}
\multicolumn{1}{c}{$a$}
&
Experiment
&
$f^{a}_{235}$
&
$f^{a}_{238}$
&
$f^{a}_{239}$
&
$f^{a}_{241}$
&
$R_{a}^{\text{exp}}$
&
\multicolumn{1}{c}{$\sigma_{a}^{\text{exp}}$ [\%]}
&
$\sigma_{a}^{\text{cor}}$ [\%]
&
$\sigma_{a}^{\text{the}}$ [\%]
&
$L_{a}$ [m]
\\
\toprule
$1$	&Bugey-4		&$0.538$	&$0.078$	&$0.328$	&$0.056$	&$0.932$	&$1.4$	&\rdelim\}{2}{20pt}[1.4]	&$2.5$	&$15$\\
$2$	&Rovno91		&$0.606$	&$0.074$	&$0.277$	&$0.043$	&$0.930$	&$2.8$	&                       	&$2.4$	&$18$\\
\midrule
$3$	&Rovno88-1I		&$0.607$	&$0.074$	&$0.277$	&$0.042$	&$0.907$	&$6.4$	&\rdelim\}{2}{20pt}[3.1] \rdelim\}{5}{20pt}[2.2]	&$2.4$	&$18$\\
$4$	&Rovno88-2I		&$0.603$	&$0.076$	&$0.276$	&$0.045$	&$0.938$	&$6.4$	&                                               	&$2.4$	&$18$\\
$5$	&Rovno88-1S		&$0.606$	&$0.074$	&$0.277$	&$0.043$	&$0.962$	&$7.3$	&\rdelim\}{3}{45pt}[3.1]                        	&$2.4$	&$18$\\
$6$	&Rovno88-2S		&$0.557$	&$0.076$	&$0.313$	&$0.054$	&$0.949$	&$7.3$	&                                               	&$2.5$	&$25$\\
$7$	&Rovno88-2S		&$0.606$	&$0.074$	&$0.274$	&$0.046$	&$0.928$	&$6.8$	&                                               	&$2.4$	&$18$\\
\midrule
$8$	&Bugey-3-15		&$0.538$	&$0.078$	&$0.328$	&$0.056$	&$0.936$	&$4.2$	&\rdelim\}{3}{20pt}[4.0]                        	&$2.5$	&$15$\\
$9$	&Bugey-3-40		&$0.538$	&$0.078$	&$0.328$	&$0.056$	&$0.942$	&$4.3$	&                                               	&$2.5$	&$40$\\
$10$	&Bugey-3-95		&$0.538$	&$0.078$	&$0.328$	&$0.056$	&$0.867$	&$15.2$	&                                               	&$2.5$	&$95$\\
\midrule
$11$	&Gosgen-38		&$0.619$	&$0.067$	&$0.272$	&$0.042$	&$0.955$	&$5.4$	&\rdelim\}{3}{20pt}[2.0] \rdelim\}{4}{20pt}[3.8]	&$2.4$	&$37.9$\\
$12$	&Gosgen-46		&$0.584$	&$0.068$	&$0.298$	&$0.050$	&$0.981$	&$5.4$	&                                               	&$2.4$	&$45.9$\\
$13$	&Gosgen-65		&$0.543$	&$0.070$	&$0.329$	&$0.058$	&$0.915$	&$6.7$	&                                               	&$2.4$	&$64.7$\\
$14$	&ILL			&$1$	&$0$	&$0$	&$0$	&$0.792$	&$9.1$	&                                               	&$2.4$	&$8.76$\\
\midrule
$15$	&Krasnoyarsk87-33	&$1$	&$0$	&$0$	&$0$	&$0.925$	&$5.0$	&\rdelim\}{2}{20pt}[4.1]	&$2.4$	&$32.8$\\
$16$	&Krasnoyarsk87-92	&$1$	&$0$	&$0$	&$0$	&$0.942$	&$20.4$	&                       	&$2.4$	&$92.3$\\
$17$	&Krasnoyarsk94-57	&$1$	&$0$	&$0$	&$0$	&$0.936$	&$4.2$	&0                      	&$2.4$	&$57$\\
$18$	&Krasnoyarsk99-34	&$1$	&$0$	&$0$	&$0$	&$0.946$	&$3.0$	&0                      	&$2.4$	&$34$\\
\midrule
$19$	&SRP-18		&$1$	&$0$	&$0$	&$0$	&$0.941$	&$2.8$	&0	&$2.4$	&$18.2$\\
$20$	&SRP-24		&$1$	&$0$	&$0$	&$0$	&$1.006$	&$2.9$	&0	&$2.4$	&$23.8$\\
\midrule
$21$	&Nucifer		&$0.926$	&$0.061$	&$0.008$	&$0.005$	&$1.014$	&$10.7$	&0	&$2.3$	&$7.2$\\
$22$	&Chooz			&$0.496$	&$0.087$	&$0.351$	&$0.066$	&$0.996$	&$3.2$	&0	&$2.5$	&$\approx 1000$\\
$23$	&Palo Verde		&$0.600$	&$0.070$	&$0.270$	&$0.060$	&$0.997$	&$5.4$	&0	&$2.4$	&$\approx 800$\\
$24$	&Daya Bay		&$0.561$	&$0.076$	&$0.307$	&$0.056$	&$0.946$	&$2.0$	&0	&$2.5$	&$\approx 550$\\
$25$	&RENO			&$0.569$	&$0.073$	&$0.301$	&$0.056$	&$0.944$	&$2.2$	&0	&$2.4$	&$\approx 411$\\
$26$	&Double Chooz		&$0.511$	&$0.087$	&$0.340$	&$0.062$	&$0.935$	&$1.4$	&0	&$2.5$	&$\approx 415$\\
\bottomrule
\end{tabular}
}
\caption{ \label{tab:rat}
List of the experiments which measured the absolute reactor antineutrino flux.
For each experiment numbered with the index $a$,
the index $k = 235, 238, 239, 241$
indicate the four fissionable isotopes
$^{235}\text{U}$,
$^{238}\text{U}$,
$^{239}\text{Pu}$, and
$^{241}\text{Pu}$,
$f^{a}_{k}$ are the fission fractions,
$R_{a}^{\text{exp}}$ is the ratio of measured and predicted rates,
$\sigma_{a}^{\text{exp}}$ is the corresponding relative experimental uncertainty,
$\sigma_{a}^{\text{cor}}$ is the relative systematic uncertainty
which is correlated in each group of experiments indicated by the braces,
$\sigma_{a}^{\text{the}}$ is the relative theoretical uncertainty
which is correlated among all the experiments,
and
$L_{a}$ is the source-detector distance.
}
\end{table}

\begin{table}[!t]
\centering
\begin{tabular}{crr}
$k$
&
\multicolumn{1}{c}{$\sigma_{f,k}^{\text{S}}$}
&
\multicolumn{1}{c}{$\sigma_{f,k}^{\text{SH}}$}
\\
\toprule
$235$ & $ 6.61 \pm 2.11\%$ & $ 6.69 \pm 2.44\%$ \\
$238$ & $10.10 \pm 8.15\%$ & $10.10 \pm 8.15\%$ \\
$239$ & $ 4.34 \pm 2.45\%$ & $ 4.40 \pm 2.88\%$ \\
$241$ & $ 5.97 \pm 2.15\%$ & $ 6.03 \pm 2.60\%$ \\
\bottomrule
\end{tabular}
\caption{ \label{tab:csf}
Cross sections per fission of the four fissionable isotopes calculated
by the Saclay (S) group ($\sigma_{f,k}^{\text{S}}$)
in Ref.~\cite{Mention:2011rk}
and those obtained from the Huber (SH) correction ($\sigma_{f,k}^{\text{SH}}$)
in Ref.~\cite{Huber:2011wv}.
The units are $10^{-43} \, \text{cm}^2 / \text{fission}$.
The index $k = 235, 238, 239, 241$
indicates the four isotopes
$^{235}\text{U}$,
$^{238}\text{U}$,
$^{239}\text{Pu}$, and
$^{241}\text{Pu}$.
}
\end{table}

\begin{table}[!t]
\centering
\begin{tabular}{c|llll}
&
\multicolumn{1}{c}{$^{235}\text{U}$}
&
\multicolumn{1}{c}{$^{238}\text{U}$}
&
\multicolumn{1}{c}{$^{239}\text{Pu}$}
&
\multicolumn{1}{c}{$^{241}\text{Pu}$}
\\
\hline
$^{235}\text{U}$  & $0.0267$ & $0$ & $0.0203$ & $0.0255$ \\
$^{238}\text{U}$  & $0$ & $0.6776$ & $0$ & $0$ \\
$^{239}\text{Pu}$ & $0.0203$ & $0$ & $0.0161$ & $0.0194$ \\
$^{241}\text{Pu}$ & $0.0255$ & $0$ & $0.0194$ & $0.0246$ \\
\end{tabular}
\caption{ \label{tab:cov}
Covariance matrix of the
cross sections per fission of the four fissionable isotopes.
}
\end{table}

\begin{table}[!t]
\centering
\begin{tabular}[t]{c}
\\
\toprule
$\chi^{2}_{\text{min}}$\\
NDF\\
GoF\\
$\Delta{m}^2_{41}$\\
$\sin^22\vartheta_{ee}$\\
$|U_{e4}|^2$\\
$\Delta\chi^{2}_{\text{NO}}$\\
$n\sigma_{\text{NO}}$\\
\bottomrule%
\end{tabular}%
\begin{tabular}[t]{c}
Rea:Rat\\
\toprule
$12.4$\\
$24$\\
$100\%$\\
$0.48$\\
$0.14$\\
$0.037$\\
$13.1$\\
$3.2$\\
\bottomrule%
\end{tabular}%
\begin{tabular}[t]{c}
Rea:Spe\\
\toprule
$73.9$\\
$82$\\
$73\%$\\
$1.7$\\
$0.050$\\
$0.013$\\
$6.4$\\
$2.1$\\
\bottomrule%
\end{tabular}%
\begin{tabular}[t]{c}
Rea:Rat+Spe\\
\toprule
$94.6$\\
$108$\\
$82\%$\\
$1.7$\\
$0.062$\\
$0.016$\\
$11.3$\\
$2.9$\\
\bottomrule%
\end{tabular}%
\begin{tabular}[t]{c}
Rea+Gal\\
\toprule
$107.1$\\
$112$\\
$61\%$\\
$3.0$\\
$0.14$\\
$0.036$\\
$16.0$\\
$3.6$\\
\bottomrule%
\end{tabular}%
\begin{tabular}[t]{c}
$\nu_{e}$Dis\\
\toprule
$163.0$\\
$174$\\
$71\%$\\
$1.7$\\
$0.066$\\
$0.017$\\
$14.1$\\
$3.3$\\
\bottomrule%
\end{tabular}%
\begin{tabular}[t]{c}
$\nu_{e}$Dis+$\beta$\\
\toprule
$163.1$\\
$176$\\
$75\%$\\
$1.7$\\
$0.066$\\
$0.017$\\
$14.0$\\
$3.3$\\
\bottomrule%
\end{tabular}%
\caption{ \label{tab:nuedis}
Results of the fits of $\nu_{e}$ and $\bar\nu_{e}$ disappearance data:
minimum $\chi^2$ ($\chi^{2}_{\text{min}}$),
number of degrees of freedom (NDF),
goodness of fit (GoF),
best fit values of
$\Delta{m}^2_{41}$,
$\sin^22\vartheta_{ee}$, and
$|U_{e4}|^2$,
$\chi^{2}$ difference $\Delta\chi^{2}_{\text{NO}}$ between the $\chi^{2}$ of no oscillations and $\chi^{2}_{\text{min}}$, and
the resulting number of $\sigma$'s ($n\sigma_{\text{NO}}$)
for two degrees of freedom corresponding to two fitted parameters
($\Delta{m}^2_{41}$ and $\sin^22\vartheta_{ee}$).
The columns correspond to the fits of the data of
reactor rates (Rea:Rat),
reactor spectra (Rea:Spe),
reactor rates and spectra (Rea:Rat+Spe),
reactor and Gallium data (Rea+Gal),
$\nu_{e}$ and $\bar\nu_{e}$ disappearance data ($\nu_{e}$Dis),
$\nu_{e}$ and $\bar\nu_{e}$ disappearance data and $\beta$ decay constraints ($\nu_{e}$Dis+$\beta$).
}
\end{table}

\begin{table}[!t]
\centering
\begin{tabular}[t]{c}
\\
\toprule
$\chi^{2}_{\text{min}}$\\
NDF\\
GoF\\
$\Delta{m}^2_{41}$\\
$|U_{e4}|^2$\\
$|U_{\mu4}|^2$\\
$\sin^22\vartheta_{e\mu}$\\
$\sin^22\vartheta_{ee}$\\
$\sin^22\vartheta_{\mu\mu}$\\
\midrule
$\Delta\chi^{2}_{\text{NO}}$\\
$\text{NDF}_{\text{NO}}$\\
$n\sigma_{\text{NO}}$\\
\midrule
$(\chi^{2}_{\text{min}})_{\text{App}}$\\
$\text{NDF}_{\text{App}}$\\
$\text{GoF}_{\text{App}}$\\
$\Delta{m}^2_{41}$\\
$\sin^22\vartheta_{e\mu}$\\
\midrule
$(\chi^{2}_{\text{min}})_{\text{Dis}}$\\
$\text{NDF}_{\text{Dis}}$\\
$\text{GoF}_{\text{Dis}}$\\
$\Delta{m}^2_{41}$\\
$|U_{e4}|^2$\\
$|U_{\mu4}|^2$\\
$\sin^22\vartheta_{e\mu}$\\
$\sin^22\vartheta_{ee}$\\
$\sin^22\vartheta_{\mu\mu}$\\
\midrule
$\Delta\chi^{2}_{\text{PG}}$\\
$\text{NDF}_{\text{PG}}$\\
$\text{GoF}_{\text{PG}}$\\
\bottomrule%
\end{tabular}%
\begin{tabular}[t]{c}
Glo16A\\
\toprule
$288.4$\\
$250$\\
$4.8\%$\\
$1.6$\\
$0.027$\\
$0.015$\\
$0.0015$\\
$0.10$\\
$0.058$\\
\midrule
$53.1$\\
$3$\\
$6.7$\\
\midrule
$94.3$\\
$84$\\
$21\%$\\
$0.61$\\
$0.0058$\\
\midrule
$180.8$\\
$163$\\
$16\%$\\
$1.7$\\
$0.025$\\
$0.011$\\
$0.0011$\\
$0.097$\\
$0.042$\\
\midrule
$13.4$\\
$2$\\
$0.13\%$\\
\bottomrule%
\end{tabular}%
\begin{tabular}[t]{c}
Glo16B\\
\toprule
$556.9$\\
$525$\\
$16\%$\\
$1.6$\\
$0.028$\\
$0.014$\\
$0.0015$\\
$0.11$\\
$0.054$\\
\midrule
$51.9$\\
$4$\\
$6.4$\\
\midrule
$94.3$\\
$84$\\
$21\%$\\
$0.61$\\
$0.0058$\\
\midrule
$448.3$\\
$439$\\
$37\%$\\
$1.7$\\
$0.025$\\
$0.0088$\\
$0.00086$\\
$0.097$\\
$0.035$\\
\midrule
$14.4$\\
$2$\\
$0.075\%$\\
\bottomrule%
\end{tabular}%
\begin{tabular}[t]{c}
Glo17\\
\toprule
$622.1$\\
$585$\\
$14\%$\\
$1.7$\\
$0.021$\\
$0.016$\\
$0.0013$\\
$0.080$\\
$0.062$\\
\midrule
$51.7$\\
$4$\\
$6.4$\\
\midrule
$94.3$\\
$84$\\
$21\%$\\
$0.61$\\
$0.0058$\\
\midrule
$510.6$\\
$499$\\
$35\%$\\
$1.7$\\
$0.017$\\
$0.0073$\\
$0.00048$\\
$0.065$\\
$0.029$\\
\midrule
$17.2$\\
$2$\\
$0.019\%$\\
\bottomrule%
\end{tabular}%
\begin{tabular}[t]{c}
PrGlo17\\
\toprule
$595.1$\\
$579$\\
$31\%$\\
$1.7$\\
$0.020$\\
$0.015$\\
$0.0012$\\
$0.079$\\
$0.058$\\
\midrule
$47.4$\\
$4$\\
$6.1$\\
\midrule
$77.3$\\
$78$\\
$50\%$\\
$0.97$\\
$0.0026$\\
\midrule
$510.6$\\
$499$\\
$35\%$\\
$1.7$\\
$0.017$\\
$0.0073$\\
$0.00048$\\
$0.065$\\
$0.029$\\
\midrule
$7.2$\\
$2$\\
$2.7\%$\\
\bottomrule%
\end{tabular}%
\caption{ \label{tab:all}
Results of the 3+1 global
Glo16A,
Glo16B,
Glo17, and
PrGlo17
fits of SBL data
discussed,
respectively,
in subsections
\ref{sub:Glo16A},
\ref{sub:Glo16B},
\ref{sub:Glo17}, and
\ref{sub:PrGlo17}.
The first group of rows gives:
the minimum $\chi^2$ ($\chi^{2}_{\text{min}}$),
the number of degrees of freedom (NDF),
the goodness of fit (GoF),
the best fit values of the mixing parameters
$\Delta{m}^2_{41}$,
$|U_{e4}|^2$,
$|U_{\mu4}|^2$,
and of the oscillation amplitudes
$\sin^22\vartheta_{e\mu}$,
$\sin^22\vartheta_{ee}$,
$\sin^22\vartheta_{\mu\mu}$.
The second group of rows gives the
$\chi^{2}$ difference $\Delta\chi^{2}_{\text{NO}}$ between the $\chi^{2}$ of no oscillations and $\chi^{2}_{\text{min}}$ and
the resulting number of $\sigma$'s ($n\sigma_{\text{NO}}$)
for $\text{NDF}_{\text{NO}}$ degrees of freedom corresponding to the number of fitted parameters.
The third and fourth group of rows give,
respectively,
the results of different 3+1 fits of appearance (App) and disappearance (Dis) data.
The fifth group of rows gives
the results for the appearance-disappearance parameter goodness of fit
\cite{Maltoni:2003cu}:
the $\chi^2$ difference
$\Delta\chi^{2}_{\text{PG}}$
and the resulting goodness of fit
$\text{GoF}_{\text{PG}}$
for $\text{NDF}_{\text{PG}}$ degrees of freedom.
}
\end{table}

\begin{table}[t]
\begin{center}
\begin{tabular}{cccc}
CL
&
$|U_{e4}|^2$
&
$|U_{\mu4}|^2$
&
$|U_{\tau4}|^2$
\\
\toprule
68.27\% ($1\sigma$)
&
$ 0.016 - 0.024 $
&
$ 0.011 - 0.018 $
&
$\lesssim$
0.0032
\\
95.45\% ($2\sigma$)
&
$ 0.013 - 0.028 $
&
$ 0.0083 - 0.022 $
&
$\lesssim$
0.018
\\
99.73\% ($3\sigma$)
&
$ 0.0098 - 0.031 $
&
$ 0.0060 - 0.026 $
&
$\lesssim$
0.039
\\
\bottomrule
\end{tabular}
\end{center}
\caption{ \label{tab:prgint}
Marginal allowed intervals of the mixing parameters
$|U_{e4}|^2$,
$|U_{\mu4}|^2$, and
$|U_{\tau4}|^2$
obtained in the pragmatic 3+1 global fit
``PrGlo17''
of SBL data.
}
\end{table}

\clearpage

\end{document}